\documentclass[10pt,a4paper]{article}
\usepackage{empheq}
\usepackage[utf8]{inputenc}
\usepackage{epigraph}
\usepackage{amsmath}
\usepackage{wrapfig}
\usepackage{epsfig}
\usepackage{amssymb}
\usepackage{graphicx}
\usepackage{slashed}
\usepackage{indentfirst}
\usepackage{jheppub}
\usepackage{hyperref}
\usepackage{url}
\usepackage{epstopdf}
\usepackage{scalerel}
\usepackage{pdfpages}
\usepackage{empheq}
\usepackage{fancyvrb}

\DeclareMathAlphabet{\mathpzc}{OT1}{pzc}{m}{it}

\newcommand{\sym}{$\mathcal{N}=4$ SYM}

\newcommand{\beq}{\begin{equation}}
\newcommand{\eeq}{\end{equation}}
\newcommand{\beqq}{\begin{equation*}}
\newcommand{\eeqq}{\end{equation*}}
\newcommand\beqa{\begin{eqnarray}}
\newcommand\eeqa{\end{eqnarray}}
\newcommand\beqaa{\begin{eqnarray*}}
\newcommand\eeqaa{\end{eqnarray*}}
\newcommand\bea{\begin{array}}
\newcommand\eea{\end{array}}
\newcommand\beaa{\begin{array}}
\newcommand\eeaa{\end{array}}

\def\XXint#1#2#3{{\setbox0=\hbox{$#1{#2#3}{\int}$ }
\vcenter{\hbox{$#2#3$ }}\kern-.5\wd0}}

\def\hid#1{}

\newcommand{\im}{{\rm Im}\;}

\def\XXint#1#2#3{{\setbox0=\hbox{$#1{#2#3}{\int}$}
\vcenter{\hbox{$#2#3$}}\kern-.5\wd0}}

\newcommand{\neqa}{\nonumber\end{eqnarray}}
\newcommand{\la}[1]{\label{#1}}

\newcommand{\eq}[1]{(\ref{#1})}

\newcommand{\hs}{\frac{\sqrt{3}}{2}}
\renewcommand{\d}{\partial}

\newcommand{\<}{{\langle}}
\renewcommand{\>}{{\rangle}}

\newcommand{\re}{\relax{\rm I\kern-.18em R}}

\renewcommand{\sp}{p\hspace{-.40em}/}

\def\su2{{SU(2)}}

\def\[{\left[}
\def\]{\right]}

\def\e{\epsilon}

\def\s{\sigma}

\def\({\left(}
\def\){\right)}
\def\[{\left[}
\def\]{\right]}

\def\<{\langle}
\def\>{\rangle}

\def\cO{{\cal O}}

\def\s*{\ *_{\!\!\!\!\!\!\!\!\!\,_{\,_\text{\scriptsize{sym}}}}}
\def\hs*{\ \hat{*}_{\!\!\!\!\!\!\!\!\!\,_{\,_\text{\scriptsize{sym}}}}}
\def\d{\partial}

\def\i2{\frac{i}{2}}

\def\bQ{{\bf Q}}
\def\bP{{\bf P}}

\def\bq{{\bf q}}

\def\spi{\relax{\rm \pi\kern-0.5em /}}
\def\sA{\relax{\rm A\kern-0.5em /}}
\def\sp{\relax{\rm p\kern-0.5em /}}
\def\sd{\relax{\rm \d\kern-0.5em /}}
\def\sk{\relax{\rm k\kern-0.5em /}}
\def\sn{\relax{\rm n\kern-0.5em /}}
\def\sl{\relax{\rm l\kern-0.5em /}}
\def\sP{\relax{\rm P\kern-0.7em /}}
\def\sBethe{\relax{\rm \Bethe\kern-0.5em /}}
\def\cN{{\cal N}}

\newcommand{\cQ}{\mathcal{Q}}

\newcommand{\bbC}{\mathbb{C}}

\renewcommand{\Im}{{\rm Im}}
\renewcommand{\Re}{{\rm Re}}

\def\d{\partial}

\VerbatimFootnotes

\makeatother

\author[a,b,c]{Mikhail Alfimov,}
\author[d,e]{Nikolay Gromov,}
\author[c]{Grigory Sizov}

\affiliation[a]{National Research University Higher School of Economics, 6 Usacheva str., 119048 Moscow, Russia}
\affiliation[b]{Lebedev Physical Institute of Russian Academy of Sciences, 53 Leninskiy prospekt, 119991 Moscow, Russia}
\affiliation[c]{Ecole Normale Superieure, 45 Rue d'Ulm, 75005 Paris, France}
\affiliation[d]{Mathematics Department, King's College London, The Strand, London WC2R 2LS, UK}
\affiliation[e]{St.Petersburg INP, Gatchina, 188300, St.Petersburg, Russia}

\emailAdd{malfimov@hse.ru}
\emailAdd{nikgromov@gmail.com}
\emailAdd{grisha.sizov@gmail.com}

\abstract{
We developed a general non-perturbative framework for the BFKL spectrum of planar $\cN=4$ SYM, based on the Quantum Spectral Curve (QSC). It allows one to study the spectrum in the whole generality, extending previously known methods to  arbitrary values of conformal spin $n$. 
We show how to apply our approach to reproduce all known perturbative results for the Balitsky-Fadin-Kuraev-Lipatov (BFKL) Pomeron eigenvalue and get new predictions. In particular, we re-derived the Faddeev-Korchemsky Baxter equation for the Lipatov spin chain with non-zero conformal spin reproducing the corresponding BFKL kernel eigenvalue. We also get new non-perturbative analytic results for the Pomeron eigenvalue in the vicinity of $|n|=1,\;\Delta=0$ point and we obtained an explicit formula for the BFKL intercept function for arbitrary conformal spin up to the 3-loop order in the small coupling expansion and partial result at the 4-loop order. In addition, we implemented the numerical algorithm of \cite{Gromov:2015wca} as an auxiliary file to this arXiv submission. From the numerical result we managed to deduce an analytic formula for the strong coupling expansion of the intercept function for arbitrary conformal spin.
}

\title{BFKL Spectrum of \sym: non-Zero Conformal Spin}

\preprint{FIAN/TD-01/18 \\ \rightline{LPTENS/18/04}}

\begin{document}

\maketitle

\newpage

\epigraph{This article is dedicated to the memory of Lev Nikolaevich Lipatov, who was a constant source of inspiration for us and who deeply influenced our research. He will be greatly missed.
}{}
\section{Introduction}
${\cal N}=4$ Super-Yang-Mills theory has been playing an important role in our understanding of Quantum Field Theories, especially in an AdS/CFT context. Due to the Kotikov-Lipatov maximal transcendentality principle \cite{Kotikov:2001sc,Kotikov:2002ab} some of the results obtained in this theory can be directly exported to more realistic planar QCD. In this paper we describe how to efficiently perform calculations in this theory for one of the key QCD observables - BFKL spectrum, using integrability at any value of the `t Hooft coupling $\lambda$, which was discovered initially by Lipatov in the LO BFKL spectrum \cite{Lipatov:1993yb}, and developed far beyond the perturbative regime in the ${\cal N}=4$ SYM in recent years. Lev Nikolaevich was one of the main driving forces behind this progress and it is deeply saddening for us to know that he left us in September 2017.

In the beginning we are going to briefly describe the meaning of the quantities studied in the present work in the context of high energy scattering. The total cross-section $\sigma(s)$ for the high-energy scattering of two colorless particles A and B in the next-to-leading logarithmic approximation can be written as \cite{Kotikov:2000pm}
\beq
\sigma(s)=\int\frac{d^2 q d^2 q'}{(2\pi)^2 q^2 q'^{2}}\Phi_A(q)\Phi_B(q')\int\limits_{a-i\infty}^{a+i\infty}\frac{d\omega}{2\pi i}\(\frac{s}{s_0}\)^{\omega}G_{\omega}(q,q')\;,
\eeq
mwhere $\Phi_i(q_i)$ are the impact factors, $G_{\omega}(q,q')$ is the $t$-channel partial wave for the gluon-gluon scattering, $s_0=|q||q'|$ and depend on the transverse momenta and $s=2p_A p_B$, where $p_A$ and $p_B$ are the 4-momenta of the particles $A$ and $B$ respectively. For the $t$-channel partial wave there holds the Bethe-Salpeter equation
\beq
\omega G_{\omega}(q,q_1)=\delta^{D-2}(q-q_1)+\int d^{D-2}q_2 K(q,q_2)G_{\omega}(q_2,q_1)\;,
\eeq
where $K(q,q_2)$ is called the BFKL integral kernel. It appears to be possible to classify the eigenvalues $\omega$ of this BFKL kernel using two quantum numbers: integer $n$ (conformal spin) and real $\nu$
\beq\label{BFKL_Pomeron_eigenvalues}
\omega=\omega(n,\nu)\;.
\eeq
The function $\omega(n,\nu)$ is called the Pomeron eigenvalue of the BFKL kernel or just the BFKL Pomeron eigenvalue and its values for different $n$ and $\nu$ constitute the BFKL spectrum. For the phenomenological applications of the BFKL kernel eigenvalues with non-zero conformal spin see \cite{Kepka:2010hu}. The object $\omega(n,\nu)$\footnote{In \cite{Kotikov:2000pm} the function $\omega$ is used with the different argument $\gamma=1/2+i\nu$.} in the planar ${\cal N}=4$ SYM will be studied in this work by means of integrability.

The study of integrable structures in 4d gauge theory has long and interesting history of development. Integrability in QCD and supersymmetric Yang-Mills theories appeared in two contexts. First, in the gauge theory, namely QCD, the Bartels-Kwiecinski-Praszalowicz (BKP) equation \cite{Bartels:1980pe,Kwiecinski:1980wb} for multi-reggeon states was reformulated by L.N.~Lipatov \cite{Lipatov:1993yb} as the model with holomorphic and antiholomorphic hamiltonians, which has a set of mutually commuting operators originating from the monodromy matrix satisfying the Yang-Baxter equation. After that L.D.~Faddeev and G.P.~Korchemsky in \cite{Faddeev:1994zg} proved this model to be completely integrable and equivalent to the spectral problem for $SL(2,\mathbb{C})$ XXX Heisenberg spin chain. Then, in the context of high-energy scattering there was considered a certain class of light-cone operators in QCD and supersymmetric Yang-Mills theories and in \cite{Braun:1998id,Belitsky:2004sf,Belitsky:2005bu,Belitsky:2006av} the problem of finding the anomalous dimensions of light-cone operators was formulated in terms of $SL(2,\mathbb{R})$ Heisenberg spin chain.

The other achievement was that the maximally supersymmetric $\cN=4$ Yang-Mills theory in 4 dimensions, which is dual to $AdS_5 \times S^5$ type IIB superstring theory was shown to be integrable \cite{Minahan:2002ve,Beisert:2010jr}. The study of the integrability structure of the latter theory allowed to explore its spectrum in the non-perturbative regime. The solution to the spectral problem was formulated in terms of the Quantum Spectral Curve (QSC) \cite{Gromov:2013pga,Gromov:2014caa} (for the recent reviews see \cite{Gromov:2017blm} and \cite{Kazakov:2018ugh}). Nevertheless, until recently it was not known how to build the bridge between the integrability in the BFKL limit and integrability found in the AdS/CFT framework. In \cite{Kotikov:2007cy} the 4-loop Asymptotic Bethe Ansatz (ABA) contribution to the anomalous dimension of the twist-2 $\mathfrak{sl}(2)$ operators was analytically continued to the non-integer spins and compared with the corresponding prediction from the BFKL Pomeron eigenvalues. This analytic continuation to non-integer spins was incorporated into the QSC formalism in \cite{Gromov:2014bva} for twist-2 operators from the $\mathfrak{sl}(2)$ sector and then in \cite{Alfimov:2014bwa} the Faddeev-Korchemsky Baxter equation \cite{Faddeev:1994zg} for Lipatov $SL(2,\bbC)$ spin chain was derived correctly reproducing the leading order (LO) BFKL Pomeron eigenvalue. In addition, QSC allowed to calculate analytically \cite{Gromov:2015vua} the previously unknown next-to-next-to-leading order (NNLO) BFKL eigenvalue in the $\cN=4$ supersymmetric Yang-Mills theory. At the same time, a very efficient numerical algorithm was constructed in \cite{Gromov:2015wca}, which allows to study not only the BFKL limit of the spectrum of the theory, but the whole anomalous dimension of a given operator for arbitrary values of the charges.

In \cite{Alfimov:2014bwa} the twist-2 $\mathfrak{sl}(2)$ operators of the form
\beq
\cO=\textrm{tr}Z D_+^S Z+(\textrm{permutations})\;,
\eeq
were considered and from the perturbative calculations in the gauge theory for the case of even integer $S$ we know the dimension of these operators $\Delta$ as a function of $S$ up to several loops order. In the QSC framework the solution of the Baxter equation for the spectrum of such operators in the case of zero conformal spin $n$ and integer even spins $S$ was obtained in \cite{Gromov:2013pga}. Then in \cite{GromovKazakovunpublished,Janik:2013nqa} there was found the solution of this Baxter equations valid for arbitrary spin $S$, which leads to the anomalous dimension of the twist-2 $\mathfrak{sl}(2)$ operators analytically continued for non-integer spin $S$. After making this analytic continuation in the BFKL regime we are able to exchange the roles of $\Delta$ and $S$ obtaining $S+1=\omega(n=0,\nu)$, where $\nu=-i\Delta/2$ and $\Delta$ is the dimension of the operator in question.
\begin{figure}[ht]
\center{
\includegraphics[width=0.7\linewidth]{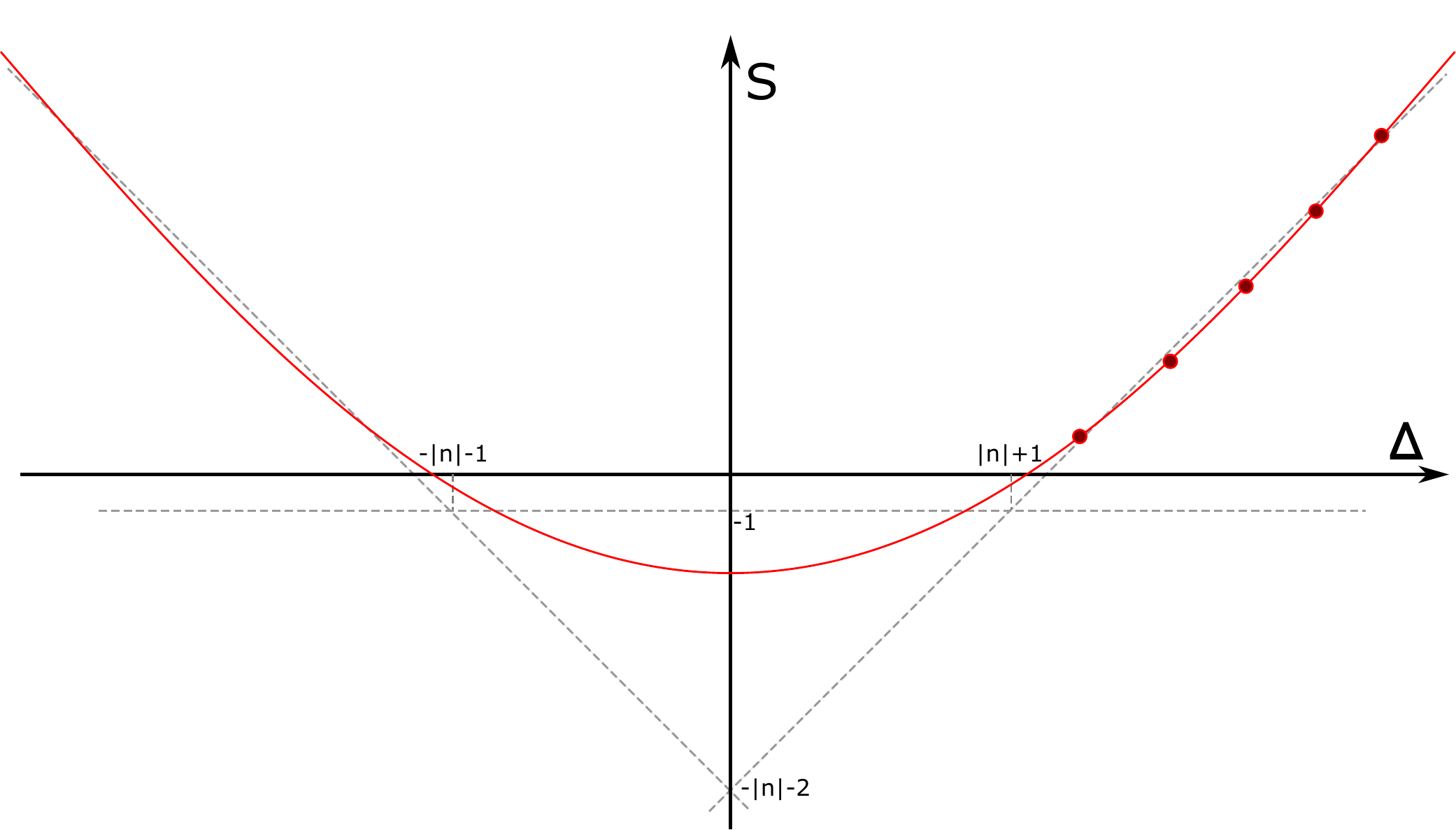}
\caption{Trajectory  of the length-2 operator for conformal spin $n=S_2$ as a function of the full dimension $\Delta$. The dots correspond to local operators $\textrm{tr}Z D_+^{S} \partial_{\perp}^{n} Z$. For the local operators $S+n$ is restricted to be even.
\label{operator_trajectories_nonzero_conformal_spin}}
}
\end{figure}

In the present work we consider the generalization allowing for an arbitrary value of the conformal spin. Namely, we consider the length-2 operators
\beq\label{operators}
\cO=\textrm{tr}Z D_+^{S_1} \partial_{\perp}^{S_2} Z+(\textrm{permutations})\;.
\eeq
For the operators \eqref{operators} we follow the same strategy as for the case of zero conformal spin. Analogously to that case we build the analytic continuation in the spins $S_1$ and $S_2$, which are identified with the spin $S$ and conformal spin $n$ respectively. Let us illustrate this analytic continuation with the Figure \ref{operator_trajectories_nonzero_conformal_spin}. The physical operators, for which the sum of non-negative integer $S=S_1$ and $n=S_2$ is even, are depicted with the dots. Then, flipping the roles of the dimension $\Delta$ and $S=S_1$ we can reach the BFKL regime described by the quantity $\omega(n=S_2,\nu)=S_1+1$, where $\nu=-i\Delta/2$.

The way to proceed with the problem in question is to first generalize the QSC approach to non-integer values of $S_1$ (as was already done in \cite{Gromov:2014bva}) and then also to non-integer values of $S_2$. We describe the technical details of this procedure in the Section \ref{ext_QSC_nonint_sp}. This allows to treat $\omega(n,\nu)$ as an analytic function of both its parameters which simplifies both analytical and numerical considerations. This gives a universal framework for studying the BFKL spectrum in full generality for all values of the parameters on equal footing within the extended QSC formalism.

Having formulated the problem as an extension of the initial QSC, a number of methods, initially developed for the local operators, became available for the BFKL problem. In particular we are enabled to employ a very powerful numerical algorithm \cite{Gromov:2015wca} after some modifications. As we take the spins $S_1$ to be continuous variable we can consider instead of the function $\Delta(S_1,S_2)$ the function $S_1(\Delta,S_2)$. Then, using the algorithm we build the operator trajectories for different values of conformal spin $S_2$ and the dependencies of the spin $S_1$ on the coupling constant $g$ for different values of conformal spin and dimension $\Delta$ (including a particular interesting intercept function corresponding to $\Delta=0$). Having the numerical results for the operator trajectories we were able to fit the numerical values of the BFKL kernel eigenvalues\footnote{See M.Alfimov's presentation at GATIS Training Event at DESY \cite{Alfimov:presentation}.}, which were confirmed in \cite{Caron-Huot:2016tzz} using a different method. 

Another method available within the QSC formalism is an efficient perturbative expansion developed in \cite{Marboe:2014gma,Marboe:2014sya,Marboe:2017dmb,Gromov:2015vua,Gromov:2017cja}. We applied this method to find the value of the Pomeron intercept for the arbitrary value of the conformal spin up to $3$ loops. Our result is in full agreement with \cite{Caron-Huot:2016tzz} at the NNLO level, but we also give a prediction for the next NNNLO order. 

Then, we found and studied in detail a particularly interesting point in the space of parameters of the BFKL Pomeron eigenvalue. This is the ``BPS" point $\Delta=0$ and $n=1$. As we have confirmed both numerically and analytically, the operator trajectory goes through the point $S=-1$, $n=1$ and $\Delta=0$ for any value of the coupling constant $g$. Studying the vicinity of this point we were able to find two non-perturbative quantities: ``slope-to-intercept function" and ``curvature function". The first function is the first derivative of $S\(\Delta,n\)$ with respect to $n$ at the point $\Delta=0$, $n=1$ and the second function is the second derivative of $S\(\Delta,n\)$ with respect to $\Delta$ at the same point. We used the methods developed in \cite{Gromov:2014bva} to compute analytically these quantities non-perturbatively to all orders in $g$.

Finally, we were able to identify the intercept function in the strong coupling expansion up to the 4th order. To obtain it we utilized the dependencies of the intercept on the coupling constant calculated by the QSC numerical method. By conducting the numerical fit of these dependencies for different values of conformal spin $n$ we predict the formula for the intercept strong coupling expansion up to the 4th order for arbitrary conformal spin.

Let us present a brief summary of the quantities we calculated. They include the NNLO intercept function \eqref{NNLO_intercept_function} and the non-rational part of the NNNLO intercept function \eqref{NNNLO_intercept_function_non_rational}. The other quantities we computed exactly to all orders in the `t Hooft coupling constant are the slope-to-intercept \eqref{slope_to_intercept} and the curvature \eqref{curvature_function} functions with the strong coupling expansions of these functions given by \eqref{slope} and \eqref{cur_str_coupling} respectively. In addition, there was written the strong coupling expansion \eqref{strongres} of the intercept function for arbitrary conformal spin $n$. We also implemented the numerical method for finding the eigenvalues at arbitrary values of the parameters in \verb"Mathematica", the corresponding files \verb"code_for_arxiv.nb" and \verb"BFKLdata.mx" can be found in the attachments to this arXiv submission. See \verb"description.txt" file for the description.

This work is organized as follows. In the Section \ref{ext_QSC_nonint_sp} we give the general introduction into the QSC approach, extending it to the situation when both spins are non-integer. The Section \ref{num_calc_QSC} describes our numerical results. The Section \ref{weak_coup_exp} contains the weak coupling analysis. In the Section \ref{near_BPS_exp} we analyze the expansion near the BPS point to find the non-perturbative quantities such as slope of intercept and curvature functions. In Section \ref{intercept_strong_coupling} we analyze the Pomeron intercept at strong coupling.

\section{Description of the QSC based framework}\label{ext_QSC_nonint_sp}

In this Section we are going to present the framework which we use to solve the QSC \cite{Gromov:2013pga,Gromov:2014caa} and whose derivation is based on the analytic and asymptotic properties of the Q-functions. First, we reformulate the QSC in terms of gluing matrix. Namely, we start from the several axioms concerning the analytic structure of the Q-system and the symmetries which preserve the QQ-relations and derive from them the so-called gluing conditions. These gluing conditions already appeared in \cite{Gromov:2014caa,Gromov:2015vua} but our approach presented below does not utilize the notion of $\mu$- and $\omega$-functions to obtain the gluing matrix. Second, using the connection between the asymptotics of the certain subset of Q-functions and the global charges together with their analytic properties, the system of constraints for the gluing matrix is derived. It appears to be possible to solve these equations in some physically interesting cases. Namely, we find the gluing matrix for the case when both AdS spins $S_1$ and $S_2$ are integers of the same parity and its form appears to be very simple and in complete agreement with the result of \cite{Gromov:2014caa}. Then we consider a more general case of non-integer AdS spins $S_1$ and $S_2$
, which is particularly interesting for the exploration of the BFKL regime. For this case we have not found the general solution for the gluing matrix, however we found the certain subclass of solutions and it appears to be applicable to our quantities of interest.
We mostly follow the original paper \cite{Gromov:2014caa}, but the discussion of the gluing matrix and the extension to the non-integer quantum numbers is new. The reader familiar with the QSC formalism could skip to Subsection \ref{sec:constr}.

\subsection{Algebraic part of the construction}\label{QSC_algebraic_construction}
QSC consists of a set of Q-functions of the complex spectral parameter $u$ and relations between them. We will restrict ourselves to the most essential parts of the construction but still  keeping the discussion self-contained. For more detailed description of the QSC see \cite{Gromov:2013pga,Gromov:2014bva} and for the pedagogical introduction see \cite{Gromov:2014caa}.

In total there are $256$ Q-functions $Q_{a_1,\dots,a_n|i_1,\dots,i_m}(u)$ totally antisymmetric in the two groups of ``bosonic" ($a$'s) and ``fermionic" ($i$'s) indices with $1 \leq n,m \leq 4$, however not all of them are independent. The main building blocks of the QSC construction are the $4+4$ ``elementary" Q-functions: $Q_{a|\emptyset}(u)$, where $a=1, \ldots, 4$, and $Q_{\emptyset|i}(u)$, where $i=1, \ldots, 4$. Setting the normalization $Q_{\emptyset|\emptyset}=1$ and starting from these $8$ Q-functions, one can recover the whole Q-system applying the QQ-relations written in \cite{Gromov:2014caa}. In particular the QQ-relation for the Q-function with one ``bosonic" and one ``fermionic" index looks as follows
\beq\label{Qai_eq}
Q_{a|i}\(u+\frac{i}{2}\)-Q_{a|i}\(u-\frac{i}{2}\)=Q_{a|\emptyset}(u) Q_{\emptyset|i}(u)\;.
\eeq
and $Q_{a|i}(u)$ is a solution of \eqref{Qai_eq}. From now on we are going to use the shorthand notation for the shift in the variable $u$: $f(u+ik/2)=f^{[k]}(u)$. In a similar way one can build all 256 Q-functions out of the basic 8 mentioned above. One should also impose the quantum unimodularity condition
\beq\label{unimodularity_condition}
Q_{1234|1234}=1.
\eeq

An important symmetry of the QSC is the Hodge-duality, which exchanges
\begin{multline}\label{Hodge_duality}
Q_{a_1,\dots,a_n|i_1,\dots,i_m} \leftrightarrow
Q^{a_1,\dots,a_n|i_1,\dots,i_m} \equiv \\
\equiv (-1)^{(4-n)m}\epsilon^{b_{n+1} \dots b_4 a_1 \dots a_n}\epsilon^{j_{m+1} \dots j_4 i_1 \dots i_m}Q_{b_{n+1},\dots,b_4|j_{m+1},\dots,j_4}\;,
\end{multline}
where in the right-hand side of \eqref{Hodge_duality} there is no summation over the repeated indices. The Hodge-dual Q-functions \eqref{Hodge_duality} with the upper indices also satisfy the same QQ-relations as the Q-functions with the lower indices.

Due to \eqref{unimodularity_condition} we are able to obtain the relations which allow to get fast to the Hodge-dual Q's
\beq\label{Qai_inverse}
Q^{a|i}Q_{a|j}=-\delta^i_j\;, \quad Q^{a|i}Q_{b|i}=-\delta^a_b\;.
\eeq
The Q-function $Q^{a|i}$ allows to write the Q-functions with one upper index in a concise form
\beq\label{uP}
Q^{a|\emptyset}=(Q^{a|i})^+ Q_{\emptyset|i}
\eeq
and
\beq\label{uQ}
Q^{\emptyset|i}=(Q^{a|i})^+ Q_{a|\emptyset}\;.
\eeq
From the condition \eqref{unimodularity_condition} it can be shown that
\beq
Q_{a|\emptyset} Q^{a|\emptyset}=0\;, \quad Q_{\emptyset|i} Q^{\emptyset|i}=0\;.
\eeq

In addition, the Q-system has a symmetry, which is called the $H$-symmetry \cite{Gromov:2014caa} and which leaves the QQ-relations intact. It corresponds to the transformations of Q-functions by $i$-periodic matrices that rotate the ``bosonic" and ``fermionic" indices separately. Their form for all Q-functions can be found in \cite{Gromov:2014caa}, but in this Section we need the explicit form of them only for the Q-functions with one index. They are
\beq\label{Q_rotations}
Q_{a|\emptyset} \rightarrow (H_B)_a^c Q_{c|\emptyset}\;, \quad Q^{a|\emptyset} \rightarrow (H_B^{-1})^a_c Q^{c|\emptyset}\;, \quad Q_{\emptyset|i} \rightarrow (H_F)_i^j Q_{\emptyset|j}\;, \quad Q^{\emptyset|i} \rightarrow (H_F^{-1})^i_j Q^{\emptyset|j},
\eeq
where $H_B(u)$ and $H_F(u)$ are $i$-periodic $4 \times 4$ matrices. The determinants of these matrices have to satisfy
\beq
\det H_B(u) \det H_F(u)=1
\eeq
for the quantum unimodularity condition \eqref{unimodularity_condition} not to change under such $H$-rotations. The important particular case of this symmetry is the rescaling of the Q-functions with one index. It acts as follows
\beq\label{Q_rescaling}
Q_{a|\emptyset} \rightarrow \alpha_a Q_{a|\emptyset}\;, \quad Q_{\emptyset|i} \rightarrow \beta_i Q_{\emptyset|i}\;, Q^{a|\emptyset} \rightarrow \frac{1}{\alpha_a} Q^{a|\emptyset}\;, \quad Q^{\emptyset|i} \rightarrow \frac{1}{\beta_i} Q^{\emptyset|i}.
\eeq

The equation \eqref{Qai_eq} allows to obtain a 4th order Baxter equation for the functions $Q_{\emptyset|i}(u)$, $i=,1\ldots,4$. In \cite{Alfimov:2014bwa} this equation was derived and looks as follows
\begin{multline}\label{Baxter_4th_order}
Q_{\emptyset|i}^{[+4]}-Q_{\emptyset|i}^{[+2]}\[D_1-Q_{a|\emptyset}^{[+2]}Q^{a|\emptyset|[+4]}D_0 \]+Q_{\emptyset|i}\[D_2-Q_{a|\emptyset}Q^{a|\emptyset[+2]}D_1+Q_{a|\emptyset}Q^{a|\emptyset[+4]}D_0 \]-\\
-Q_{\emptyset|i}^{[-2]}\[\bar{D}_1+Q_{a|\emptyset}^{[-2]}Q^{a|\emptyset[-4]}\bar{D}_0 \]+Q_{\emptyset|i}^{[-4]}=0\;,
\end{multline}
where
\begin{align}\label{D_determinants}
& D_k=\det\limits_{1 \leq i,j \leq 4} Q^{\emptyset|j[4-2i+2\delta_{i,k+1}]}\;, \;k=0,1\;, \\
& D_2=\det\limits_{1 \leq i,j \leq 4} Q^{\emptyset|j[4-2i+2\delta_{i,1}+\delta_{i,2}]}\;, \notag \\
& \bar{D}_k=\det\limits_{1 \leq i,j \leq 4} Q^{\emptyset|j[-4+2i-2\delta_{i,k+1}]}\;, \;k=0,1\;. \notag
\end{align}
It is also possible to show from the same equation as \eqref{Qai_eq} for the Q-functions with upper indices that the functions $Q^{\emptyset|i}(u)$, $i=1,\ldots,4$ satisfy the 4th order Baxter equation, which looks as \eqref{Baxter_4th_order} but with $Q_{a|\emptyset}$ exchanged with $Q^{a|\emptyset}$. For the sake of conciseness we do not write this Baxter equation explicitly.

After finishing the description of the algebraic structure of the QSC essential for the formulation of the QSC equations in the next Subsection we describe the analyticity properties of the Q-functions, which constitute the crucial part of our QSC framework.

\subsection{Analytic part of the construction}\label{QSC_analytic_construction}

To describe the analytic structure of the Q-system we have to first define the analyticity properties of the basic set of Q-functions: $Q_{a|\emptyset}(u)$, $a=1,\ldots,4$ and $Q_{\emptyset|i}(u)$, $i=1,\ldots,4$. The only singularities of these functions are quadratic branch points which come in pairs at the positions $\pm 2g+ik$, where $k \in \mathbb{Z}$. For each pair of branch points we can choose either short cut on the interval $[-2g+ik,2g+ik]$ or a long cut $(-\infty+ik,-2g+ik] \cup [2g+ik,ik+\infty)$, where $k$ is some integer. In what follows the sheet of the Q-function with only the short cuts is called physical and the function on this sheet is denoted by $\hat{Q}(u)$, while on the sheet, where all the cuts are long, is called mirror and the function is designated by $\check{Q}(u)$ on it. The continuation of any function $f(u)$ under the cut on the real axis is denoted by $\tilde f(u)$. The branch points of all the functions we will consider are quadratic, i.e. $\tilde{\tilde{f}}(u) = f(u)$.

In what follows we will denote the functions $Q_{a|\emptyset}(u)$, $Q^{a|\emptyset}(u)$, $Q_{\emptyset|i}(u)$ and $Q^{\emptyset|i}(u)$, with prescribed analytical properties, as $\bP_a(u)$, $\bP^a(u)$, $\bQ_i(u)$ and $\bQ^i(u)$ respectively. To proceed let us write the asymptotics of the Q-functions with one index. We know that all the Q-functions including $\bP_a$, $\bP^a$, $\bQ_i$ and $\bQ^i$ have the power-like asymptotics at large $u$, which for the basic 8 Q-functions can be taken from \cite{Gromov:2014caa}
\beq\label{PQ_asymptotics}
\bP_a \simeq A_a u^{-\tilde{M}_a}\;, \quad \bP^a \simeq A^a u^{\tilde{M}_a-1}\;, \quad \bQ_i \simeq B_i u^{\hat{M}_i-1}\;, \quad \bQ^i \simeq B^i u^{-\hat{M}_i}\;,
\eeq
where $\tilde{M}_a$, $a=1,\ldots,4$ and $\hat{M}_i$, $i=1,\ldots,4$ are functions of the values of the 6 Cartan generators of the $\mathfrak{psu}(2,2|4)$ symmetry algebra of the \sym: integer $J_1,J_2,J_3$ ($\tilde{M}_a$) and $\Delta,S_1,S_2$ ($\hat{M}_i$), which are specified below
\begin{align}\label{global_charges_all}
& \tilde{M}_a=\left\{\frac{1}{2}\(J_1+J_2-J_3+2\), \frac{1}{2}\(J_1-J_2+J_3\), \frac{1}{2}\(-J_1+J_2+J_3+2\),\frac{1}{2}\(-J_1-J_2-J_3\) \right\}, \notag \\
& \hat{M}_i=\left\{\frac{1}{2}\(\Delta-S_1-S_2+2\), \frac{1}{2}\(\Delta+S_1+S_2\), \frac{1}{2}\(-\Delta-S_1+S_2+2\), \frac{1}{2}\(-\Delta+S_1-S_2\) \right\}.
\end{align}

\begin{figure}[ht]
{\center
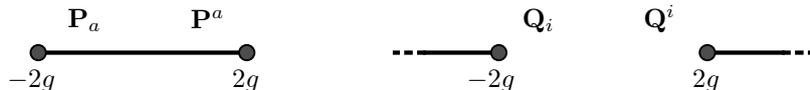
\caption{Analytic structure of the $\bP$- and $\bQ$-functions on their defining sheet.
\label{PQ_def_sheet}}
}
\end{figure}

As we know from the classical integrability of the dual superstring $\sigma$-model (see, for example, \cite{Gromov:2017blm}), the $\bP$- and $\bQ$-functions at least have the quadratic branch points at $u=\pm 2g$. From the asymptotics \eqref{PQ_asymptotics} and \eqref{global_charges_all} we can expect the cuts of $\bP$-functions to be short\footnote{For some values of the Cartan charges $J_1$, $J_2$ and $J_3$ of the \sym \ symmetry algebra $\mathfrak{psu}(2,2|4)$ there could appear a quadratic branch cut going to infinity. However, the $\bP$-functions usually come in bilinear combinations and these cuts cancel each other in these combinations.}. The minimal choice for the functions $\bP_a(u)$ and $\bP^a(u)$, $a=1,\ldots,4$ is to have only one short cut on the real axis. From the asymptotics of the $\bQ$-functions we can see that they have a nontrivial monodromy around infinity, thus we have to assume the cuts of these functions to be long. So again the minimal choice for $\bQ_i(u)$ and $\bQ^i(u)$, $i=1,\ldots,4$ would be to have only one long cut on the real axis. The analytic structure of $\bP$- and $\bQ$-functions is illustrated on the Figure \ref{PQ_def_sheet}. Notice, that because the functions $\bQ_i$ and $\bQ^i$ have the long cuts in the complex plane, their asymptotics prescribed from the large $u$ limit of the superstring $\sigma$-model hold in the upper half-plane and in the lower half-plane they can be different, therefore the third and fourth formulas from \eqref{PQ_asymptotics} are valid for $\Im\;u>0$\footnote{It should be noted that in the case, when at least one of the spins $S_1$ and $S_2$ is non-integer, the asymptotics of the $\bQ$-functions in the lower half-plane on the sheet with the long cuts can be not power-like but instead become some power times an exponential factor.}.

It is convenient for us to introduce some short-hand notations as in \cite{Gromov:2014caa}: UHP -- upper half-plane, LHP -- lower half-plane, UHPA -- upper half-plane analytic and LHPA -- lower half-plane analytic. Besides, we are going to use LHS and RHS for left and right-hand side respectively.

As we have introduced the analytic structure of the basic set of Q-functions, let us proceed with the consideration of the other ones. We define the function $\cQ_{a|i}$ as an UHPA solution of the equation \eqref{Qai_eq}
\beq\label{Qai_eq_UHPA}
\cQ_{a|i}^+-\cQ_{a|i}^-=\bP_a \bQ_i\;, \quad \textrm{Im}\;u>0
\eeq
with the asymptotic
\beq\label{Qai_asymptotic}
\cQ_{a|i} \simeq -i\frac{A_a B_i}{-\tilde{M}_a+\hat{M}_i}u^{-\tilde{M}_a+\hat{M}_i}.
\eeq
In what follows we will denote the UHPA Q-functions of the Q-system obtained from $\bP_a$, $\bQ_i$ and $\cQ_{a|i}$ by the application of the QQ-relations by curly $\cQ$ (as it was done in \cite{Gromov:2014caa} to underline that these Q-functions have certain analytic properties and where the corresponding Q-system was called fundamental). For the Hodge-dual Q-functions, which are UHPA as well and satisfy the same QQ-relations, we will also use curly $\cQ$ to depict its Q-functions.

Substitution of \eqref{PQ_asymptotics} and \eqref{Qai_asymptotic} into $\bQ_i=-\cQ_{a|i}^+\bP^a$ and $\bP_a=-\cQ_{a|i}^+ \bQ^i$ expressed from \eqref{uP} and \eqref{uQ} themselves leads us to the systems of equations for $A_a A^a$ and $B_i B^i$ respectively, which can be solved \cite{Gromov:2014caa,Gromov:2017blm} and give the result
\beq\label{AA_BB_general}
A_{a_0} A^{a_0}=i\frac{\prod\limits_{j=1}^4 \(\tilde{M}_{a_0}-\hat{M}_j\)}{\prod\limits_{b=1 \atop b \neq a_0} \(\tilde{M}_{a_0}-\tilde{M}_b\)}\;, \quad B_{i_0} B^{i_0}=-i\frac{\prod\limits_{a=1}^4 \(\hat{M}_{i_0}-\tilde{M}_a\)}{\prod\limits_{j=1 \atop j \neq i_0}^4 \(\hat{M}_{i_0}-\hat{M}_j\)}\;, \quad a_0,i_0=1,\ldots,4\;,
\eeq
where there is no summation over the indices $a_0$ and $i_0$ implied. For further convenience we introduce the shorthand notations $A_{a_0}A^{a_0}={\bf A}_{i_0}$ and $B_{i_0}B^{i_0}={\bf B}_{i_0}$.

Each function $\cQ_{a|i}(u)$ is analytic for $\textrm{Im}\;u>-1/2$ due to the fact that both $\bP_a(u)$ and $\bQ_i(u)$ are UHPA. As it was mentioned in the Subsection \ref{QSC_algebraic_construction} with the usage of the QQ-relations we can restore the remaining Q-functions thus building the UHPA Q-system. The Hodge-duality \eqref{Hodge_duality} does not change the analytic properties, therefore the Hodge-dual Q-system with the upper indices has the same analytic properties, i.e. UHPA.

Now we are going to turn to the analytic structure of the functions $\bQ_i(u)$, $i=1,\ldots,4$. Let us remember the formulas \eqref{uP}. One can notice that the functions $\bQ_i$ and $\bQ^i$ with $-\cQ_{a|i}^+ \bP^a$ and $\cQ^{a|i+} \bP_a$ respectively coincide only in the UHP, because their analytic structure in the LHP is different. Indeed, we can see that if we rewrite the QQ-relations for $\cQ_{a|i}$ using \eqref{uP}
\beq
\cQ_{a|i}^-=\(1+\bP_a \bP^b\)\cQ_{b|i}^+\;,
\eeq
it is possible to find the values of $\cQ_{a|i}$ for $\Im\;u<-1/2$. In the strip $-k-1<\Im\;u<-k$ for $k=0,1,2,\ldots$ the functions $\cQ_{a|i}^-$ are given by
\beq\label{Qai_shift}
\cQ_{a|i}^-=\prod\limits_{l=0}^k \(1+\bP_a \bP^b\)^{[2l]} \cQ_{a|i}^{[2k+1]}\;.
\eeq
From \eqref{Qai_shift} we can see that the functions $\cQ_{a|i}$ have the infinite number of short cuts at the horizontal lines with $\Im\;u=-k-1/2$ for $k=0,1,2,\ldots$ and due to \eqref{Qai_inverse} $\cQ^{a|i}$ has the same structure of cuts in the complex plane. Therefore, the functions $-\cQ_{a|i}^+ \bP^a$ and $\cQ^{a|i+} \bP_a$ have also the infinite ladder of cuts at $\Im\;u=-k$ for $k=0,1,2,\ldots$, which clearly do not coincide with the analytic structure of $\bQ_i$ and $\bQ^i$ in the LHP, who are LHPA. 

However, we can resolve this difficulty by interpreting the analytic continuation of $-\cQ_{a|i}^+ \bP^a$ and $\cQ^{a|i+}\bP_a$ under their short cut on the real axis from above as $\bQ_i$ and $\bQ^i$ respectively. To formulate this clearly let us use the hats and checks introduced in the beginning of the present Subsection, which denote the values of the $\bP$- and $\bQ$-functions on the different sheets. In these notations first of all the equations \eqref{Qai_eq_UHPA} and \eqref{Qai_shift} determine $\hat{\cQ}_{a|i}$ on the physical sheet with the short cuts. Then, the values of the $\hat{\bQ}_i$ and $\hat{\bQ}^i$ on their physical sheet with the short cuts are given by
\beq
\hat{\bQ}_i=-\hat{\cQ}_{a|i}^+ \hat{\bP}^a\;, \quad \hat{\bQ}^i=\hat{\cQ}^{a|i+} \hat{\bP}_a
\eeq
and coincide with the $\bQ$-functions on the mirror sheet with the long cut on the real axis in the UHP
\beq\label{UHP_Q}
\hat{\bQ}_i=\check{\bQ}_i\;, \quad \hat{\bQ}^i=\check{\bQ}^i\;, \quad \Im\;u>0\;.
\eeq
Whereas in the LHP we interpret the analytic continuation of $\hat{\bQ}_i$ and $\hat{\bQ}^i$ under their cut on the real axis as $\check{\bQ}_i$ and $\check{\bQ}^i$ in the LHP
\beq\label{LHP_Q}
\tilde{\hat{\bQ}}_i=\check{\bQ}_i\;, \quad \tilde{\hat{\bQ}}^i=\check{\bQ}^i\;, \quad \Im\;u<0\;,
\eeq
where tilde denotes the analytic continuation under the cut on the real axis.

Looking at the obtained picture from above, we conclude that there is no fundamental reason to choose the generated Q-system to be UHPA. Indeed, there exists a transformation of complex conjugation, which preserves the QQ-relations but interchanges UHPA with LHPA. Its explicit form is written in \cite{Gromov:2014caa}
\beq\label{cjg_transform}
\cQ_{a_1,\ldots,a_n|i_1,\ldots,i_m}(u) \rightarrow (-1)^{\frac{(m+n)(m+n-1)}{2}}\bar{\cQ}_{a_1,\ldots,a_n|i_1,\ldots,i_m}(u)\;.
\eeq
The transformation \eqref{cjg_transform} generates the Q-system which is LHPA and satisfies the same QQ-relations as the initial UHPA Q-system. It should be noted that the Hodge-dual Q-system also admits such a transformation
\beq\label{cjg_transform_Hodge}
\cQ^{a_1,\ldots,a_n|i_1,\ldots,i_m}(u) \rightarrow (-1)^{\frac{(m+n)(m+n-1)}{2}}\bar{\cQ}^{a_1,\ldots,a_n|i_1,\ldots,i_m}(u)\;.
\eeq

Let us now remember the analyticity properties of the function $\bQ^i$ from \eqref{UHP_Q} and \eqref{LHP_Q}. We see that the functions $\tilde{\hat{\bQ}}^i$ are LHPA and therefore have the same analyticity properties as the functions $\bar{\hat{\bQ}}_i$ and $\bar{\hat{\bQ}}^i$. From the strong coupling limit of the superstring $\sigma$-model and its classical integrability (see the pedagogical explanation of this in \cite{Gromov:2017blm}) we know that for the case of integer spins $S_1$ and $S_2$ each function $\tilde{\hat{\bQ}}^i$, $i=1,\ldots,4$ coincides with the certain function from the set $\bar{\hat{\bQ}}_j$, $j=1,\ldots,4$ in this limit. Thus, summarizing all this, we impose the equality of the LHPA functions $\tilde{\hat{\bQ}}^i$ and $\bar{\hat{\bQ}}_j$ up to some matrix $M^{ij}(u)$
\beq\label{gluing_condition_lower}
\tilde{\hat{\bQ}}^i(u)=M^{ij}(u)\bar{\hat{\bQ}}_j(u)\;.
\eeq
From now on we will call \eqref{gluing_condition_lower} the gluing condition and $M^{ij}(u)$ the gluing matrix, whose properties we will analyze below. We formulated QSC in the form \eqref{gluing_condition_lower} because this form is convenient to analytically continue the QSC solution for the case of non-integer spins $S_1$ and $S_2$. The transformation \eqref{cjg_transform} generates the Q-system which is LHPA and satisfies the same QQ-relations as the initial UHPA Q-system. As there is no principal difference between UHPA and LHPA Q-systems and they describe the same spectral problem and due to the unitarity of the \sym \ theory, the UHPA and LHPA Q-systems have to be related by the symmetries of the Q-system, namely, the combination of the Hodge duality and $H$-symmetry\footnote{The presence of the Hodge duality can explained from the consideration of the classical limit of the superstring $\sigma$-model, which says that the analytic continuation of the Q-functions with the lower indices $\tilde{\hat{\bQ}}_i(u)$ are related to the $\bQ$-functions with the upper indices $\hat{\bQ}^i$, not the lower ones.}. Thus, we can interpret the gluing matrix $M^{ij}$ as an $i$-periodic matrix of the $H$-transformation, which, in particular, relates the $\bQ$-functions with one lower and one upper ``fermionic" index on the mirror sheet
\beq\label{QQbar_transform}
\check{\bQ}^i=M^{ij} \bar{\check{\bQ}}_j\;, \quad \check{\bQ}_i=\(M^{-t}\)_{ij} \bar{\check{\bQ}}^j\;,
\eeq
where $-t$ means that the inverse matrix is transposed.

Using the analyticity properties of the functions $\check{\bQ}^i$ and $\bar{\check{\bQ}}_j$ we are able to establish some properties of the matrix $M^{ij}$. Utilizing the $i$-periodicity of the matrix $M^{ij}$, it is possible to express its elements in terms of $\check{\bQ}^i$ and $\bar{\check{\bQ}}_j$. From the $i$-periodicity of $M^{ij}(u)$, \eqref{QQbar_transform} and remembering the QQ-relation for Q-function with 4 ``fermionic" indices $Q_{\emptyset|1234}=\det\limits_{1 \leq k,l \leq 4} Q_{\emptyset|k}^{[5-2l]}$ one can show that
\beq\label{M_equation}
M^{ij} \bar{\check{\cQ}}_{\emptyset|1234}=\det\left(\begin{array}{cccc}
\bar{\check{\bQ}}_1^{[+3]} & \bar{\check{\bQ}}_1^{[+1]} & \bar{\check{\bQ}}_1^{[-1]} & \bar{\check{\bQ}}_1^{[-3]} \\
\cdots \\
\check{\bQ}^{i[+3]} & \check{\bQ}^{i[+1]} & \check{\bQ}^{i[-1]} & \check{\bQ}^{i[-3]} \\
\cdots \\
\bar{\check{\bQ}}_4^{[+3]} & \bar{\check{\bQ}}_4^{[+1]} & \bar{\check{\bQ}}_4^{[-1]} & \bar{\check{\bQ}}_4^{[-3]}
\end{array}\right){\scriptscriptstyle \leftarrow j}\textrm{\tiny -th row}\;.
\eeq

First of all let us show that the matrix elements $M^{ij}$ do not have any branch points. To see this let us notice that for $\Im\;u<-3/2$ the Q-function on the LHS and the determinant on the RHS of \eqref{M_equation} are analytic and do not have branch points. Therefore in the same region and due to the $i$-periodicity $M^{ij}$ has to be free from the branch points in the whole complex plane.

Second, in principle, $M^{ij}$ can have poles. As $M^{ij}$ is $i$-periodic, the existence of a pole, for example, in the point $u_0$ automatically leads to the infinite number of poles in the points $u_0+ik$, where $k \in \mathbb{Z}$. However, at least in the region $\Im\;u<-3/2$ the RHS of \eqref{M_equation} is analytic, thus the poles of $M^{ij}$ have to be compensated by the zeroes of $\bar{\check{\cQ}}_{\emptyset|1234}$ in the same points or, in other words, there exists $k_0$ such that $\bar{\check{\cQ}}_{\emptyset|1234}(u_0+ik)=0$ for $k \leq k_0$. In its turn this means that the number of zeroes of $\bar{\check{\cQ}}_{\emptyset|1234}$ is infinite and these zeroes accumulate at infinity. We know that $\bar{\check{\cQ}}_{\emptyset|1234}$ has power-like asymptotic, then there exist such $a$ and $b$ that
\beq\label{Q1234_asymptotic}
\bar{\check{\cQ}}_{\emptyset|1234}(u)-a u^b=\cO\(u^{b-1}\)\;, \quad u \rightarrow \infty\;.
\eeq
However, evaluating the LHS of \eqref{Q1234_asymptotic} at $u=u_0+ik$ for $k \leq k_0$ leads to a contradiction with the RHS of \eqref{Q1234_asymptotic}. From this contradiction we conclude that $M^{ij}$ cannot have any poles.

Summarizing what was said we see that the gluing matrix $M^{ij}$ is analytic in the whole complex plain. For the physical state, which means that the spins $S_1$ and $S_2$ are integer and the asymptotics of the functions $\tilde{\bQ}^i$ are the power-like, analyticity and $i$-periodicity of the gluing matrix $M^{ij}$ leads us to the conclusion that it is constant in this case.

Now let us return back to the equations \eqref{QQbar_transform}. As we know that the matrix $M^{ij}$ is free of any singularities, then analytically continuing both sides of the equations \eqref{QQbar_transform} to the sheet with the short cuts we obtain
\beq\label{gluing_conditions}
\tilde{\hat{\bQ}}^i=M^{ij} \bar{\hat{\bQ}}_j\;, \quad \tilde{\hat{\bQ}}_i=\(M^{-t}\)_{ij} \bar{\hat{\bQ}}^j\;.
\eeq

Analytic properties of the $\bQ$-functions allow us to establish one important property of the gluing matrix. Analytically continuing both sides of the first equation from \eqref{gluing_conditions}, using the fact that due to the quadratic nature of the branch points analytic continuation and complex conjugation commute with each other, and then applying the second equation from \eqref{gluing_conditions} we derive
\beq\label{Qtilde_twice}
\hat{\bQ}^i=M^{ij}\(\bar{M}^{-t}\)_{jk}\hat{\bQ}^k\;.
\eeq
Noticing that \eqref{Qtilde_twice} is true for any point $u$ and applying the same trick as above and using the analyticity properties of the gluing matrix, we arrive to the conclusion that
\beq
M^{ij}\(\bar{M}^{-t}\)_{jk}=\delta^i_k\;
\eeq
and the gluing matrix is hermitian
\beq\label{glm_hermiticity}
\bar{M}^{ij}(u)=M^{ji}(u)\;.
\eeq

In what follows we are going mainly to deal with the Q-functions on the physical sheet therefore from now on we omit the hats and checks above the designations of the Q-functions implying that all the Q-functions are considered on the sheet with the short cuts if the opposite is not mentioned specifically.

Now we are ready to find the other constraints on the gluing matrix which follow from the conjugation and parity symmetries of the Q-system. To do this we will need the 4th order Baxter equation for the functions $\bQ_i(u)$ to see if the certain properties of the $\bP_a(u)$ and $\bP^a(u)$ can allow us to relate $\bQ_i(u)$, $\bar{\bQ}_i(u)$ and $\bQ_i(-u)$. In the two subsequent Subsections we analyze the implications of the conjugation and parity properties respectively.

\subsection{Complex conjugation symmetry}

Let us now concentrate on the conjugation properties of the $\bP$- and $\bQ$-functions assuming the charges $\Delta$, $S_1$ and $S_2$ to be real. In \cite{Gromov:2014caa} from the reality of the energy, Y- and T-functions and the fact that the complex conjugation supplemented by the certain sign factor is the symmetry of the Q-system it is shown that complex conjugation is equivalent to some $H$-symmetry transformation already mentioned in the Subsection \ref{QSC_algebraic_construction}. The matrix $h_B$ of this transformation was proven in \cite{Gromov:2014caa} to be constant due to analytic properties and power-like asymptotic of the $\bP$-functions. Then there was found a transformation which allows to make all $\bP$-functions with lower indices purely real and thus the $\bP$-functions with upper indices pure imaginary. However, in our calculations we use the different normalization and make the other $H$-rotation by multiplying $\bP_3$ and $\bP_4$ by $i$ and thus $\bP^1$ and $\bP^2$ also by $i$ and obtain
\beq\label{conjugation_P_functions}
\bar{\bP}_{1,2}=\bP_{1,2}\;, \quad \bar{\bP}_{3,4}=-\bP_{3,4}\;, \quad \bar{\bP}^{1,2}=-\bP^{1,2}\;, \quad \bar{\bP}^{3,4}=\bP^{3,4} \notag
\eeq
or, in other words
\beq\label{P_conjugation}
\bar{\bP}_a=C_a^b \bP_b\;, \quad \bar{\bP}^a=-C^a_b \bP^b\;, \quad C=\textrm{diag}\{1,1,-1,-1\}\;.
\eeq

Given the conjugation properties \eqref{P_conjugation}, we see that the Baxter equation \eqref{Baxter_4th_order}, written for the Q-functions with prescribed analytic properties
\begin{multline}\label{Baxter_4th_order1}
\bQ_i^{[+4]}-\bQ_i^{[+2]}\[D_1-\bP_a^{[+2]}\bP^{a[+4]}D_0 \]+\bQ_i\[D_2-\bP_a\bP^{a[+2]}D_1+\bP_a\bP^{a[+4]}D_0 \]-\\
-\bQ_i^{[-2]}\[\bar{D}_1+\bP_a^{[-2]}\bP^{a[-4]}\bar{D}_0 \]+\bQ_i^{[-4]}=0
\end{multline}
with $D_k$ and $\bar{D}_k$ given by \eqref{D_determinants} with $Q^{a|\emptyset}=\bP^a$, remains the same, but for $\bar{\bQ}_i(u)$ now. Thus, the functions $\bar{\bQ}_i$ satisfy the equation \eqref{Baxter_4th_order1} too. As the functions $\bar{\bQ}_i(u)$ constitute a basis in the space of solutions of the Baxter equation \eqref{Baxter_4th_order1} this means that there has to exist an $i$-periodic matrix $\Omega_i^j(u)$ such that
\beq\label{Q_complex_conjugation}
\bar{\bQ}_i(u)=\Omega_i^j(u)\bQ_j(u).
\eeq
As in \cite{Gromov:2016rrp} from \eqref{P_conjugation} and \eqref{uP} this matrix can be found to be
\beq\label{Omega_matrix_definition}
\Omega_i^j=\bar{\cQ}_{a|i}^- C^a_b \cQ^{b|j-}\;.
\eeq
It is $i$-periodic $\Omega_i^{j++}=\Omega_i^j$ (see Appendix \ref{QSC_construction_app} for the proof) and using this it is not hard to show that
\beq
\Omega_i^j \bar{\Omega}_j^k=\bar{\cQ}_{a|i}^- C^a_b \cQ^{b|j-} \cQ_{c|j}^+ C^c_d \bar{\cQ}^{d|k+}=\bar{\cQ}_{a|i}^- C^a_b \cQ^{b|j-} \cQ_{c|j}^- C^c_d \bar{\cQ}^{d|k-}=\delta_i^k\;,
\eeq
which means that $\Omega^{-1}=\bar{\Omega}$ and
\beq\label{Q_complex_conjugation_inverse}
\bQ_i(u)=\bar{\Omega}_i^j(u) \bar{\bQ}_j(u)\;.
\eeq
The matrix $\bar{\Omega}_i^j$ also relates $\bar{\bQ}^j$ and $\bQ^i$
\beq\label{Q_complex_conjugation2}
\bar{\bQ}^j=-\bar{\Omega}_i^j \bQ^i
\eeq
and vice versa
\beq\label{Q_complex_conjugation_inverse2}
\bQ^j=-\Omega_i^j \bar{\bQ}^i\;.
\eeq

To determine the consequences of the conjugation symmetry for the gluing matrix we substitute \eqref{Q_complex_conjugation} into the first gluing condition from \eqref{gluing_conditions} and obtain
\beq\label{QtildeQ_relation}
\tilde{\bQ}^i=M^{ij}\Omega_j^k \bQ_k\;.
\eeq
Let us analyze the matrix $M^{ij}\Omega_j^k$ more closely. As it is a product of two $i$-periodic matrices it has also to be $i$-periodic. We remember that according to its definition \eqref{Omega_matrix_definition} the matrix $\Omega^k_j(u)$ has an infinite ladder of short cuts. Using the result of \cite{Gromov:2016rrp} we get the discard of $\Omega^k_j(u)$
\beq\label{Omega_discard}
\tilde{\Omega}^k_j-\Omega^k_j=-\bar{\tilde{\bQ}}_j \tilde{\bQ}^k+\bar{\bQ}_j \bQ^k\;.
\eeq
Multiplying both sides of \eqref{Omega_discard} by $M^{ij}$ and utilizing the first gluing condition from \eqref{gluing_conditions} we derive the equation\footnote{Notice that the RHS of the equation \eqref{MOmega_discard} coincides with the RHS of the equation of the $\bQ\omega$-system $\tilde{\omega}^{ij}-\omega^{ij}=-\bQ^i \tilde{\bQ}^j+\bQ^j \tilde{\bQ}^i$.}
\beq\label{MOmega_discard}
M^{ij}\tilde{\Omega}^k_j-M^{ij}\Omega^k_j=-\bQ^i \tilde{\bQ}^k+\bQ^k \tilde{\bQ}^i\;.
\eeq
We note that the RHS of \eqref{MOmega_discard} is antisymmetric in the indices $i$ and $k$, thus we conclude that the function $M^{ij}\Omega^k_j+M^{kj}\Omega^i_j$ has no cuts on the real axis. As this function is $i$-periodic it follows that $M^{ij}\Omega^k_j+M^{kj}\Omega^i_j$ is analytic in the whole complex plane.

Let us introduce a new notation
\beq\label{omega_definition}
\omega^{ik} \equiv M^{ij}\Omega_j^k\;,
\eeq
where $\omega^{ik}$ are the $i$-periodic functions on the sheet with the short cuts. Remembering \eqref{QtildeQ_relation} and applying \eqref{omega_definition} on the sheet with the short cuts we obtain the equation
\beq\label{omega_discard_physical}
\tilde{\hat{\omega}}^{ik}-\hat{\omega}^{ik}=\(\delta^i_j \hat{\bQ}^k \hat{\bQ}_l-\delta^k_j \hat{\bQ}^i \hat{\bQ}_l\)\hat{\omega}^{jl}\;.
\eeq
On the other hand, the functions $\omega^{ik}$ are $i$-periodic on the sheet with the short cuts, thus on the sheet with the long cuts their analytic continuation under the cut on the real axis is given by the simple formula
\beq
\tilde{\check{\omega}}^{ik}=\(\check{\omega}^{ik}\)^{++}\;.
\eeq
Rewriting \eqref{omega_discard_physical} on the sheet with the long cuts gives us
\beq\label{omega_discard_mirror}
\(\check{\omega}^{ik}\)^{++}-\check{\omega}^{ik}=\(\delta^i_j \check{\bQ}^k \check{\bQ}_l-\delta^k_j \check{\bQ}^i \check{\bQ}_l\)\check{\omega}^{jl}\;.
\eeq
If we return several steps back, we can derive from the QQ-relations the equation for the function $\cQ^{ab|ij}$, which looks almost exactly like \eqref{omega_discard_mirror}
\beq\label{Q4ind_eq}
\check{\cQ}^{ab|ij+}-\check{\cQ}^{ab|ij-}=\(\delta^i_k \check{\bQ}^j \check{\bQ}_l-\delta^j_k \check{\bQ}^i \check{\bQ}_l\)\check{\cQ}^{ab|kl-}\;.
\eeq
Recalling the notion of $\mu$-functions introduced in the QSC framework in \cite{Gromov:2013pga,Gromov:2014caa}, which are $i$-periodic on the sheet with the long cuts, we multiply both sides by $\check{\mu}_{ab}$, which leads us to
\beq
\(\check{\mu}_{ab}\check{\cQ}^{ab|ij-}\)^{++}-\(\check{\mu}_{ab}\check{\cQ}^{ab|ij-}\)=\(\delta^i_k \check{\bQ}^j \check{\bQ}_l-\delta^j_k \check{\bQ}^i \check{\bQ}_l\)\(\check{\mu}_{ab}\check{\cQ}^{ab|kl-}\)\;.
\eeq
Therefore, the functions $\check{\omega}^{ij}$ and $\check{\mu}_{ab}\check{\cQ}^{ab|ij-}$ satisfy the same equation. As the functions $\check{\mu}_{ab}\check{\cQ}^{ab|ij-}$ are antisymmetric in $i$ and $j$ due to the antisymmetry of the Q-function $\cQ^{ab|ij}$, it is natural to impose the constraint that $\omega^{ik}$ is also antisymmetric and
\beq\label{omega_antisymmetry}
\omega^{ik}=M^{ij}\Omega_j^k=-M^{kj}\Omega_j^i=-\omega^{ki}\;.
\eeq
In the following Subsections we are going to exploit \eqref{omega_antisymmetry} to constrain the gluing matrix for different spins $S_1$ and $S_2$.

\subsection{Parity symmetry}

Now we are going to describe the parity properties of the Q-system. For a large class of  states the $\bP$-functions possess the certain parity. Such states include the states with the charges $J_1=2$, $J_2=J_3=0$, which we consider in the rest of the paper and also the ground state with the charges $J_1=3$ and $J_2=J_3=0$,
which is relevant for the BFKL Odderon eigenvalue (see \cite{Beccaria:2011uz,Brower:2014wha}).
Thus, for the case $J_1=2$, $J_2=J_3=0$ we have
\begin{align}\label{global_charges}
& \tilde{M}_a=\left\{2, 1, 0, -1 \right\}, \\
& \hat{M}_i=\left\{\frac{1}{2}\(\Delta-S_1-S_2+2\), \frac{1}{2}\(\Delta+S_1+S_2\), \frac{1}{2}\(-\Delta-S_1+S_2+2\), \frac{1}{2}\(-\Delta+S_1-S_2\) \right\}. \notag
\end{align}

As we understood the analytic structure of $\bP$- and $\bQ$-functions, taking into account the asymptotics of these functions expressed in terms of the charges \eqref{global_charges}, it is natural to assume the existence of the certain symmetry between the Q-functions with lower and upper indices, which also changes sign of $\Delta$. This symmetry takes a particularly simple form for the following choice of the normalization of the $\bP$-
\beq\label{Adu_symmetry_fixed}
A_a=\(1,1,-{\bf A}_3,{\bf A}_4\)\;, \quad A^a=\({\bf A}_1,{\bf A}_2,-1,1\)\;,
\eeq
and $\bQ$-functions
\beq\label{Bdu_symmetry_fixed}
B_i=\(-{\bf B}_1,1,-{\bf B}_3,-1\)\;, \quad B^i=\(-1,{\bf B}_2,-1,-{\bf B}_4\)\;,
\eeq
which can be set with the usage of the rescaling symmetry \eqref{Q_rescaling}. We obtain\footnote{The change of the sign $\Delta \rightarrow -\Delta$ is a symmetry of the equation and it should map one solution to another solution. One can check that $\bP'^a=\chi^{ab}\bP_b$, $\bP'_a=\(\chi^{-1}\)_{ab} \bP^b$, $\bQ'^i=\eta^{ij}\bQ_j$, $\bQ'_i=\(\eta^{-1}\)_{ij}\bQ^j$ and $M'^{ij}=\eta^{ik}\(M^{-t}\)_{kl}\eta^{lj}$ is also a solution to the QQ-relations and the gluing conditions but with $\Delta$ flipped to $-\Delta$. As it can be seen explicitly in the Appendix \ref{weak_coupling_QSC} in the weak coupling limit these two solutions coincide therefore our solution is mapped onto itself. Given the starting point the recursive procedure described in the Section \ref{weak_coup_exp} is non-ambiguous we conclude that this property holds to all orders in the coupling constant.}
\beq\label{DeltaminusDelta_symmetry}
\bP^a(\Delta,u)=\chi^{ab}\bP_b(-\Delta,u)\;, \quad \bQ^i(\Delta,u)=\eta^{ij}\bQ_j(-\Delta,u)\;,
\eeq
where
\beq\label{chi}
\chi^{ab}=\(\begin{array}{cccc}
0 & 0 & 0 & -1 \\
0 & 0 & 1 & 0 \\
0 & -1 & 0 & 0 \\
1 & 0 & 0 & 0
\end{array}\)\;, \quad
\eta^{ij}=\(\begin{array}{cccc}
0 & -1 & 0 & 0 \\
1 & 0 & 0 & 0 \\
0 & 0 & 0 & 1 \\
0 & 0 & -1 & 0
\end{array}\)\;
\eeq
and $\chi^{ab}$ is the same matrix for the left-right symmetric states as in \cite{Gromov:2013pga,Gromov:2014caa}.

For the operators we examine ($J_1=2$ and $J_2=J_3=0$) in the following Sections the $\bP$-functions have the certain parity. Their parity is dictated by the asymptotics of the $\bP$-functions \eqref{PQ_asymptotics}
\beq\label{P_parity}
\bP_a(-u)=(-1)^{a+1}\bP_a(u)\;, \quad \bP^a(-u)=(-1)^a \bP^a(u)\;.
\eeq

The symmetry \eqref{P_parity} is a symmetry of the Baxter equation \eqref{Baxter_4th_order1}, thus $\bQ_i(-u)$ is also a solution of \eqref{Baxter_4th_order1}. Utilizing the same logic as in the case of the complex conjugation, we conclude that there exists an $i$-periodic matrix $\Theta_i^j(u)$ (see the proof in Appendix \ref{QSC_construction_app}) such that
\beq\label{Q_minus_u}
\bQ_i(-u)=\Theta_i^j(u)\bQ_j(u)\;.
\eeq
It is possible also to find the matrix with such a property. Utilizing again \eqref{uP}, we obtain
\beq\label{Theta_definition}
\Theta_i^j(u)=(-1)^{a+1}\cQ_{a|i}^-(-u) \cQ^{a|j-}(u)\;,
\eeq
where the summation over $a$ is implied. The matrix $\Theta$ has the property $\Theta_i^j(u) \Theta_j^k(-u)=\delta_i^k$ and thus $\Theta^{-1}(u)=\Theta(-u)$. We can write
\beq\label{Q_minus_u_inverse}
\bQ_i(u)=\Theta_i^j(-u)\bQ_j(-u).
\eeq
The matrix $\Theta_i^j(-u)$ also relates $\bQ^j(-u)$ and $\bQ^i(u)$
\beq\label{Q_minus_u2}
\bQ^j(-u)=\Theta_i^j(-u) \bQ^i(u)
\eeq
and vice versa
\beq\label{Q_minus_u_inverse2}
\bQ^j(u)=\Theta_i^j(u) \bQ^i(-u).
\eeq

Analogously to the consideration of parity symmetry, we can find the discard of $\Theta^k_j$ on the cut which is situated on the real axis
\beq
\tilde{\Theta}^k_j(u)-\Theta^k_j(u)=-\tilde{\bQ}_j(-u)\tilde{\bQ}^k(u)+\bQ_j(-u)\bQ^k(u)\;.
\eeq

Using the matrix $\Theta_i^j(u)$ from \eqref{Theta_definition}, the gluing conditions \eqref{QtildeQ_relation} and \eqref{Q_minus_u_inverse} we are able to introduce another gluing matrix
\beq\label{gliung_conditions_u_minus_u}
\tilde{\bQ}^i(u)=L^{ij}(u)\bQ_j(-u)\;, \quad \tilde{\bQ}_i=\(L^{-t}\)_{ij}(u)\bQ^j(-u)\;,
\eeq
where
\beq\label{L_M_relation}
L^{il}(u)=M^{ij}(u)\Omega_j^k(u)\Theta_k^l(-u)\;,
\eeq
and as the $\bQ$-functions on the both sides of the gluing conditions are LHPA, then by the same arguments as for $M^{ij}(u)$ the matrix $L^{ij}(u)$ is also analytic in the whole complex plane. Analogously to the case of the gluing matrix $M^{ij}(u)$ by going under the cut twice we derive the property
\beq\label{L_transpose}
L^{ji}(u)=L^{ij}(-u)\;.
\eeq

Together with \eqref{glm_hermiticity}, \eqref{omega_antisymmetry} and \eqref{L_M_relation} condition \eqref{L_transpose} constitutes the set of equations which are used to calculate the gluing matrix for different values of the spins $S_1$ and $S_2$.

Since for the states in question the $\bP$-functions have the certain parity, this has some consequences for the asymptotic expansion of the $\bQ$-functions. As it is explained in detail in the Section \ref{num_calc_QSC} with the description of the numerical algorithm the certain parity of the $\bP$-functions leads to the form \eqref{Qai_asymptotic_expansion_parity} of the asymptotic expansion of $\cQ_{a|i}$
\beq\label{Qai_asymptotic_expansion_parity3}
\cQ_{a|i}(u) \simeq u^{-\tilde{M}_a+\hat{M}_i}\sum\limits_{l=0}^{+\infty}\frac{B_{a|i,2l}}{u^{2l}}\;.
\eeq
By applying analogous arguments to the QQ-relation for the Hodge dual function $\cQ^{a|i}$, we conclude that the asymptotic expansion of $\cQ^{a|i}$ is also given by
\beq\label{Qai_asymptotic_expansion_parity2}
\cQ^{a|i}(u) \simeq u^{\tilde{M}_a-\hat{M}_i}\sum\limits_{l=0}^{+\infty}\frac{B^{a|i,2l}}{u^{2l}}\;.
\eeq
Then we remember that $\bQ_i=-\cQ_{a|i}^+ \bP^a$ and $\bQ^i=\cQ^{a|i+} \bP_a$, which after the substitution of \eqref{P_x_expansions_twist2}, \eqref{Qai_asymptotic_expansion_parity3} and \eqref{Qai_asymptotic_expansion_parity2} lead us to the asymptotic expansions at infinity of the $\bQ$-functions
\beq\label{Q_asymptotic_expansions}
\bQ_i(u) \simeq u^{\hat{M}_i-1}\(B_i+\sum\limits_{l=1}^{+\infty}\frac{B_{i,2l}}{u^{2l}}\)\;, \quad \bQ^i(u) \simeq u^{-\hat{M}_i}\(B^i+\sum\limits_{l=1}^{+\infty}\frac{B^{i,2l}}{u^{2l}}\)\;.
\eeq
In what follows we will regard to the $\bQ$-functions with the asymptotic expansions \eqref{Q_asymptotic_expansions} as having the ``pure" asymptotic, as these expansions contain the powers of $u$, which differ only by an integer number. The asymptotic expansions \eqref{Q_asymptotic_expansions} will be important in determining the structure of the matrices $\Omega_i^j(u)$ and $\Theta_i^j(u)$, which are analyzed below.

\subsection{Constraining the gluing matrix}\la{sec:constr}

In the present Subsection we are going to derive the set of equations for the elements of the gluing matrix originating from the conditions found in the previous Subsections. To remind briefly the QSC framework we are using let us recall the constraints on the gluing matrices known by now. The non-degenerate matrices $M^{ij}(u)$ and $L^{ij}(u)$ satisfy the following set of constraints
\begin{align}\label{M_L_constraints}
& \bar{M}^{ji}(u)=M^{ij}(u)\;, \quad M^{ij}(u)\Omega_j^k(u)=-M^{kj}(u)\Omega_j^i(u)\;, \quad \(\Omega^{-1}\)_i^j(u)=\bar{\Omega}_i^j(u)\;, \\
& L^{il}(u)=M^{ij}(u)\Omega_j^k(u)\Theta_k^l(-u)\;, \quad L^{li}(-u)=L^{il}(u)\;, \quad \(\Theta^{-1}\)_j^k(u)=\Theta_j^k(-u)\;. \notag
\end{align}
Now we are able to consider the gluing matrix for the case of different $AdS$ spins $S_1$ and $S_2$ solving the set of constraints \eqref{M_L_constraints}.

\subsubsection{Integer \texorpdfstring{$S_1$}{Lg} and \texorpdfstring{$S_2$}{Lg}}

Let us start our study from the situation when all the charges except for the dimension $\Delta$ are integer. More precisely, in the present Subsection we address the case when the spins $S_1$ and $S_2$ have the same parity. This is motivated by the fact that for $S_2=0$ the physical states have even non-negative $S_1$\footnote{This is due to the cyclicity constraint on the states of the $\mathfrak{sl}(2)$ Heisenberg spin chain, which is only consistent with the symmetric distribution of roots leading to $S_1$ even.}. To analyze the constraints \eqref{M_L_constraints} more closely we need to find the properties of the matrices $\Omega_i^j(u)$ and $\Theta_i^j(u)$.

In what follows we will need the asymptotics of the matrices $\Omega_i^j$ and $\Theta_i^j(u)$. To analyze them let us remember \eqref{Q_complex_conjugation} and \eqref{Q_minus_u}. As the asymptotics of the $\bQ$-functions on the both sides of \eqref{Q_complex_conjugation} and \eqref{Q_minus_u} are power-like and the $\Omega$-matrix consists of the $i$-periodic functions, the series expansions of $\Omega_i^j(u)$ and $\Theta_i^j(u)$ for $|\Re\;u| \gg 1$ are given by
\beq\label{Omega_Theta_matrix_asymptotic_expansion}
\Omega_i^j(u)=\sum\limits_{k=0}^{+\infty}\(\Omega^{(k)}_{\pm}\)_i^j e^{\mp 2\pi k u}\;, \quad \Theta_i^j(u)=\sum\limits_{k=0}^{+\infty}\(\Theta^{(k)}_{\pm}\)_i^j e^{\mp 2\pi k u}\;.
\eeq
where the signs correspond to expansion at $+\infty$ and $-\infty$ respectively. It should be noted that \eqref{Omega_Theta_matrix_asymptotic_expansion} does not have any growing terms on the RHS, because this would violate the power-like asymptotic of the $\bQ$-functions.

The asymptotics of the $\bQ$-functions are pure and the asymptotic expansion is given by \eqref{Q_asymptotic_expansions}. Looking at the values of $\hat{M}_i$, $i=1,\ldots,4$ one may think that because $\hat{M}_1-\hat{M}_2$ and $\hat{M}_3-\hat{M}_4$ are integers, there could potentially appear a mixing of $\bQ_1(u)$ with $\bQ_2(u)$ and $\bQ_3(u)$ with $\bQ_4(u)$ as this does not violate the purity of the asymptotics. However as the spins $S_1$ and $S_2$ have the same parity and
\beq
\hat{M}_1-\hat{M}_2=1\mod\;\; 2\;, \quad \hat{M}_3-\hat{M}_4=1\mod\;\; 2\;,
\eeq
the functions $\bQ_1$, $\bQ_2$ and $\bQ_3$, $\bQ_4$ cannot mix, because their asymptotic expansions \eqref{Q_asymptotic_expansions} contain only even powers of $u$ in the round brackets. Therefore the matrices $(\Omega^{(0)}_{\pm})_i^j$ and $(\Theta^{(0)}_{\pm})_i^j$ have to be diagonal. Let us now find them.

First, we consider the matrix $\Omega_i^j(u)$ and remember \eqref{Q_complex_conjugation}. If $\Re \;u$ tends to $+\infty$, then, as $\bar{\bQ}_i \simeq \bar{B}_i u^{\hat{M}_i-1}$, the diagonal element $(\Omega^{(0)}_+)_i^i$ is equal to $\bar{B}_i/B_i$. But if $\Re\;u$ tends to $-\infty$ the situation is a little more subtle. The functions $\bQ_i(u)$ have an infinite ladder of short cuts going down from the real axis, while the functions $\bar{\bQ}_i(u)$ have the same ladder of cuts going up. Then taking the limit of $\Re \;u$ to $-\infty$ we have to go to $-\infty$ along the semicircle in the UHP for $\bQ_i(u)$, i.e. $\bQ_i(u) \simeq B_i e^{i\pi(\hat{M}_i-1)}(-u)^{\hat{M}_i-1}$ and along the semicircle in the LHP for $\bar{\bQ}_i(u)$, i.e. $\bar{\bQ}_i(u) \simeq \bar{B}_i e^{-i\pi(\hat{M}_i-1)}(-u)^{\hat{M}_i-1}$, therefore we see that the diagonal element of $(\Omega^{(0)}_-)_i^i$ is equal to $\bar{B}_i/B_i e^{-2i\pi \hat{M}_i}$. To sum up, we obtain
\beq\label{Omega_matrix_asymptotic}
\Omega_i^j(u)=\left\{\begin{array}{c}
\delta_i^j e^{2i\phi_{B_j}}+(\Omega^{(1)}_+)_i^j e^{-2\pi u}+\cO\(e^{-4\pi u}\), \quad \Re\;u \gg 1\;, \\
\delta_i^j e^{2i\phi_{B_j}-2i\pi\hat{M}_j}+(\Omega^{(1)}_-)_i^j e^{2\pi u}+\cO\(e^{4\pi u}\), \quad \Re\;u \ll -1\;,
\end{array}\right.
\eeq
where $e^{2i\phi_{B_i}}=\bar{B}_i/B_i$.

Second, analyzing the matrix $\Theta_i^j(u)$ from \eqref{Q_minus_u} is analogous. Thus, applying the arguments from the previous paragraph, we see that at $\Re \;u$ tending to $+\infty$ we have to go around the semicircle in the UHP and $\bQ_i(-u) \simeq B_i e^{i\pi(\hat{M}_i-1)}u^{\hat{M}_i-1}$, while at $\Re \;u$ tending to $-\infty$ we have $\bQ_i(-u) \simeq B_i (-u)^{\hat{M}_i-1}$. Then we obtain
\beq
\label{Theta_matrix_asymptotic}
\Theta_i^j(u)=\left\{\begin{array}{c}
-\delta_i^j e^{i\pi\hat{M}_j}+(\Theta^{(1)}_+)_i^j e^{-2\pi u}+\cO\(e^{-4\pi u}\), \quad \Re\;u \gg 1\;, \\
-\delta_i^j e^{-i\pi\hat{M}_j}+(\Theta^{(1)}_-)_i^j e^{2\pi u}+\cO\(e^{4\pi u}\), \quad \Re\;u \ll -1\;.
\end{array}\right.
\eeq

As it was explained for example in \cite{Gromov:2017blm} in the strong coupling limit the asymptotics of the functions $\tilde{\bQ}^i(u)$ are some powers of $u$, then the only possible ansatz for the gluing matrix $M^{ij}(u)$ is to assume it to be a constant matrix. Thus for the case in question we obtain from \eqref{omega_antisymmetry} and \eqref{Omega_matrix_asymptotic} rather simple conditions
\begin{align}
& M^{ji}=-M^{ij}e^{2i(\phi_{B_j}-\phi_{B_i})}, \label{omega_antisymmetry_integer_charges1} \\
& M^{ji}=-M^{ij}e^{2i(\phi_{B_j}-\phi_{B_i})+2i\pi(\hat{M}_i-\hat{M}_j)}. \label{omega_antisymmetry_integer_charges2}
\end{align}
Combining the two conditions \eqref{omega_antisymmetry_integer_charges1} and \eqref{omega_antisymmetry_integer_charges2} we obtain the following
\beq\label{int_ch_cond1}
M^{ij}\(e^{2i\pi\(\hat{M}_i-\hat{M}_j\)}-1\)=0.
\eeq

Let us see now which additional restrictions do we have in the case $J_2=J_3=0$ and $J_1=2$. First of all from our assumptions about the asymptotics of the functions $\bQ^i(u)$ we understand that the gluing matrix $L^{ij}$ given by \eqref{L_M_relation} has to be constant, i.e. $L^{ij}(u)=L^{ij}$ is a symmetric matrix. Therefore, from the \eqref{L_M_relation}, \eqref{Omega_matrix_asymptotic} and \eqref{Theta_matrix_asymptotic} we immediately find
\beq
L^{ij}=-M^{ij}e^{2i\phi_{B_j}-i\pi\hat{M}_j}\;.
\eeq
Then, using \eqref{omega_antisymmetry_integer_charges1} and the symmetry of $L^{ij}$, we derive
\beq\label{int_ch_cond2}
M^{ij}\(e^{i\pi\(\hat{M}_i-\hat{M}_j\)}+1\)=0\;.
\eeq
It is easy to see that if \eqref{int_ch_cond2} is true then \eqref{int_ch_cond1} is also true. We have to calculate the differences between the charges $\hat{M}_i$ to determine which elements of the matrix $M^{ij}$ are non-vanishing. It appears that only $\hat{M}_1-\hat{M}_2$ and $\hat{M}_3-\hat{M}_4$ are integers
\begin{align}\label{M_int_ch_nonzero}
& \hat{M}_1-\hat{M}_2=-S_1-S_2+1\;, \\
& \hat{M}_3-\hat{M}_4=-S_1+S_2+1\;. \notag
\end{align}
Thus for the case of integer spins $S_1$ and $S_2$ we are left with the spins $S_1$ and $S_2$ with the same parity, which is consistent with our initial setup. Therefore, only the matrix elements $M^{12}=\bar{M}^{21}$ and $M^{34}=\bar{M}^{43}$ are non-zero. Then, in the case of integer spins $S_1$ and $S_2$ of the same parity we obtain the following gluing matrix
\beq\label{gluing_matrix_spins_same_parity}
M^{ij}=\(
\begin{array}{cccc}
0 & M^{12} & 0 & 0 \\
\bar{M}^{12} & 0 & 0 & 0 \\
0 & 0 & 0 & M^{34} \\
0 & 0 & \bar{M}^{34} & 0
\end{array}
\).
\eeq
Using also \eqref{omega_antisymmetry_integer_charges1} and \eqref{omega_antisymmetry_integer_charges2}, which are equivalent for $S_1$ of $S_2$ of the same parity as $\hat{M}_1-\hat{M}_2$ and $\hat{M}_3-\hat{M}_4$ equal 1 modulo 2, we are able to fix the phases of the non-zero matrix elements of \eqref{gluing_matrix_spins_same_parity}
\beq
M^{12}=\left|M^{12}\right|e^{i\(\pm\frac{\pi}{2}+\phi_{B_1}-\phi_{B_2}\)}\;, \quad M^{34}=\left|M^{34}\right|e^{i\(\pm\frac{\pi}{2}+\phi_{B_3}-\phi_{B_4}\)}\;.
\eeq

Now let us start the consideration of the case when at least one of the spins is not integer as this is particularly interesting for the BFKL limit.

\subsubsection{Non-integer \texorpdfstring{$S_1$}{Lg} and \texorpdfstring{$S_2$}{Lg}}

First of all, from the asymptotics \eqref{global_charges} we immediately see that if at least one of the charges $S_1$ or $S_2$ is non-integer, then not to violate the purity of the asymptotic expansions \eqref{Q_asymptotic_expansions} the matrices $(\Omega^{(0)}_{\pm})_i^j$ and $(\Theta^{(0)}_{\pm})_i^j$ cannot mix different $\bQ$-functions and have to be diagonal. Therefore, these matrices are given by \eqref{Omega_Theta_matrix_asymptotic_expansion}. This means, that in the case of at least one non-integer spin under the assumption that the gluing matrix is constant we obtain the same constraint \eqref{int_ch_cond2}. However, as $S_1$ or $S_2$ or both spins are non-integer, all the differences $\hat{M}_i-\hat{M}_j$ are non-integer in general, therefore we conclude that $M^{ij}=0$. Then we have to modify the ansatz for $M^{ij}(u)$.

The matrix $M^{ij}(u)$ is analytic and $i$-periodic, so the minimal choice would be to add the terms proportional to $e^{2\pi u}$ and $e^{-2\pi u}$
\beq\label{gluing_matrix_ansatz_noninteger}
M^{ij}(u)=M^{ij}_1+M^{ij}_2 e^{2\pi u}+M^{ij}_3 e^{-2\pi u}
\eeq
and this is consistent with what we know from the consideration of the BFKL limit for which $S_1$ approaches $-1$ and $S_2=0$ (see \cite{Alfimov:2014bwa}). From the previous conditions \eqref{glm_hermiticity} it follows that the matrices $M^{ij}_{1,2,3}$ are hermitian.

Substituting \eqref{gluing_matrix_ansatz_noninteger} into \eqref{omega_antisymmetry} we obtain the following conditions for the matrix $M^{ij}(u)$
\begin{align}
& M^{ji}_2=-M^{ij}_2 e^{2i\(\phi_{B_j}-\phi_{B_i}\)}\;, \label{omega_antisymmetry_noninteger_charges}\\
& M^{ji}_3=-M^{ij}_3 e^{2i\(\phi_{B_j}-\phi_{B_i}\)+2i\pi\(\hat{M}_i-\hat{M}_j\)}\;, \notag
\end{align}
where the summation over the repeated indices is not implied. For $i=j$ we immediately see from \eqref{omega_antisymmetry_noninteger_charges} that
\beq
M^{ii}_2=M^{ii}_3=0\;.
\eeq

Let us remember that for the case in question $J_2=J_3=0$ and $J_1=2$. The matrix $L^{ij}(u)$ is given by the formula \eqref{L_M_relation}. Taking the limits $u \rightarrow \pm\infty$ and remembering the expansions \eqref{Omega_matrix_asymptotic} and \eqref{Theta_matrix_asymptotic} we have to assume the existence of the exponential contributions to $L^{ij}(u)$
\beq
L^{ij}(u)=L^{ij}_1+L^{ij}_2 e^{2\pi u}+L^{ij}_3 e^{-2\pi u}\;,
\eeq
where the matrix $L^{ij}_{1}$ is symmetric and $L^{ji}_2=L^{ij}_3$ due to \eqref{L_transpose} and the latter two of them are given by
\begin{align}
& L^{il}_2=M^{ij}_2 (\Omega^{(0)}_+)_j^k (\Theta^{(0)}_-)_k^l=-M^{il}_2 e^{2i\phi_{B_l}-i\pi\hat{M}_l}\;, \\
& L^{il}_3=M^{ij}_3 (\Omega^{(0)}_-)_j^k (\Theta^{(0)}_+)_k^l=-M^{il}_3 e^{2i\phi_{B_l}-i\pi\hat{M}_l}\;.
\end{align}
Exploiting the symmetry $L^{ji}_3=L^{ij}_2$ and the relation from \eqref{omega_antisymmetry_noninteger_charges} we derive
\beq
M^{ij}_3=-M^{ij}_2 e^{i\pi(\hat{M}_j-\hat{M}_i)}\;. \label{commutativity_condition}
\eeq

As we observed in the case of integer spins $S_1$ and $S_2$ the determinant of the gluing matrix is constant. According to \eqref{gluing_matrix_ansatz_noninteger} in the case of at least one non-integer spin this determinant is not guaranteed to be integer. However, if we assume for a moment that the determinant of \eqref{gluing_matrix_ansatz_noninteger} contains exponents, the form of the second gluing condition from \eqref{gluing_conditions} will contain exponents in the denominator. But as there is no preference to upper and lower indices,
which get exchanged under $\Delta\to-\Delta$ symmetry,
we have to assume that both gluing conditions \eqref{gluing_conditions} include the exponents $e^{2\pi u}$ only in the numerator of $M^{ij}(u)$, therefore we impose a new constraint
\beq\label{constant_determinant_constraint}
\det_{1 \leq i,j \leq 4}M^{ij}(u)=\textrm{const}\;.
\eeq

Let us now show that staring from some simple ansatz for the gluing matrix $M^{ij}(u)$ we are able to solve the constraints \eqref{commutativity_condition} and \eqref{constant_determinant_constraint}. We saw from the implementation of the numerical algorithm described in the Section \ref{num_calc_QSC} that in the case when both spins $S_1$ and $S_2$ are non-integer it is sufficient for convergence of the numerical procedure to allow presence of exponents in $M^{13}(u)$ and $M^{14}(u)$ only. Thus, we have the following ansatz for the hermitian matrices $M^{ij}_{2,3}$
\beq
M^{ij}_{2,3}=\left(\begin{array}{cccc}
0 & 0 & M_{2,3}^{13} & M_{2,3}^{14} \\
0 & 0 & 0 & 0 \\
M_{2,3}^{31} & 0 & 0 & 0 \\
M_{2,3}^{41} & 0 & 0 & 0 \\
\end{array}\right)\;,
\eeq
where the non-zero matrix elements are subject to the relations \eqref{omega_antisymmetry_noninteger_charges} and \eqref{commutativity_condition}.

From the constraint \eqref{omega_antisymmetry} we find the equation
\beq
e^{2i\phi_{B_2}}M_1^{22}=e^{2i\phi_{B_2}-2i\pi\hat{M}_2}M_1^{22}=0\;,
\eeq
from which we have
\beq\label{M_diagonal_constraint}
M_1^{22}=0\;.
\eeq

Application of the constraint \eqref{constant_determinant_constraint}, i.e. the demand of the absence of the powers of $e^{2\pi u}$ in the determinant of the gluing matrix leads us to the following equations
\begin{align}\label{MM_constraint}
& M_1^{23}M_2^{14}=M_1^{24}M_2^{13}\;, \\
& M_1^{24}M_2^{13}\(e^{i\pi(S_2-S_1)}-1\)=0\;. \notag
\end{align}
As the spins $S_1$ and $S_2$ are non-integer and we assume their difference to be non-integer too and from \eqref{MM_constraint} we obtain
\beq\label{MM_constraint2}
M_1^{23}M_2^{14}=M_1^{24}M_2^{13}=0\;.
\eeq
In the case in question $M_2^{13}$ and $M_2^{14}$ are not equal to zero, therefore we are left with the equality
\beq\label{MM_constraint3}
M_1^{23}=M_1^{24}=0\;.
\eeq

Summarizing the results \eqref{M_diagonal_constraint} and \eqref{MM_constraint3} we obtain\footnote{In the case when the spin $S_2$ is integer, the gluing matrix simplifies to \eqref{gluing_matrix_integer_conformal_spin} and we have $M^{14}(u)=M^{44}(u)=0$ as we will see it in the Section \ref{num_calc_QSC} describing the applied QSC numerical algorithm.}
\beq\label{gluing_conditions_non_integer_spins_solution}
M=\left(\begin{array}{cccc}
M_1^{11} & M_1^{12} & M_1^{13} & M_1^{14} \\
\bar{M}_1^{12} & 0 & 0 & 0 \\
\bar{M}_1^{13} & 0 & M_1^{33} & M_1^{34} \\
\bar{M}_1^{14} & 0 & \bar{M}_1^{34} & M_1^{44} \\
\end{array}\right)+ \\
\left(\begin{array}{cccc}
0 & 0 & M_2^{13} & M_2^{14} \\
0 & 0 & 0 & 0 \\
\bar{M}_2^{13} & 0 & 0 & 0 \\
\bar{M}_2^{14} & 0 & 0 & 0 \\
\end{array}\right)e^{2\pi u}
+\left(\begin{array}{cccc}
0 & 0 & M_3^{13} & M_3^{14} \\
0 & 0 & 0 & 0 \\
\bar{M}_3^{13} & 0 & 0 & 0 \\
\bar{M}_3^{14} & 0 & 0 & 0 \\
\end{array}\right)e^{-2\pi u},
\eeq
where the elements of the matrices $M_{2,3}^{ij}$ are subject to \eqref{omega_antisymmetry_noninteger_charges} and \eqref{commutativity_condition}. As we have the relation \eqref{commutativity_condition} it is sufficient to write only the phases of the non-zero matrix elements of the matrix $M_2^{ij}$ extracted from \eqref{omega_antisymmetry_noninteger_charges}
\beq
M_2^{13}=\left|M_2^{13}\right|e^{i\(\pm\frac{\pi}{2}+\phi_{B_1}-\phi_{B_3}\)}\;, \quad M_2^{14}=\left|M_2^{14}\right|e^{i\(\pm\frac{\pi}{2}+\phi_{B_1}-\phi_{B_4}\)}\;.
\eeq

Let us point out that the construction presented above will provide an analytic continuation to all values of $S_2$ from the integer values $S_2\geq 0$. However,
this analytic continuation breaks down the symmetry $S_2\to -S_2$, which is naively present in the QSC, as one can see from the asymptotic \eqref{global_charges_all}.
The analytic continuation, which describes perfectly positive integer $S_2$ will produce poles at negative integer $S_2$. 
This could look a bit puzzling, but the resolution of this paradox is in the existence of the second solution for the mixing matrix which is obtained by relabeling indices in accordance with $S_2\to -S_2$. 
In practice result must be even in $S_2$ and it is enough to consider $S_2\geq 0$ so it is sufficient to use the mixing matrix presented above.

To sum up the contents of the present Section, we have to point out several things. First, we formulated the algebraic structure of the Q-system by writing down the QQ-relations and 4th order Baxter equation originating from them. Second, the analytic structure of the Q-system was motivated from the solution of the classically integrable dual superstring $\sigma$-model and the QQ-relations. The symmetries of the Q-system allowed us to introduce the gluing conditions for which we managed to impose several constraints. These constraints were partially solved for different values of the spins $S_1$ and $S_2$. We examined the case when both spins are integer and non-integer. In the next Section we are going to appreciate an importance of the derived gluing conditions and see how they appear in the QSC numerical algorithm.

\section{Numerical solution}\label{num_calc_QSC}

The equations of QSC are especially well-suited for numerical analysis: simple analytical properties of the $\bP$-functions allow to parametrize them in terms of a truncated Laurent series and then constrain these coefficients by the gluing condition. Numerical algorithms for solving QSC equations were developed and applied in \cite{Gromov:2015wca,Gromov:2015dfa,Gromov:2016rrp,Hegedus:2016eop}. In a non-symmetric case, such as BFKL with $S_2=n\ne0$, the procedure has to be modified in a way which we will describe here.
We attached a \verb"Mathematica" notebook named \verb"code_for_arxiv.nb" implementing the algorithm, which we used to obtain the results described in this Section.

Let us start by briefly reminding the main steps of the numerical algorithm. A comprehensive description of the algorithm for the left-right symmetric case can be found in \cite{Gromov:2015wca,Gromov:2017blm}. Here we will point out the main features we have to take into account in the case without left-right symmetry. As in the left-right symmetric case, for the $\bP$-functions there is a sheet with only one cut in the complex plane where the following parametrisation is valid
\begin{align}\label{P_x_expansions_twist2}
& \bP_a(u)=x^{-\tilde{M}_a}\(g^{-\tilde{M}_a}A_a\(1+\frac{\delta_{a,4}}{x^2}\)+\sum\limits_{k=1}^{+\infty}\frac{c_{a,k}}{x^{2k}}\)\;, \\
& \bP^a(u)=x^{\tilde{M}^a-1}\(g^{\tilde{M}_a-1}A^a\(1+\frac{\delta_{a,1}}{x^2}\)+\sum\limits_{k=1}^{+\infty}\frac{c^{a,k}}{x^{2k}}\)\;, \notag
\end{align}
where $x(u)=\frac{u+\sqrt{u-2g}\sqrt{u+2g}}{2g}$.
The expansions \eqref{P_x_expansions_twist2} contain only the even powers of Zhukovsky variable $x(u)$ because for the state in question the $\bP$-functions possess the certain parity determined by their asymptotics from \eqref{PQ_asymptotics} and \eqref{global_charges}. However, the coefficients $c_{a,k}$ and $c^{a,k}$ are not independent and are subject to the conditions following from
\beq\label{PP_condition}
\bP_a \bP^a=0\;.
\eeq
In the left-right symmetric case the condition \eqref{PP_condition} was satisfied automatically.

Since $\cQ_{a|i}$ is analytic in the UHP and has a power-like behaviour at $u\rightarrow \infty$, its  asymptotic expansion for sufficiently large $\im u$ in the UHP can be written as
\beq\label{Qai_asymptotic_expansion}
\cQ_{a|i}(u) \simeq u^{-\tilde{M}_a+\hat{M}_i}\sum\limits_{k=0}^{+\infty}\frac{B_{a|i,k}}{u^k}\;,
\eeq
where
\beq
B_{a|i,0}=-i\frac{A_a B_i}{-\tilde{M}_a+\hat{M}_i}\;.
\eeq
Plugging the ansatz \eqref{P_x_expansions_twist2} and the expansion of $\cQ_{a|i}$ into the equation
\beq\label{Qai_coeff_P_coeff}
\cQ_{a|i}^+-\cQ_{a|i}^-=-\bP_a \bP^b \cQ_{b|i}^+,
\eeq
we are able to fix the coefficients $B_{a|i,k}$ in terms of the operator charges and the coefficients $c_{a,k}$ and $c^{a,k}$. The fact that the $\bP$-functions have the certain parity and they are given by \eqref{P_x_expansions_twist2} leads to the disappearance of the odd coefficients $B_{a|i,2l+1}=0$ for $l=0,1,2,\ldots$ in \eqref{Qai_asymptotic_expansion} and we obtain
\beq\label{Qai_asymptotic_expansion_parity}
\cQ_{a|i}(u) \simeq u^{-\tilde{M}_a+\hat{M}_i}\sum\limits_{l=0}^{+\infty}\frac{B_{a|i,2l}}{u^{2l}}\;.
\eeq
After doing this, using the same finite difference equation in the form
\beq
\cQ_{a|i}^-=\(\delta_a^b+\bP_a \bP^b\)\cQ_{a|i}^+
\eeq
we find the numerical value of $\cQ_{a|i}$ in the vicinity of the real axis.

One remaining ingredient of the iterative numerical procedure is the loss function ~--- a function which is zero for the exact solution and which should decrease as each iteration brings us closer to the exact solution. We have the following loss function
\beq\label{loss_function}
S=\sum\limits_{i,j} |F^i(u_j)|^2\;,
\eeq
which is zero when the gluing condition is satisfied.
Here
\beq
F^i(u)=\cQ^{a|i+}(u)\tilde{\bP}_a(u)+M^{ij}(u)\bar{\cQ}^-_{b|i}(u)\bar{\bP}^b(u)
\eeq
and $\{u_i\}$ is a set of points on the interval $[-2g;2g]$. Every function $F^i(u)$ depends on the charges $S$, $\Delta$, $n$, the coefficients $c_{a,k}$ and $c^{a,k}$ and the coefficients of the gluing matrix. As a starting point for the numerical algorithm one can use the weak coupling data from Appendix \ref{weak_coupling_QSC}.

In the present work we are interested in the case of non-integer spin $S_1=S$. As it was already shown in the Section \ref{ext_QSC_nonint_sp} in this situation we cannot keep all the gluing conditions \eqref{gluing_matrix_spins_same_parity}. 
However, we found that only two gluing conditions $F^2$ and $F^4$ are sufficient to constrain all the coefficients $c$ and are still valid even for non-integer $S_1$ providing thus a natural way to analytically continue to non-integer spins. 
In terms of \eqref{loss_function} this means that the sum in that formula goes only over $i=2,4$. After the loss function and all the constraints are formulated, the algorithm searches for the parameters which minimize the loss function subject to the constraints using a numerical optimisation procedure (Levenberg-Marquardt algorithm). 
Then, using the obtained numerical values of the $\bQ$-functions, we are able to restore the ansatz for the gluing matrix, which tells us which elements of the gluing matrix contain the exponential terms and which of them are equal to zero.
This allows to verify the modification proposed in the Section \ref{ext_QSC_nonint_sp} for the gluing matrix \eqref{gluing_matrix_integer_conformal_spin} (in agreement with \cite{Gromov:2015vua}).
Thus for integer $S_2=n$ this leads to the gluing matrix for non-integer $S_1$ given by \eqref{gluing_conditions_non_integer_spins_solution} with
\beq\label{gluing_matrix_integer_conformal_spin}
M^{14}(u)=M^{44}(u)=0\;.
\eeq
In the situation when $S_2=n$ is not integer the gluing matrix \eqref{gluing_matrix_integer_conformal_spin} needs further modification. 
To achieve this we use the fact that relaxing the conditions \eqref{gluing_matrix_integer_conformal_spin}
is sufficient to make the numerical procedure convergent. Restoring again the gluing matrix, for general real value of the spin $S_2$ (or conformal spin $n$ in high-energy scattering terminology) we are left with the gluing matrix which coincides with \eqref{gluing_conditions_non_integer_spins_solution}.

\begin{figure}[ht]
\center{
\includegraphics[width=0.7\linewidth]{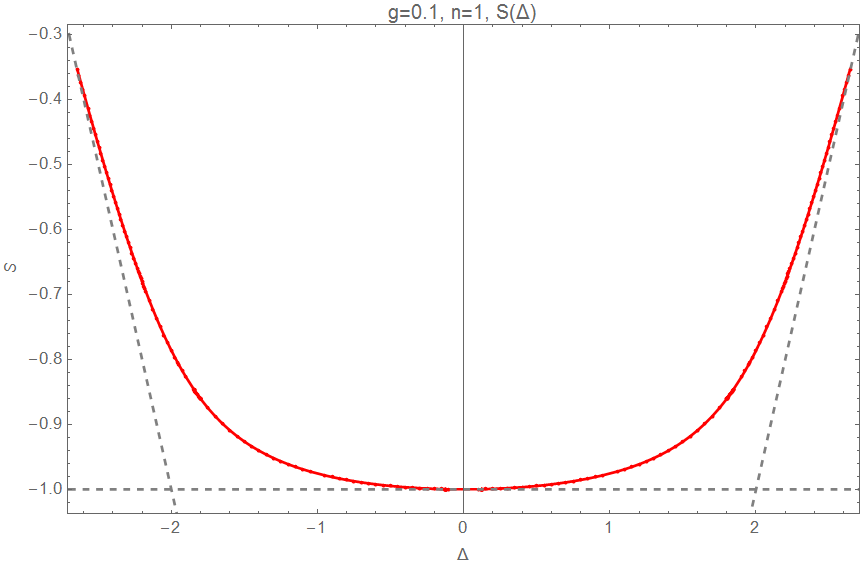}
\caption{Trajectory of the length-2 operator for conformal spin $n=1$ and coupling constant $g=0.1$.
\label{trajectory}}
}
\end{figure}

Using the proposed numerical algorithm, we managed to calculate several numerical quantities for the cases when $n$ is non-zero and even non-integer. On the Figure \ref{trajectory} one can find the length-2 operator trajectory for $n=1$.

\begin{figure}[h]
\center{
\includegraphics[width=0.7\linewidth]{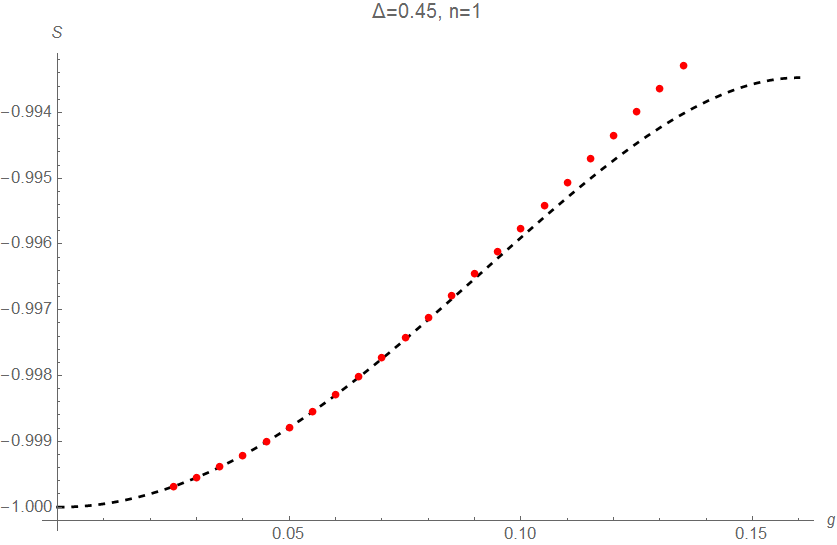}
\caption{Dependence of $S$ on the coupling constant for $\Delta=0.45$ and $n=1$. Red dots depict the numerical result and dashed line depicts the sum of analytical perturbative results at LO and NLO orders.
\label{S(g)}}
}
\end{figure}

It is also possible to numerically calculate the dependence of the spin $S$ on the coupling constant $g$ for the fixed dimension $\Delta$. On the Figure \ref{S(g)} you can see the dependence $S(g)$ for $\Delta=0.45$ and $n=1$ in comparison with the same result calculated perturbatively as the sum of LO and NLO BFKL eigenvalues.

Additionally, this numerical scheme allows us to compare the numerical values of BFKL kernel eigenvalues with the known perturbative eigenvalues at LO and NLO orders. In the Table \ref{Fitting} the numerical values of the BFKL kernel eigenvalue fitted from the plots of the Figure \ref{S(g)} are written in the first four orders together with perturbative results in the first two orders calculated for $n=1$ and $\Delta=0.45$. In the LO, NLO and NNLO order we observe the agreement with 22, 20 and 16 digits precision respectively.

\begin{table}
\centering{
\begin{tabular}{ | c | c | c | }
\hline
&  Numerical fit & Exact perturbative \\
\hline
LO & 0.50919539836118337091859 & 0.509195398361183370691860 \\
\hline
NLO & -9.9263626361061612225 & -9.9363626361061612225 \\
\hline
NNLO & 151.9290181554014 & 151.9290181554014 \\
\hline
NNNLO & -2136.77907308 & ? \\
\hline
\end{tabular}
}
\caption{BFKL kernel eigenvalues calculated numerically up to NNNLO and perturbatively up to NNLO for the conformal spin $n=1$ and dimension $\Delta=0.45$.}
\label{Fitting}
\end{table}

\begin{figure}[h]
\center{
\includegraphics[width=0.7\linewidth]{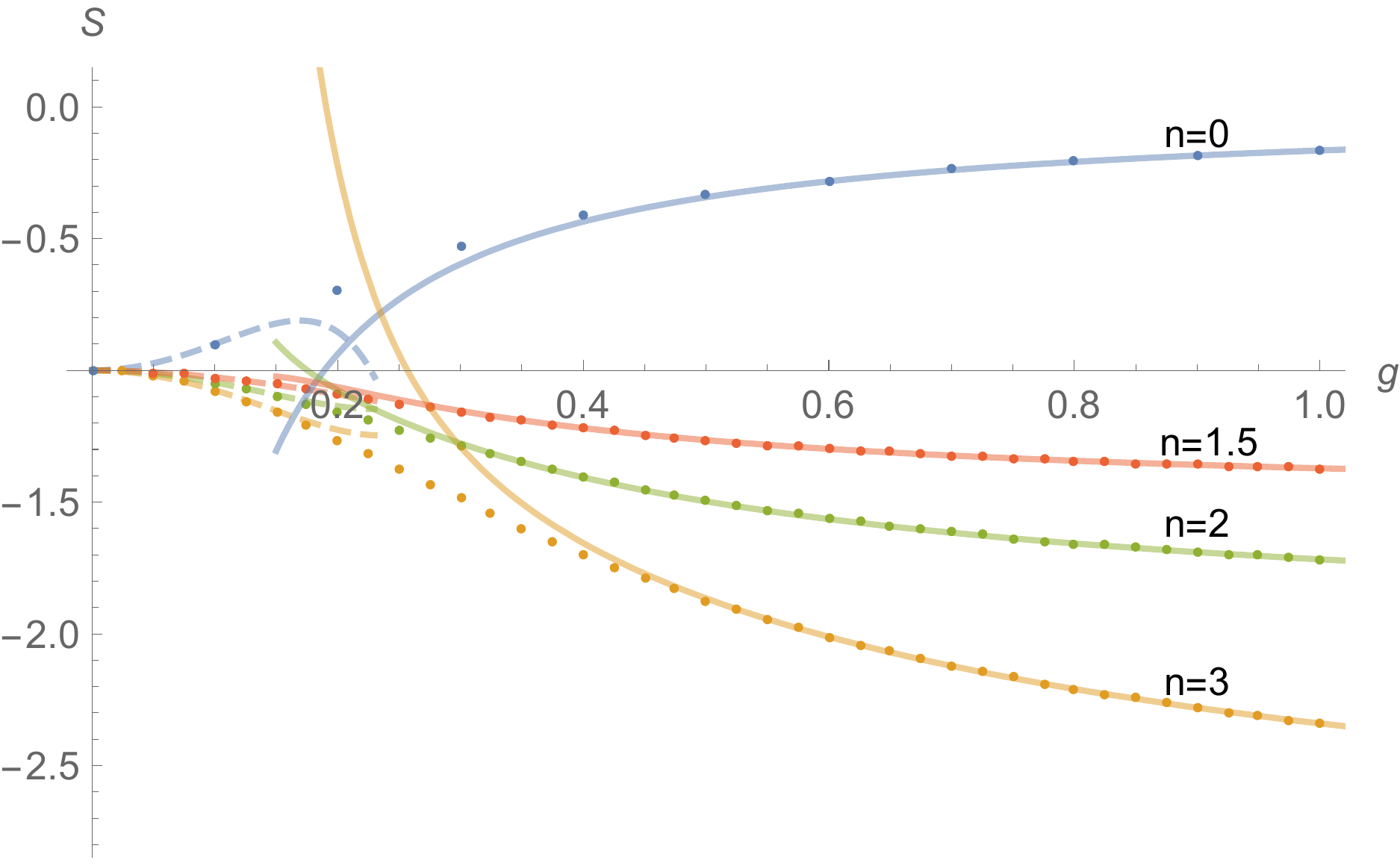}
\caption{Intercept $S(0,n)$ as the function of the coupling constant $g$ for conformal spins $n=0$, $n=3/2$, $n=2$ and $n=3$ (dots), weak coupling expansion of the intercept (dashed lines) and strong coupling expansion (continuous lines).
\label{intercept_numerics}}
}
\end{figure}

From now on let us concentrate on the numerical calculation of the intercept function. On the Figure \ref{intercept_numerics} one can find the dependencies of the intercept on the coupling constant $g$ for the different values of conformal spin $n$. The dashed lines are plotted according to the intercept function from Section \ref{weak_coup_exp} calculated in the small coupling regime. The continuous lines correspond to the strong coupling expansion of the intercept function from Section \ref{intercept_strong_coupling}, which was fitted from numerical data obtained in the present Section.

In the next Section we are going to analyze the weak coupling expansion of the intercept function. To achieve this we apply the iterative method first applied in \cite{Gromov:2015vua}. 

\section{Weak coupling expansion}\label{weak_coup_exp}

In this Section we explore the function $S(\Delta,n)$ perturbatively at weak coupling for arbitrary integer conformal spin $n$. In particular, we are interested in the BFKL intercept $j(n)=S(0,n)+1$. The calculation of this quantity consists of two steps.

First, we apply the QSC iterative procedure introduced in \cite{Gromov:2015vua} to the calculation of the intercept function for some integer $n$\footnote{The reason we have to take specific $n$ in our analytic calculations is that for arbitrary $n$ the leading order solution is already quite complicated hypergeometric function. It would be really great to extend the method of \cite{Gromov:2015vua} to be able to deal with this class of functions iteratively. This would allow one to derive the result for arbitrary $n$ at once.}. To do this we adopt this procedure to the case without the left-right symmetry. We repeat the main points of the iterative algorithm introduced in \cite{Gromov:2015vua} and describe the functions which are used in it for the case $n \neq 0$.

In the second part we formulate an ansatz for the weak coupling expansion of the intercept function for arbitrary value of conformal spin in terms of binomial harmonic sums and fix the coefficients of this ansatz using the values of the intercept at several integer $n$, which we calculate solving the QSC iteratively order by order. This approach appears to be successful in the NNLO order in the coupling constant allowing us to find the intercept function at this order, but at NNNLO order we were not able to fix the rational part of the result for arbitrary $n$ (see the details in the Subsection \ref{int_func_fit}). This is due to the lack of the generalized-``reciprocity" at the NNNLO order. It would be very interesting to understand why and how the reciprocity in $n$ is violated, which would allow to obtain the rational part with a greatly reduced basis of functions. We postpone this quesiton to the future investigation.

Let us also mention that the method we explain here should be also applicable for non-zero $\Delta$ when $\Delta+n$ takes odd integer values. The case of $n=0$ was considered in \cite{Gromov:2015vua}, and it was sufficient to take a few values of $\Delta$ in order to fix the NNLO dimension. This would be also interesting to investigate in the future.

\subsection{Description of the iterative procedure}\label{iterative_procedure}

The iterative procedure increasing the number of orders is essentially the same as described in \cite{Gromov:2015vua}, but modified to account for non left-right symmetric case. The procedure is based on applying a version of variation of parameters method applied to the equation
\beq\label{Qai_eq_m}
\cQ_{a|i}^+-\cQ_{a|i}^-=-\bP_a \bP^b \cQ_{b|i}^-\;,
\eeq
which is a simple consequence of \eqref{Qai_eq} and \eqref{uP}. Indeed, suppose the function $\cQ_{a|i}^{(0)}$ solves the equation \eqref{Qai_eq_m} up to a discrepancy $dS_{a|i}$
\beq
\cQ_{a|i}^{(0)+}-\cQ_{a|i}^{(0)-}+\bP_a \bP^b \cQ_{b|i}^{(0)-}=dS_{a|i}\;.
\eeq
The exact solution can be represented as the zero-order solution plus a correction. We expand the correction in the basis over the components of the zero-order solution
\beq\label{Qai_corrected}
\cQ_{a|i}=\cQ_{a|i}^{(0)}+b_i^{j+} \cQ_{a|j}^{(0)}\;.
\eeq
If the discrepancy is of the order of some small parameter $\epsilon$ to the power $m$, i.e. $dS_{a|i}=\cO(\e^m)$, then we can obtain the equation for the coefficients $b_i^k$ with the doubled precision
\beq\label{b_eq}
b_i^{k++}-b_i^k=(dS_{a|i}+b_i^j dS_{a|j})\cQ^{(0)a|k+}=dS_{a|i}\cQ^{(0)a|k+}+\cO(\e^{2m})\;.
\eeq
The discrepancy thus becomes two orders smaller with each iteration step.
  
For the problem in question we restrict ourselves to the situation when $\Delta=0$, $n$ is an integer number and we perform the expansion in the parameter $\e=g^2$. To start solving the finite difference equation \eqref{b_eq}, we have to find the zero-order in $g$ solution $\cQ_{a|i}^{(0)}$. The way to find these functions is to consider the 4-th order Baxter equation \eqref{Baxter_4th_order1} with the $\bP$-functions in the coefficients of it in the LO given by \eqref{P_BFKL_LO}, \eqref{A3A4_BFKL_LO}, \eqref{coefficients_analyticity_constraints} and \eqref{gen_LR_sym} with $\Delta=0$. For example,
the solution of this equation with the pure asymptotics \eqref{Q_asymptotic_expansions} in the LO for $n=7$ is
\begin{align}\label{Q_func_LO}
    & \bQ^{(0)}_1=\frac{3}{175}\(-iu^4 \eta_2+\frac{i}{5}u^2 \eta_2+u^3+\frac{i}{2}u^2-\frac{11}{30}u-\frac{i}{10}\)\;, \quad \bQ^{(0)}_2=u^2\;, \quad \bQ^{(0)}_3=u^4-\frac{u^2}{5}\;, \\
    & \bQ^{(0)}_4=-\frac{9}{1715}\(\(-iu^4+\frac{8iu^2}{49}\) \eta_2+\frac{9i}{49}\(u^4-\frac{u^2}{5}\) \eta_4+\right. \notag \\
    & \qquad \qquad \qquad \qquad \qquad \qquad \qquad \qquad \qquad \qquad \left.+u^3+\frac{iu^2}{2}-\frac{115}{294}u-\frac{17i}{98}+\frac{33}{490u}+\frac{9 i}{490u^2}\)\;, \notag
\end{align}
where $\eta_s(u)$ are examples of the so-called $\eta$-functions, whose definition is given below. As we checked by solving the 4th order Baxter equation \eqref{Baxter_4th_order1} for different values of the conformal spin $n$, for odd values of $n$, as for the case $n=0$ described in  \cite{Gromov:2015vua}, the $\bQ$-functions in the LO in $g$ at $\Delta=0$ can be expressed as linear combinations of the $\eta$-functions with the coefficients being Laurent polynomials\footnote{Laurent polynomial is a polynomial with both positive and negative powers of the variable plus a constant term.} in $u$ plus a Laurent polynomial in $u$ without $\eta$-function multiplying it as in \eqref{Q_func_LO}. The $\eta$-functions were introduced in \cite{Leurent:2013mr} and then used in \cite{Marboe:2014gma,Gromov:2015vua,Gromov:2015dfa} with their generalized version in \cite{Gromov:2016rrp} and for ABJM theory in \cite{Anselmetti:2015mda} with an application in \cite{Lee:2017mhh} as
\beq\label{eta_func_def}
\eta_{s_1,\ldots,s_k}(u)=\sum\limits_{n_1>n_2>\ldots>n_k \geq 0}\frac{1}{(u+in_1)^{s_1} \ldots (u+in_k)^{s_k}}\;,
\eeq
where $s_i \geq 1$, $i=1,\ldots,k$. Unfortunately, for non-zero even $n$ we were not able to determine the class of functions to which $\bQ^{(0)}_i$ belong, therefore from this moment we restrict ourselves to the odd values of $n$. Then, applying the equation
\beq\label{Qai_interations_LO}
\cQ^{(0)+}_{a|i}-\cQ^{(0)-}_{a|i}=\bP^{(0)}_a \bQ^{(0)}_i\;,
\eeq
where $\bP^{(0)}_a$ are given by \eqref{P_BFKL_LO} with $\Delta=0$, we can find the functions $\cQ_{a|i}$ in the leading order in the coupling constant.

To describe the solution to \eqref{Qai_interations_LO} we have to explain some properties of $\eta$-functions. This class of functions is particulary convenient because it is closed under all relevant for us operations. First, a product of two $\eta$-functions can be expressed as a linear combination of $\eta$-functions using the so-called ``stuffle" relations \cite{Duhr:2014woa}. Second, the solution of the equation of the form \beq\label{eq_findiff}
f(u+i)-f(u) = u^n \eta_{s_1,\dots, s_k} 
\eeq
can be expressed as a linear combination of $\eta$-functions with the coefficients being Laurent polynomials in $u$. These two properties make $\eta$-functions very useful when solving the QSC perturbatively at weak coupling \cite{Leurent:2013mr,Marboe:2014gma,Gromov:2015vua}. The described properties of the $\eta$-functions lead us to the conclusion that at least $\cQ^{(0)}_{a|i}$ are also expressed as linear combinations of the $\eta$-functions with the coefficients being Laurent polynomials in $u$ plus a Laurent polynomial in $u$. Then, recalling that the $\bP$-functions in the NLO in the coupling constant are Laurent polynomials in $u$ as well (see, for example, the formulas \eqref{P_BFKL_NLO} and \eqref{P_BFKL_NNLO}), we see that the discrepancy $dS_{a|i}$ and the product in the RHS of \eqref{b_eq} are also of the form of a linear combination of the $\eta$-functions with the Laurent polynomial coefficients plus a Laurent polynomial.

We call the operation inverting the linear operator in the LHS of \eqref{eq_findiff} ``periodization" (for a more precise definition see \ref{sec:NLOsolution}). It is easy to see that the ``periodization" operation solves the equation \eqref{b_eq} if the zero-order approximation entering the RHS is expressed as a linear combination of $\eta$-functions with the coefficients being Laurent polynomials in $u$ plus a Laurent polynomial in $u$. We were able to find such representation for odd values of $n$, but not for even ones. After the zero-order solution is found, we iterate it as described above, applying the operation of periodization to the RHS of \eqref{b_eq} in order to find the coefficients $b_i^k$. Because of the two properties of $\eta$-functions mentioned above, at each iteration the solution is again obtained in the form of a linear combination of $\eta$-functions with Laurent polynomial coefficients plus a Laurent polynomial.

After finding the corrected $\cQ_{a|i}$ \eqref{Qai_corrected} at the given iteration step we still have some unfixed coefficients in it including the quantity of our interest $S(0,n)$. To find them we calculate the $\bQ$-functions from \eqref{uQ} and \eqref{uP} and apply the gluing conditions for the case of integer conformal spin $n$, i.e. \eqref{gluing_conditions_non_integer_spins_solution} with  \eqref{gluing_matrix_integer_conformal_spin} satisfied. As it was explained in the Section \ref{ext_QSC_nonint_sp} in these gluing conditions the $\bQ$-functions has to possess the pure asymptotics \eqref{Q_asymptotic_expansions}. Therefore we find the combinations of the $\bQ$-functions with the pure asymptotics. It appears to be sufficient to use only 2 of 4 gluing conditions, namely
\begin{align}
& \tilde{\bQ}^2=\bar{M}_1^{12}\bar{\bQ}_1\;, \\
& \tilde{\bQ}^4=\bar{M}_1^{34}\bar{\bQ}_3\;, \notag
\end{align}
taking the $\bQ$-functions on the cut on the real axis. This procedure allows to fix the remaining unknown coefficients including the function $S(0,n)$. The described method allowed us to find the values of the intercept functions for odd conformal spins in the range from $n=1$ to $n=91$ up to NNNLO order in the coupling constant. These data will be used in the next Subsection, where we put forward the ansatz for the structure of the intercept function for arbitrary value of conformal spin.

\subsection{Multiloop expansion of the intercept function for arbitrary conformal spin}\label{int_func_fit}

Using the procedure described in the previous Subsection \ref{iterative_procedure} we have calculated the expansion of the BFKL eigenvalue intercept for odd $n$ up to $n=91$ in the weak coupling limit up to the order $g^8$ (NNNLO). These data are valuable by themselves, as they can serve as a test for future higher-order or non-perturbative calculations. What is more important, however, is that it allowed us to find NNLO and partially NNNLO BFKL eigenvalue intercept as a function of the conformal spin $n$. 

We start by noticing that LO and NLO BFKL intercept can be represented as a linear combination of nested harmonic sums of uniform transcendentality. Indeed, the LO and NLO BFKL Pomeron eigenvalues themselves can be expressed (see, for example, \cite{Kotikov:2002ab} and Appendix \ref{app:NLOBFKL}, where this calculation is explained in details) through the nested harmonic sums described, for example, in \cite{Costa:2012cb}
\beq\label{harmonic_sums}
S_{a_1,a_2,\dots,a_n}(x)=\sum\limits_{y=1}^{x} \frac{\textrm{sign}(a_1)^y}{y^{|a_1|}}S_{a_2,\dots,a_n}(y)\;, \quad S_\emptyset(x)=1\;,
\eeq
where $x$ is a positive integer. The indices $a_i$ are non-zero integer numbers and the transcendentality of the given nested harmonic sum is defined as the sum $\sum\limits_{i=1}^{n}|a_i|$. Note that if one of the indices $a_i$ is negative, the formula \eqref{harmonic_sums} holds only for even integer $x$. In the literature \cite{Kotikov:2005gr,Kazakov:1987jk,Lopez:1980dj,Kotikov:2004er,Blumlein:2009ta} there was described the analytic continuation of the harmonic sums in question from the positive integer even $x$. To work with such a continuation we utilize the Mathematica package \verb"Supppackage" applied in \cite{Gromov:2015vua}. One can take $\Delta=0$ in these eigenvalues, which after some simple algebra gives
\begin{align}\label{intercepts}
& j_{LO}(n)=8S_1\(\frac{n-1}{2}\)\;, \\
& j_{NLO}(n)=4S_3\(\frac{n-1}{2}\)+4S_{-3}\(\frac{n-1}{2}\)-8S_{-2,1}\(\frac{n-1}{2}\)+\frac{2\pi^2}{3}S_1\(\frac{n-1}{2}\)\;. \notag
\end{align}
Here and below the transcendentality is computed as follows: the transcendentality of a product is assumed to be equal to the sum of transcendentalities of the factors and transcendentality of a rational number is 0. Transcendentality of $\log 2$ is 1 and transcendentality of $\zeta_k$ is $k$. Since $\zeta_k$ for even $k$ is proportional to $\pi^2$, it is easy to see that transcendentality of $\pi$ is $1$.

To conduct the calculations with harmonic sums one can use the \verb"HarmonicSums" package for Mathematica \cite{Ablinger:2010kw,Ablinger:2013hcp,Ablinger:2013cf,Ablinger:2011te,Blumlein:2009ta,Remiddi:1999ew,Vermaseren:1998uu} or the \verb"Supppackage" utilized by the authors of \cite{Gromov:2015vua}. It should be noted, that in the present work we utilize the same conventions for the analytic continuation of harmonic sums as in the latter work \cite{Gromov:2015vua}. As we see, the argument of all the harmonic sums in \eqref{intercepts} is $(n-1)/2$. This leads one to an idea of trying to find NNLO and NNNLO intercepts as analogous linear combinations of harmonic sums with transcendental coefficients of uniform transcendentality. The coefficients of the linear combination can be constrained using the data generated by the iterative procedure. But the number of harmonic sums of certain transcendentality grows fast as transcendentality increases. Fortunately, one can drastically reduce the number of harmonic sums in the ansatz by conjecturing a certain property of the result we call reciprocity. 
 
The property in question \cite{Dokshitzer:2005bf,Dokshitzer:2006nm,Basso:2006nk,Alday:2015eya} is parallel to the Gribov-Lipatov reciprocity \cite{Gribov:1972ri,Gribov:1972rt} and was observed in the weak coupling expansion of the scaling dimensions of the twist operators. Let us remind the statement of the reciprocity: if one defines an auxiliary function ${\cal P}$ \cite{Dokshitzer:2005bf,Dokshitzer:2006nm,Basso:2006nk} such that the anomalous dimension $\gamma$ of the operator with the spin $M$ satisfies in all orders in the coupling constant
\beq\label{Pdef}
\gamma(M)={\cal P}\(M+\frac{\gamma(M)}{2}\)
\eeq
then the inverse Mellin transform of $\cal P$ has the property
\beq\label{reciprocity}
\{\mathcal{M}^{-1}\mathcal{P}\}(x)=-x\{\mathcal{M}^{-1}\mathcal{P}\}\(\frac{1}{x}\)\;.
\eeq
The asymptotic expansion of the function ${\cal P}(M)$ for large $M$ then acquires a nice property: it consists only of the powers and possibly logarithms of $M(M+1)$ thus possessing the symmetry $M \rightarrow -1-M$.

Thus the function $ {\cal P}(M)$ is much more convenient to work with than $\gamma(M)$ itself: the function $\gamma(M)$ can be expressed through the nested harmonic sums, while $ {\cal P}$, on the other hand, can be expressed through a much smaller class of functions which satisfy the property \eqref{reciprocity}. Such functions were identified and used in \cite{Dokshitzer:2006nm,Beccaria:2007bb,Beccaria:2009eq,Beccaria:2009vt} as reciprocity-respecting harmonic sums. However, in our calculations we use another basis of the functions satisfying \eqref{reciprocity}, which was applied in the works \cite{Lukowski:2009ce,Velizhanin:2010cm,Velizhanin:2010vw,Velizhanin:2011pb,Velizhanin:2013vla,Marboe:2014sya,Marboe:2016igj}. These functions are called the binomial harmonic sums and for integer $M$ they are defined as (see \cite{Vermaseren:1998uu})
\beq\label{binomial}
\mathbb{S}_{i_1,\dots,i_k}(M)=(-1)^M \sum\limits_{j=1}^M(-1)^j \(\begin{array}{c}
M\\
j
\end{array}\)
\(\begin{array}{c}
M+j\\
j
\end{array}\)
S_{i_1,\dots,i_k}(j)\;.
\eeq 
Note that we consider only the positive indices $i_l$, $l=1,\ldots,k$ in the definition \eqref{binomial}. Those are exactly the sums whose asymptotic expansion is even at infinity after the argument is shifted by $1/2$.

All this is directly applicable to our case and we are able to formulate an ansatz for the NNLO intercept function. From the LO and NLO expressions \eqref{intercepts} we see that their asymptotic expansions at large $n$ are even in $n$. Since we are using the harmonic sums of the argument $M=(n-1)/2$, we need to keep only the harmonic sums invariant under the transformation $M\rightarrow -1-M$ or $n \rightarrow -n$ in our notations. Those are exactly the binomial sums \eqref{binomial}. The expressions \eqref{intercepts} for LO and NLO intercepts can be easily expressed through them
\begin{align}
& j_{LO}=4\mathbb{S}_1\;, \\
& j_{NLO}= 8\(\mathbb{S}_{2,1}+\mathbb{S}_3\)+\frac{4\pi^2}{3}\mathbb{S}_1\;, \notag
\end{align}
where the arguments of the sums are again $(n-1)/2$.
 
In order to find the NNLO intercept we make an ansatz in a form of a linear combination of binomial harmonic sums with transcendental coefficients. The maximal transcendentality principle, formulated by L.N. Lipatov and A.V. Kotikov \cite{Kotikov:2001sc,Kotikov:2002ab}, holds for the intercept as well: every term in the sum should be of the total transcenedentality $5$. The terms of the sum can of course be multiplied by arbitrary rational coefficients which do not affect the transcendentality.  Having constructed the ansatz in this way, we can now constrain its coefficients by the iterative data: we evaluate the ansatz (a linear combination of binomial nested harmonic sums) at several integer values of $n$ and match the result to the data obtained from the numerical procedure for the corresponding $n$. Equating the coefficients in front of each unique product of transcendental constants in these two expressions, we get a linear system for the rational coefficients of the ansatz. Solving it we obtain a surprisingly simple expression
\beq\label{NNLO_intercept_function}
j_{NNLO}=32\(\mathbb{S}_{1,4}-\mathbb{S}_{3,2}-\mathbb{S}_{1,2,2}-\mathbb{S}_{2,2,1}-2 \mathbb{S}_{2,3}\)-\frac{16\pi^2}{3} \mathbb{S}_3-\frac{32\pi^4}{45} \mathbb{S}_1\;.
\eeq
The result \eqref{NNLO_intercept_function} for the intercept function for arbitrary $n$ can be compared with the other known quantities. First of them is the NNLO BFKL Pomeron eigenvalue for the conformal spin $n=0$ calculated in \cite{Gromov:2015vua}. Taking in this eigenvalue $\Delta=0$ and comparing it with \eqref{NNLO_intercept_function} for $n=0$ we see perfect agreement. Second, for non-zero conformal spins the formulas for the Pomeron trajectories were found in \cite{Caron-Huot:2016tzz}, from which we can extract the intercept for given $n$. We also checked that the result of that work coincides with our result \eqref{NNLO_intercept_function} for several first non-negative conformal spins $n$, thus representing an independent confirmation of the correctness of our calculation.

The same procedure can be repeated in the NNNLO. The values of the NNNLO intercept for several first odd values of the conformal spin $n$ are given in the Appendix \ref{NNNLO_intercept_app}. Again, as for NNLO, an ansatz in a form of a linear combination of binomial harmonic sums with transcendental coefficients of uniform transcendentality 7 can be constructed and we attempted to fit it to the iterative data. However, we found that the basis of binomial harmonic sums is insufficient to fit the data. This signals that reciprocity understood as parity under $n$ to $-n$ seems to be broken down in this case. The reasons for this are unclear and will be the subject of further work. However, we managed to fit the certain part of the NNNLO data.

For each odd $n$ we calculated (see Appendix \ref{NNNLO_intercept_app} for several first conformal spins $n$) the value of the NNNLO intercept is a linear combination of the transcendental constants consisting of $\pi$, $\zeta_3$, $\zeta_5$ and $\zeta_7$ with rational coefficients and a rational number (see the file \verb"intercept_values_Nodd.mx" with the data for the odd conformal spins from $n=1$ to $n=91$ in the arXiv submission of this paper). Let us restrict ourselves to the values of the conformal spin in our data equal to $n=4k+1$ with $k=0,1,\ldots,22$. In these points the values of the intercept functions are given by
\begin{multline}\label{4kplus1_data}
j_{NNNLO}(4k+1)=\pi^2 j_{NNNLO}^{\pi^2}+\pi^4 j_{NNNLO}^{\pi^4}+\pi^6 j_{NNNLO}^{\pi^6}+\pi^2 \zeta_3 j_{NNNLO}^{\pi^2 \zeta_3}+ \\
+\zeta_3 j_{NNNLO}^{\zeta_3}+\zeta_5 j_{NNNLO}^{\zeta_5}+j_{NNNLO}^{\textrm{rat.}}\;,
\end{multline}
where all coefficients in front of the transcendental constants on the RHS of \eqref{4kplus1_data} are rational functions of $k$\;. Each coefficient in the RHS of \eqref{4kplus1_data} is conjectured to be be a linear combination with rational coefficients of the binomial harmonic sums with the transcendentality, supplementing the transcendentality of the corresponding coefficient to $7$. We were able to fit all the contributions except for $j_{NNNLO}^{\zeta_3}$ and $j_{NNNLO}^{\textrm{rat.}}$, which, as other harmonic sums, take rational values at the points $n=4k+1$ for integer $k \geq 0$. However, we found that the term $j_{NNNLO}^{\zeta_3}$ cannot be fitted with the ansatz consisting of the binomial harmonic sums \eqref{binomial}. This motivated us to try to fit this contribution with the nested harmonic sums \eqref{harmonic_sums}. This appeared to be really the case and we managed to fit this part with the ordinary harmonic sums, which means that the reciprocity, i.e. the symmetry $n \rightarrow -n$ in the asymptotic expansion, is violated. For the last, rational contribution $j_{NNNLO}^{\textrm{rat.}}$, we also found that it is not described by the binomial harmonic sums. Unfortunately, fitting this contribution with the ordinary harmonic sums did not lead us to completely fixing this contribution due to the lack of data. Therefore combining the obtained results we write down the non-rational part of the answer for the points $n=4k+1$, which is the sum of the terms in the RHS of \eqref{4kplus1_data} except for $j_{NNNLO}^{\textrm{rat.}}$
\begin{multline}\label{NNNLO_intercept_function_non_rational}
j_{NNNLO}^{\textrm{non-rat.}}(4k+1)= \\
=-\frac{32\pi^2}{3}\(3\mathbb{S}_{1,4}-3\mathbb{S}_{2,3}-\mathbb{S}_{3,2}+\mathbb{S}_{1,1,3}-2\mathbb{S}_{1,2,2}+\mathbb{S}_{2,2,1}-\mathbb{S}_{3,1,1}\)+\frac{16\pi^4}{15}\(4 \mathbb{S}_3-\mathbb{S}_{2,1}\)+\frac{56\pi^6}{135}\mathbb{S}_1+ \\
+\frac{32\pi^2 \zeta_3}{3}\mathbb{S}_{1,1}+224\zeta_5 \mathbb{S}_{1,1}-128\zeta_3 \(S_{-3,1}+2S_{-2,2}-5S_{1,-3}-15S_{1,3}-4S_{2,-2}-12S_{2,2}-\right. \\
\left.-15S_{3,1}-4S_{-2,1,1}+2S_{1,-2,1}+8S_{1,1,-2}+12S_{1,1,2}+12S_{1,2,1}+12S_{2,1,1}+S_{-4}+9S_4\)\;.
\end{multline}
One can find the full values of the NNNLO intercept function including the rational terms for the conformal spins $n=4k+1$ from $1$ to $89$ in the arXiv submission of this paper in the file named \verb"intercept_values_Nodd.mx". As the part of the harmonic sum consisting of the nested harmonic sums of the transcendentality $7$, which constitutes the rational part at the points $n=4k+1$, was not fitted we are unable to write an expression for the NNNLO intercept working for all conformal spins leaving this task for future studies.

Let us briefly summarize the results of the present Section. In the first Subsection applying the QSC iterative algorithm we found the values of the intercept up to NNNLO order in the coupling constant in the certain range of odd values of the conformal spin. In the second Subsection we saw that the LO and NLO intercept functions satisfy the reciprocity symmetry, which allowed us to rewrite them in terms of the binomial harmonic sums and using the ansatz in terms of these sums in the NNLO order, fix the answer for the NNLO intercept for arbitrary conformal spin. In the NNNLO order the reciprocity breaks down, nevertheless, for the conformal spins $n=4k+1$ we managed to describe the non-rational part of the NNNLO intercept function in terms of binomial and ordinary nested harmonic sums.

\section{Near-BPS all loop expansion}\label{near_BPS_exp}

In this Section we are going to analyze the QSC equations near the BPS point $\Delta=0$, $S_1=-1$ and $S_2=1$. It appears that it is possible to calculate two non-perturbative quatities in this point by the methods of QSC.
Another BPS-point $S_2=0,\;S_1=0$ was analyzed in detail in \cite{Basso:2011rs,Gromov:2012eg,Gromov:2014bva}.
In this Section we follow closely the Near-BPS expansion method by \cite{Gromov:2014bva}.

\subsection{Slope of the intercept near the BPS point}\label{sec:Slope}

A particularly important role in BFKL computations is played by the intercept function $j(n)=S(0,n)+1$, where $S=S_1$ and $n=S_2$. As we mentioned above, the point $\Delta=0,\; n=1$ is BPS, by which we mean that it is fixed for any `t Hooft coupling. The group-theoretical argument explaining this phenomenon should be based on the shortening condition.
From the QSC perspective the BPS points are the points where $A^a A_a=0$
simultaneously for all $a=1,\dots,4$.
In this Subsection we study small deviations from this BPS point and calculate the slope of $j(n)$ with respect to $n$ in the point $n=1$ in all orders in the coupling constant $g$. 

\subsubsection{LO solution}

The fact that at the BPS point $A^a A_a=0$ usually leads to more powerful condition $\bP_a=\bP^a=0$, which is known to lead to considerable simplifications \cite{Gromov:2014bva} (see also \cite{Gromov:2012eu} for a similar simplification in the TBA equations). Based on that we also expect that in our situation the Q-system simplifies a lot near the BPS point. Let us show that, indeed, $\cQ_{a|i}$ takes a very simple form for $S_2=1,\Delta=0$.

Recall that the functions $\cQ_{a|i}$ satisfy the equation 
\beq\label{dif_eq2}
\cQ_{a|i}^+-\cQ_{a|i}^-=\bP_a \bQ_i\;.
\eeq
Let us look at how the right and left hand sides of the equation \eqref{dif_eq2} behave as $S_2$ approaches 1. Scaling of $\bP_a$ and $\bQ_i$ can be deduced from their leading coefficients $A_a$ and $B_i$. For the convenience of the calculations we perform the rescaling of the $\bP$-functions. The $H$-symmetry \eqref{Q_rotations} and its particular case rescaling symmetry \eqref{Q_rescaling} explained in \cite{Gromov:2014caa} allow us to set some of the coefficients from \eqref{P_x_expansions_twist2} to some fixed values. To describe them we introduce a scaling parameter $\nu=\sqrt{S_2-1}$, which is real for $S_2 \geq 1$. For the coefficients of the $\bP$-functions we obtain
\beq\label{Adu_symmetry_fixed_sltint}
A_a=\(\nu,\nu,-\frac{{\bf A}_3}{\nu},\frac{{\bf A}_4}{\nu}\)\;, \quad c_{3,1}=0\;, \quad A^a=\(\frac{{\bf A}_1}{\nu},\frac{{\bf A}_2}{\nu},-\nu,\nu\)\;, \quad c^{2,1}=0\;.
\eeq
For the $\bQ$-functions in their turn
\beq\label{Bdu_symmetry_fixed_sltint}
B_i=\(-\frac{{\bf B}_1}{\nu},\nu,-\frac{{\bf B}_3}{\nu},-\nu\)\;, \quad B^i=\(-\nu,\frac{{\bf B}_2}{\nu},-\nu,-\frac{{\bf B}_4}{\nu}\)\;.
\eeq
As follows from the $AA$-, $BB$-relations \eqref{AA_BB_general}, \eqref{Adu_symmetry_fixed_sltint} and \eqref{Bdu_symmetry_fixed_sltint} in the small $\nu$ limit
\beq\label{AB_low_ind}
A_a=\(\nu,\nu,\frac{1+\theta}{2}i\nu,\frac{1-\theta}{2}i\nu\)+\mathcal{O}\(\nu^3\)\;, \; B_i=\(-\frac{1+\theta}{2}i\nu,\nu,\frac{1-\theta}{2}i\nu,-\nu\)+\mathcal{O}\(\nu^3\)\;.
\eeq
Notice that then from \eqref{AA_BB_general}, \eqref{Adu_symmetry_fixed_sltint} and \eqref{Bdu_symmetry_fixed_sltint} we derive that $A^a$ and $B^i$ all scale as $\nu$ and are given by the formulas
\beq\label{AB_upp_ind}
A^a=\(-\frac{1-\theta}{2}i\nu,\frac{1+\theta}{2}i\nu,-\nu,\nu\)+\mathcal{O}\(\nu^3\)\;, \; B^i=\(-\nu,-\frac{1+\theta}{2}i\nu,-\nu,-\frac{1-\theta}{2}i\nu\)+\mathcal{O}\(\nu^3\)\;,
\eeq
where
\beq\label{sltint_definition}
\theta=\left.\frac{\partial S_1}{\partial S_2}\right|_{\Delta=0 \atop S_2=1}=\left.\frac{\partial j}{\partial n}\right|_{n=1}
\eeq
is the slope-to-intercept function, which is the quantity of interest in the present Subsection. This means that the right-hand side in \eqref{dif_eq3} is small and the functions $\cQ_{a|i}$ are i-periodic in the LO. Since they are also analytic in the upper half plane, they are analytic everywhere. Recall that the asymptotics of $\cQ_{a|i}$ are given by
\beq\label{Qai_asymp2}
\cQ_{a|j} \sim -i\frac{A_a B_j}{-\tilde{M}_a+\hat{M}_j}u^{-\tilde{M}_a+\hat{M}_j}\;, \quad u \rightarrow \infty\;.
\eeq
After plugging the global charges into the expressions \eqref{global_charges} for $\tilde{M}_a$ and $\hat{M}_i$ taken at the BPS point $\Delta=0$, $S_1=-1$ and $S_2=0$ we see that all components of $\cQ_{a|i}$ are either zero or scale like constants at infinity in the LO. Constant at infinity entire function is constant everywhere, so in the LO 
\beq\label{Qai_LO_sltint}
\cQ_{a|i}=\left(
\begin{array}{cccc}
0 & 0 & 1 & 0 \\
1 & 0 & 0 & 0 \\
0 & 1 & 0 & 0 \\
0 & 0 & 0 & 1 \\
\end{array}
\right)+\mathcal{O}(\nu)\;.
\eeq

When $\Delta=0$ and $S_2=n$ is arbitrary, the left-right symmetry is restored, which simplifies the solution a lot. Despite the normalizations \eqref{Adu_symmetry_fixed_sltint} and \eqref{Bdu_symmetry_fixed_sltint} differ from \eqref{Adu_symmetry_fixed} and \eqref{Bdu_symmetry_fixed} by a rescaling symmetry one can see that the left-right symmetry is restored (which is also confirmed by the weak coupling data for arbitrary $\Delta$ and $S_2$ from Appendix \ref{weak_coupling_QSC}). At this point the symmetry \eqref{DeltaminusDelta_symmetry} takes the form
\beq
\bP^a=\chi^{ab}\bP_b, \quad \bQ^i=\eta^{ij}\bQ_j,
\eeq
where $\chi^{ab}$ and $\eta^{ij}$ are given by \eqref{chi} and $\chi^{ab}$ is the same matrix as for the left-right symmetric states as in \cite{Gromov:2013pga,Gromov:2014caa}. As we consider the case when both spins $S_1$ and $S_2$ are not integer, let us recall the gluing matrix for this case \eqref{gluing_conditions_non_integer_spins_solution} taking into account the hermiticity of the gluing matrix
\begin{align}\label{gluing_conditions_slope_to_intercept}
& \tilde{\bQ}^1=M_1^{11} \bar{\bQ}_1+M_1^{12} \bar{\bQ}_2+\(M_1^{13}+M_2^{13}e^{2\pi u}+M_3^{13}e^{-2\pi u}\)\bar{\bQ}_3+\(M_1^{14}+M_2^{14}e^{2\pi u}+M_3^{14}e^{-2\pi u}\)\bar{\bQ}_4\;, \notag \\
& \tilde{\bQ}^2=\bar{M}_1^{12} \bar{\bQ}_1\;, \\
& \tilde{\bQ}^3=\(\bar{M}_1^{13}+\bar{M}_2^{13}e^{2\pi u}+\bar{M}_3^{13}e^{-2\pi u}\)\bar{\bQ}_1+M_1^{33} \bar{\bQ}_3+M_1^{34} \bar{\bQ}_4\;, \notag \\
& \tilde{\bQ}^4=\(\bar{M}_1^{14}+\bar{M}_2^{14}e^{2\pi u}+\bar{M}_3^{14}e^{-2\pi u}\)\bar{\bQ}_1+\bar{M}_1^{34} \bar{\bQ}_3+M_1^{44} \bar{\bQ}_4 \notag
\end{align}
and keeping in mind that $M_1^{11}$, $M_1^{33}$ and $M_1^{44}$ are real.

As now we have the matrix $\eta$ from \eqref{chi}, which relates the $\bQ$-functions with lower and upper indices, it is possible to obtain an additional constraint on the gluing matrix. Plugging this relation into the first gluing condition \eqref{gluing_conditions} we derive
\beq
\tilde{\bQ}_i=\eta_{ik}M^{kl}\eta_{lj}\bar{\bQ}^j\;,
\eeq
which after comparison with the second gluing condition from \eqref{gluing_conditions} leads us to $(M^{-t})_{ij}=\eta_{ik}M^{kl}\eta_{lj}$, and after multiplication of the both sides by $(M^t)^{mj}$ we obtain an additional constraint for the gluing matrix
\beq\label{M_constraint_sltint}
\eta_{ik}M^{kl}\eta_{lm}(M^t)^{mj}=\delta_i^j\;.
\eeq

Substitution of \eqref{gluing_matrix_ansatz_noninteger} into \eqref{M_constraint_sltint} leads us to the following equations
\begin{align}
& \eta_{ik}M_{2,3}^{kl}\eta_{lm}(M^t_{2,3})^{mj}=0\;, \label{M_constraint_sltint1}\\
& \eta_{ik}M_{1}^{kl}\eta_{lm}(M^t_{2,3})^{mj}+\eta_{ik}M_{2,3}^{kl}\eta_{lm}(M^t_1)^{mj}=0\;, \label{M_constraint_sltint2}\\
& \eta_{ik}M_1^{kl}\eta_{lm}(M_1^t)^{mj}+\eta_{ik}M_2^{kl}\eta_{lm}(M_3^t)^{mj}+\eta_{ik}M_3^{kl}\eta_{lm}(M_2^t)^{mj}=\delta_i^j\;. \label{M_constraint_sltint3}
\end{align}
It should be noted that the first equation \eqref{M_constraint_sltint1} is satisfied for the gluing matrix \eqref{gluing_conditions_non_integer_spins_solution} (the same as in  \eqref{gluing_conditions_slope_to_intercept}).

To start solving the constraint \eqref{M_constraint_sltint} order by order in $\nu$ we use the following expansion of the gluing matrix, which is motivated by the scaling of the $\bQ$-functions \eqref{AB_low_ind} and \eqref{AB_upp_ind}
\beq\label{M_scaling_sltint}
M^{ij}(u)=\sum\limits_{k=0}^{+\infty}M^{(k)ij}(u)\nu^{2k}\;.
\eeq
where
\beq\label{Mk_exp}
M^{(k)ij}(u)=M_1^{(k)ij}+M_2^{(k)ij}e^{2\pi u}+M_3^{(k)ij}e^{-2\pi u}\;.
\eeq

In the LO in $\nu$ the constraints \eqref{omega_antisymmetry_noninteger_charges}, if we take into account \eqref{AB_low_ind} and $S=-1+\theta (n-1)+\cO((n-1)^2)$, lead us to
\begin{align}\label{omega_antisymmetry_noninteger_charges_LO_sltint}
& M_2^{(0)31}=-M_2^{(0)13}\;, \quad M_2^{(0)41}=M_2^{(0)14}\;, \\
& M_3^{(0)31}=-M_3^{(0)13}\;, \quad M_3^{(0)41}=M_3^{(0)14}\;. \notag
\end{align}
Together with the hermiticity of the gluing matrix \eqref{omega_antisymmetry_noninteger_charges_LO_sltint} means that $M_{2,3}^{(0)13}$ are pure imaginary, while $M_{2,3}^{(0)14}$ are real. In addition, from \eqref{commutativity_condition} in the LO in $\nu$ we obtain
\beq\label{commutativity_condition_LO_sltint}
M_3^{(0)13}=M_2^{(0)13}\;, \quad M_3^{(0)14}=-M_2^{(0)14}\;.
\eeq

Now we have to apply the constraint \eqref{M_constraint_sltint} in the LO in $\nu$. Substitution of \eqref{omega_antisymmetry_noninteger_charges_LO_sltint} and \eqref{commutativity_condition_LO_sltint} into \eqref{M_constraint_sltint2} allows us to fix
\beq\label{intermediate_fix}
M_1^{(0)33}=M_1^{(0)44}=0\;, \quad M_1^{(0)12}=M_1^{(0)34}=M_1^{(0)43}\;,
\eeq
given that $M_2^{(0)13}$ and $M_2^{(0)14}$ are non-zero. Combination of \eqref{M_constraint_sltint3}, \eqref{omega_antisymmetry_noninteger_charges_LO_sltint}, \eqref{commutativity_condition_LO_sltint} and \eqref{intermediate_fix} makes it possible for us to derive the solution
\beq\label{intermediate_fix2}
M_1^{(0)12}=M_1^{(0)34}=-1\;, \quad M_1^{(0)31}=-M_1^{(0)13}\;, \quad M_1^{(0)41}=M_1^{(0)14}\;.
\eeq
Also there exists a solution with the opposite sign of $M_1^{(0)12}$ and $M_1^{(0)34}$, but as we will see below, it is not relevant for us. Summarizing \eqref{omega_antisymmetry_noninteger_charges_LO_sltint}, \eqref{commutativity_condition_LO_sltint}, \eqref{intermediate_fix} and \eqref{intermediate_fix2}, we are able to write down the solution
\begin{multline}\label{glm_sltint_LO}
M^{(0)ij}=\begin{pmatrix}
M_1^{(0)11} & -1 & M_1^{(0)13} & M_1^{(0)14} \\
-1 & 0 & 0 & 0 \\
-M_1^{(0)13} & 0 & 0 & -1 \\
M_1^{(0)14} & 0 & -1 & 0
\end{pmatrix}+ \\
+2M_2^{(0)13}\begin{pmatrix}
0 & 0 & 1 & 0 \\
0 & 0 & 0 & 0 \\
-1 & 0 & 0 & 0 \\
0 & 0 & 0 & 0
\end{pmatrix}\cosh(2\pi u)+2M_2^{(0)14}\begin{pmatrix}
0 & 0 & 1 & 0 \\
0 & 0 & 0 & 0 \\
1 & 0 & 0 & 0 \\
0 & 0 & 0 & 0
\end{pmatrix}\sinh(2\pi u)
\end{multline}
where $M_1^{(0)11}$ and $M_{1,2}^{(0)14}$ are real and $M_{1,2}^{(0)13}$ are pure imaginary.

Let us start solving these equations in the LO. We have already found $\cQ_{a|i}$ and thus
\beq
\cQ^{a|i}=-(\cQ_{a|i})^{-t}\;.
\eeq
As the scaling of the $\bP$- and $\bQ$-functions is determined by the scaling \eqref{AB_low_ind} and \eqref{AB_upp_ind} of the leading coefficient at large $u$, we are left with the following expansions of these quantities in the small $\nu$ limit
\beq
    \bP_a=\nu\sum\limits_{k=0}^{+\infty}\bP^{(k)}_a \nu^{2k}\;, \quad \bP^a=\nu\sum\limits_{k=0}^{+\infty}\bP^{(k)a} \nu^{2k}\;, \quad \bQ_i=\nu\sum\limits_{k=0}^{+\infty}\bQ^{(k)}_i \nu^{2k}\;, \quad \bQ^i=\nu\sum\limits_{k=0}^{+\infty}\bQ^{(k)i} \nu^{2k}\;.
\eeq
Then, the correspondence between $\bQ$- and $\bP$-functions \eqref{uP} and \eqref{uQ} in the LO taking into account \eqref{Qai_LO_sltint} looks as follows
\begin{align}\label{LO_QP_rel_sltint}
& \bQ^{(0)}_1=-\bP^{(0)2}=-\bP^{(0)}_3=\bQ^{(0)2}\;, \\
& \bQ^{(0)}_2=-\bP^{(0)3}=\bP^{(0)}_2=-\bQ^{(0)1}\;, \notag \\
& \bQ^{(0)}_3=-\bP^{(0)1}=\bP^{(0)}_4=-\bQ^{(0)4}\;, \notag \\
& \bQ^{(0)}_4=-\bP^{(0)4}=-\bP^{(0)}_1=\bQ^{(0)3}\;. \notag
\end{align}
First of all let us substitute \eqref{LO_QP_rel_sltint} into the gluing conditions \eqref{gluing_conditions_slope_to_intercept} written in the LO with the gluing matrix \eqref{glm_sltint_LO} and use the conjugacy properties of the $\bP$-functions \eqref{conjugation_P_functions}
\begin{align}\label{gluing_conditions_slope_to_intercept_LO}
& \tilde{\bP}^{(0)}_1=-\bP^{(0)}_1+\(M_1^{(0)13}+2M_2^{(0)13}\cosh(2\pi u)\)\bP^{(0)}_3\;, \\
& \tilde{\bP}_2^{(0)}=\(M_1^{(0)14}+2M_2^{(0)14}\sinh(2\pi u)\)\bP_1^{(0)}+\bP_2^{(0)}-M_1^{(0)11}\bP_3^{(0)}+\(M_1^{(0)13}+2M_2^{(0)13}\cosh(2\pi u)\)\bP_4^{(0)}\;, \notag \\
& \tilde{\bP}^{(0)}_3=\bP^{(0)}_3\;, \notag \\
& \tilde{\bP}^{(0)}_4=-\(M_1^{(0)14}+2M_2^{(0)14}\sinh(2\pi u)\)\bP^{(0)}_3-\bP^{(0)}_4\;. \notag
\end{align}
According to \eqref{P_x_expansions_twist2} we see that in the LO $\bP^{(0)}_3$ is simply a constant. Furthermore the constant is fixed by the leading coefficient \eqref{AB_low_ind} in the large $u$ asymptotics
\beq\label{P3_LO_sltint}
\bP^{(0)}_3=\frac{i}{2}\(1+\theta\)\;.
\eeq

To find the solutions of the other equations from \eqref{gluing_conditions_slope_to_intercept_LO} we have to introduce the following notations for the expansion of the hyperbolic functions
\beq
\cosh(2\pi u)=\cosh_+^u+I_0+\cosh_-^u\;, \quad \sinh(2\pi u)=\sinh_+^u+\sinh_-^u\;,
\eeq
where
\beq
\cosh^u_{\pm}=\sum\limits_{k=1}^{+\infty}I_{2k}(4\pi g)x^{\pm 2k}(u)\;, \quad \sinh^u_{\pm}=\sum\limits_{k=1}I_{2k-1}(4\pi g)x^{\pm(2k-1)}(u)\;.
\eeq
In what follows we will usually omit the superscript with the variable $u$ if the context does not imply the usage of $\cosh_{\pm}$ and $\sinh_{\pm}$ with different arguments and the expression $4\pi g$ in the argument of the generalized Bessel function for the sake of conciseness.

Substituting  $\bP_3$ from \eqref{P3_LO_sltint} into the first equation of \eqref{gluing_conditions_slope_to_intercept_LO}, we have
\beq\label{eqP1_sltint}
\tilde{\bP}^{(0)}_1+\bP^{(0)}_1=\(M_1^{(0)13}+2M_2^{(0)13}I_0+2M_2^{(0)13}\(\cosh_++\cosh_-\)\)\frac{i}{2}\(1+\theta\)\;.
\eeq
We need to find the solution for $\bP^{(0)}_1$ as power series in $x$ with the leading asymptotic $A_1/(gx)^2$, where $A_1$ is given by \eqref{AB_low_ind}. Expanding the right-hand side as power series in $x$ it is easy to see that the unique solution is 
\beq\label{d03132_sltint}
M_1^{(0)13}=-2M_2^{(0)13} I_0\;, \quad M_2^{(0)13}=-\frac{i}{(1+\theta)g^2 I_2}
\eeq
and
\beq\label{P1_LO_sltint}
\bP^{(0)}_1=\frac{\cosh_-}{g^2 I_2}\;.
\eeq
Substituting again $\bP_3$ from \eqref{P3_LO_sltint} and $\bP_1$ from \eqref{P1_LO_sltint} into the fourth equation of \eqref{gluing_conditions_slope_to_intercept_LO}, we have
\beq\label{eqP4_sltint}
\tilde{\bP}^{(0)}_4+\bP^{(0)}_4=-\(M_1^{(0)14}+M_2^{(0)14}\(\sinh_++\sinh_-\)\)i(1+\theta)\;,
\eeq
from which and \eqref{AB_low_ind} together with the oddity of $\bP_4$ it follows that $M_1^{(0)14}=0$ and the solution for $\bP^{(0)}_4$ is
\beq\label{P4_LO_sltint}
\bP^{(0)}_4=\frac{1-\theta}{2}ig\(x-\frac{1}{x}\)-(1+\theta)iM_2^{(0)14}\sinh_-\;.
\eeq
Now let us substitute \eqref{P3_LO_sltint}, \eqref{P1_LO_sltint}, \eqref{P4_LO_sltint}, and \eqref{d03132_sltint} into the second equation of \eqref{gluing_conditions_slope_to_intercept_LO}. It gives the following equation
\begin{multline}\label{eqP2_sltint2}
\tilde{\bP}_2^{(0)}-\bP_2^{(0)}=-\frac{1+\theta}{2}iM_1^{(0)11}+\frac{1-\theta}{1+\theta}g\(x-\frac{1}{x}\)\(\cosh_++\cosh_-\)+ \\
+2M_2^{(0)14}\(\sinh_+ \cosh_--\sinh_- \cosh_+\)\;.
\end{multline}
As the LHS of \eqref{eqP2_sltint2} does not contain the even powers of $x$ for the RHS we have
\beq
M_1^{(0)11}=0\;.
\eeq
The equation \eqref{eqP2_sltint2} takes the form
\beq\label{eqP2_sltint3}
\tilde{\bP}_2^{(0)}-\bP_2^{(0)}=\frac{1-\theta}{1+\theta}\frac{1}{gI_2}\(x-\frac{1}{x}\)\(\cosh_++\cosh_-\)+\frac{2M_2^{(0)14}}{g^2 I_2}\(\sinh_+ \cosh_--\cosh_+ \sinh_-\)\;,
\eeq
which still depends on the unknown coefficient $M_2^{(0)14}$, which we will fix in the next Section, but going to the next order.

So far, starting from the gluing conditions for non-integer conformal spin $n$ \eqref{gluing_conditions_slope_to_intercept} we managed to solve the constraints on it in the LO getting \eqref{glm_sltint_LO} and then, using the connection between the $\bQ$- and $\bP$-functions in the LO formulated the system of equations \eqref{gluing_conditions_slope_to_intercept_LO} for the $\bP$-functions in the LO. Then from this system we found all the $\bP$-functions in the LO except for $\bP_2$ for which we derived the equation \eqref{eqP2_sltint3}, still containing one unknown constant. In the next Subsection using this equation we are going to show how to find the slope-to-intercept function.

\subsubsection{Result for the slope-to-intercept function}

From now on let us consider the equations \eqref{Qai_eq} for the $\cQ_{a|i}$ functions in the NLO. We start with $\cQ_{3|3}$
\begin{multline}\label{Q33_NLO_sltint}
\cQ^{(1)+}_{3|3}-\cQ^{(1)-}_{3|3}=\bP^{(0)}_3 \bQ^{(0)}_3= \\
=\bP^{(0)}_3 \bP^{(0)}_4=\frac{1+\theta}{2}i\(\frac{1-\theta}{2}i\(u-\frac{2g}{x}\)-(1+\theta)iM_2^{(0)14}\sinh_-\)\;.
\end{multline}
As $\cQ^{(0)}_{3|3}=0$, then $\cQ^{(1)}_{3|3}$ should not contain $\log u$ in its large $u$ asymptotic, which can only appear from the expansion of non-integer powers $A u^\lambda\simeq A+A\lambda \log u+\dots$. Thus, in the large $u$ expansion of the RHS of \eqref{Q33_NLO_sltint} the coefficient in front of $1/u$ (which would produce $\log u$ in $\cQ_{3|3}^{(1)}$) has to be equal to 0, which is guaranteed by
\beq\label{a0_32_sltint}
M_2^{(0)14}=-\frac{1-\theta}{1+\theta}\frac{g}{I_1}\;.
\eeq

The substitution of \eqref{a0_32_sltint} into \eqref{eqP2_sltint3} gives
\beq\label{eqP2_sltint4}
\tilde{\bP}^{(0)}_2-\bP^{(0)}_2=\frac{1-\theta}{1+\theta}\(\frac{1}{gI_2}\(x-\frac{1}{x}\)\(\cosh_++\cosh_-\)-\frac{2}{gI_1 I_2}\(\sinh_+ \cosh_--\cosh_+ \sinh_-\)\).
\eeq
We do not need to solve completely \eqref{eqP2_sltint4}, because only the first coefficient of the expansion of $\bP_2$ in the powers of $x$ is sufficient to find the slope-to-intercept function
\beq
\bP^{(0)}_2=-\frac{1-\theta}{1+\theta}\left(1+\frac{2}{I_1 I_2}\sum\limits_{k=1}^{+\infty}(-1)^k I_k I_{k+1}\right)\frac{1}{gx}+\ldots\;.
\eeq
Remembering the leading coefficient of the large $u$ asymptotic of $\bP_2$ from \eqref{AB_low_ind} we obtain the equation
\beq\label{eq_slope_to_intercept_function}
-\frac{1-\theta}{1+\theta}\left(1+\frac{2}{I_1 I_2}\sum\limits_{k=1}^{+\infty}(-1)^k I_k I_{k+1}\right)=1\;.
\eeq
The previous equation \eqref{eq_slope_to_intercept_function} fixes $\theta$, which is now equal to
\beq\label{slope_to_intercept}
\boxed{\theta(g)=1+\frac{I_1 I_2}{\sum\limits_{k=1}^{+\infty}(-1)^k I_{k} I_{k+1}}}\;,
\eeq
and constitutes our result for the slope-to-intercept function.

The weak coupling expansion of the obtained result \eqref{slope_to_intercept} is given by
\beq\label{sltint_weak_coupling}
\theta(g)=-\frac{2\pi^2}{3}g^2+\frac{4\pi^4}{9}g^4-\frac{28\pi^6}{135}g^6+\frac{8\pi^8}{405}g^8+\mathcal{O}\(g^{10}\)\;.
\eeq
Since the slope-to-intercept function by definition is the derivative of the intercept function with respect to the conformal spin $n$ at $n=1$ we can immediately compare the first few coefficients in the weak coupling expansion \eqref{slope_to_intercept} with the derivative of \eqref{intercepts} and \eqref{NNLO_intercept_function}. This will also show that these expressions provide the formulas compatible with our analytic continuation in $n$ away from the integer values. To find the derivatives of the binomial harmonic sums with respect to the argument we apply the \verb"SuppPackage" used in \cite{Gromov:2015vua}, which expresses the nested harmonic sums \eqref{harmonic_sums} in terms of the $\eta$-functions \eqref{eta_func_def} and allows to find the derivatives of these sums. The derivatives of the intercept function \eqref{intercepts} and \eqref{NNLO_intercept_function} can be calculated and we find that that they are in full agreement with  \eqref{sltint_weak_coupling}
\beq
\left.\frac{dj}{dn}\right|_{n=1}=-\frac{2\pi^2}{3}g^2+\frac{4\pi^4}{9}g^4-\frac{28\pi^6}{135}g^6+\mathcal{O}\(g^8\)\;,
\eeq
confirming our result \eqref{slope_to_intercept}. 

In the next Section we compute the strong coupling expansion of our result for the slope-to-intercept function. As we will see the calculation is less straightforward than at weak coupling, even thought the result is still quite simple.

\subsubsection{Strong coupling expansion of the slope-to-intercept function}

To obtain the strong coupling expansion of the slope-to-intercept function \eqref{slope_to_intercept} first we calculate the following expansion at strong coupling
\begin{multline}\label{Ik_series}
\frac{I_{k}(4\pi g) I_{k+1}(4\pi g)}{I_1(4\pi g) I_2(4\pi g)}=1-\frac{1}{\lambda^{1/2}}\(k^2+k-2\)+\frac{1}{2 \lambda}\(k^4+2 k^3-4 k^2-5 k+6\)- \\ 
-\frac{1}{24 \lambda^{3/2}}\(4 k^6+12 k^5-26 k^4-72 k^3+67 k^2+105 k-90\)+ \\
+\frac{1}{24 \lambda^2}\(k^8+4 k^7-10 k^6-44 k^5+35 k^4+148 k^3-71 k^2-153 k+90\)- \\
-\frac{1}{1920 \lambda^{5/2}}\(16 k^{10}+80 k^9-240 k^8-1440 k^7+1128 k^6+\right. \\
\left.+8760 k^5-2140 k^4-20720 k^3+3261 k^2+15345 k-4050\)+\mathcal{O}\(\frac{1}{\lambda^3}\),
\end{multline}
where
\beq\label{lambda_constant}
\lambda=(4\pi g)^2\;.
\eeq
If we just sum the series \eqref{Ik_series} multiplying it by $(-1)^k$, it appears to be divergent. 
However, simple $\zeta$-regularization gives the right result, as we verified numerically with high precision. Namely, multiplying this expression by $k^{\delta}$ and understanding the result as the limit $\delta\rightarrow0$, we get the following answer
\begin{multline}\label{auxiliary_expansion}
\sum\limits_{k=1}^{+\infty}\frac{(-1)^k I_{k}(4\pi g) I_{k+1}(4\pi g)}{I_1(4\pi g) I_2(4\pi g)}= \\
=-\frac{1}{2}-\frac{3}{4(4\pi g)}-\frac{3}{4(4\pi g)^2}-\frac{9}{32(4\pi g)^3}+\frac{9}{8(4\pi g)^4}+\frac{2331}{512(4\pi g)^5}+\mathcal{O}\left(\frac{1}{g^6}\right)\;.
\end{multline}
Then, substituting \eqref{auxiliary_expansion} into the expression for the slope-to-intercept function \eqref{slope_to_intercept}, we obtain the strong coupling expansion for it
\beq\la{slope}
\theta=-1+\frac{3}{\lambda^{1/2}}-\frac{3}{2\lambda}-\frac{9}{8\lambda^{3/2}}-\frac{9}{4\lambda^2}-\frac{711}{128\lambda^{5/2}}+\mathcal{O}\left(\frac{1}{\lambda^3}\right)\;.
\eeq
This expansion \eqref{slope} will be useful for us in the Section \ref{intercept_strong_coupling} when we are able to compare it with the derivative of our formula for the strong coupling expansion of the intercept function for arbitrary conformal spin $n$ taken at the point $n=1$, which is based on intensive numerical analysis.

\subsection{Curvature function near the BPS point}

As it was mentioned above in the Subsection \ref{sec:Slope} we are considering the expansion in the vicinity of the BPS point $\Delta=0$, $S_1=-1$ and $S_2=1$. In the previous Subsection we expanded in the powers of $S_2-1$, however, 
to find the curvature we keep $S_2=1$ and expand in the powers of $\Delta$. The scheme of solving the QSC equations in this case is similar but with one difference. 
Since the function $S(\Delta,1)$ is an even function of $\Delta$
we have to expand to the NLO order in $\Delta$ to get a non-trivial result.

As in the previous Section we will utilize a simplification in the  $\cQ_{a|i}$ function to solve the Q-system explicitly.

\subsubsection{LO solution}

Recall that the functions $\cQ_{a|i}$ satisfy the equation
\beq\label{dif_eq3}
\cQ_{a|i}^+-\cQ_{a|i}^-=\bP_a \bQ_i\;.
\eeq
Let us look at how the right and left hand sides of the equation behave as $\Delta$ approaches 0. Behaviour of $\bP_a$ and $\bQ_i$ can be deduced from their leading coefficients $A_a$ and $B_i$. Analogously to the case of the slope-to-intercept function the $H$-symmetry \eqref{Q_rotations} and its particular case rescaling symmetry \eqref{Q_rescaling} explained in \cite{Gromov:2014caa} allow us to set some of the coefficients from \eqref{P_x_expansions_twist2} to some fixed values. To describe them we introduce a scaling parameter $\e=\sqrt{\Delta}$, which is real for $\Delta \geq 0$. For the coefficients of the $\bP$-functions we obtain
\beq\label{Adu_symmetry_fixed_curv}
A_a=\(\e,\e,-\frac{{\bf A}_3}{\e},\frac{{\bf A}_4}{\e}\)\;, \quad c_{3,1}=0\;, \quad A^a=\(\frac{{\bf A}_1}{\e},\frac{{\bf A}_2}{\e},-\e,\e\)\;, \quad c^{2,1}=0\;.
\eeq
For the $\bQ$-functions in their turn
\beq\label{Bdu_symmetry_fixed_curv}
B_i=\(-\frac{{\bf B}_1}{\e},\e,-\frac{{\bf B}_3}{\e},-\e\)\;, \quad B^i=\(-\e,\frac{{\bf B}_2}{\e},-\e,-\frac{{\bf B}_4}{\e}\)\;.
\eeq
As follows from the $AA$-, $BB$-relations \eqref{AA_BB_general}, \eqref{Adu_symmetry_fixed_curv} and \eqref{Bdu_symmetry_fixed_curv} in the small $\e$ limit
\beq\label{AB_low_ind_curv}
A_a=\(\epsilon,\epsilon,\frac{1+\alpha}{2}i\epsilon,\frac{1-\alpha}{2}i\epsilon\)+\mathcal{O}\(\epsilon^3\)\;, \; B_i=\(\frac{1-\alpha}{2}i\epsilon,\epsilon,-\frac{1+\alpha}{2}i\epsilon,-\epsilon\)+\mathcal{O}\(\epsilon^3\)\;,
\eeq
where
\beq
\alpha=\left.\frac{\partial S_1}{\partial \Delta}\right|_{\Delta=0 \atop S_2=1}\;.
\eeq
Note that due to the parity symmetry $\Delta\to-\Delta$ we expect $\alpha=0$,
which will be the consistency check of our calculation. We only get a non-trivial result in the NLO order.

Then from \eqref{AA_BB_general}, \eqref{Adu_symmetry_fixed_curv} and \eqref{Bdu_symmetry_fixed_curv} we derive that $A^a$ and $B^i$ all scale as $\e$ and are given by the formulas
\beq\label{AB_upp_ind_curv}
A^a=\(\frac{1+\alpha}{2}i\epsilon,-\frac{1-\alpha}{2}i\epsilon,-\epsilon,\epsilon\)+\mathcal{O}\(\epsilon^3\)\;, \; B^i=\(-\epsilon,-\frac{1+\alpha}{2}i\epsilon,-\epsilon,-\frac{1-\alpha}{2}i\epsilon\)+\mathcal{O}\(\epsilon^3\)\;.
\eeq
This means that the right-hand side in \eqref{dif_eq2} is small and the functions $\cQ_{a|i}$ are i-periodic in the LO in $\e$. Since they are also analytic in the upper half plane, they are analytic everywhere. Recall that the asymptotics of $\cQ_{a|i}$ are given by
\beq\label{Qai_asymp3}
\cQ_{a|j} \sim -i\frac{A_a B_j}{-\tilde{M}_a+\hat{M}_j}u^{-\tilde{M}_a+\hat{M}_j}\;, \quad u \rightarrow \infty\;.
\eeq
After plugging the global charges into the expressions \eqref{global_charges} for $\tilde M$ and $\hat M$ we see that all components of $\cQ_{a|i}$ are either zero in the LO or scale like constants at infinity. Constant at infinity entire function is constant everywhere, so again in the LO 
\beq\label{Qai_LO_curv}
\cQ_{a|i}=\left(
\begin{array}{cccc}
0 & 0 & 1 & 0 \\
1 & 0 & 0 & 0 \\
0 & 1 & 0 & 0 \\
0 & 0 & 0 & 1 \\
\end{array}
\right)+\mathcal{O}(\epsilon)
\eeq
the result is the same as in the case of the slope-to-intercept function.

As in the case of the slope-to-intercept function we can exploit the symmetry between the $\bP$- and $\bQ$-functions with lower and upper indices. Since $\Delta$ is different from 0 we remember the symmetry \eqref{DeltaminusDelta_symmetry}. Despite the normalizations \eqref{Adu_symmetry_fixed_curv} and \eqref{Bdu_symmetry_fixed_curv} differ from \eqref{Adu_symmetry_fixed} and \eqref{Bdu_symmetry_fixed} by a rescaling one can see that the symmetry \eqref{DeltaminusDelta_symmetry} takes the form (which is also confirmed by the weak coupling data for arbitrary $\Delta$ and $S_2$ from Appendix \ref{weak_coupling_QSC})
\beq\label{LR_sym_curv}
\bP^a(\Delta,u)=\chi_c^{ab}\bP_b(-\Delta,u)\;, \quad \bQ^i(\Delta,u)=\eta_c^{ij}\bQ_j(-\Delta,u)\;,
\eeq
where we set $\e(-\Delta)=\sqrt{-\Delta}=i\sqrt{\Delta}=i\e$ choosing the branch of the square root with the cut going from 0 to $+\infty$ and
\beq
\chi_c^{ab}=\(\begin{array}{cccc}
0 & 0 & 0 & -i \\
0 & 0 & i & 0 \\
0 & i & 0 & 0 \\
-i & 0 & 0 & 0
\end{array}\)\;, \quad
\eta_c^{ij}=\(\begin{array}{cccc}
0 & i & 0 & 0 \\
i & 0 & 0 & 0 \\
0 & 0 & 0 & -i \\
0 & 0 & -i & 0
\end{array}\)\;.
\eeq

Usage of \eqref{LR_sym_curv} and the substitution of \eqref{AB_low_ind_curv} and \eqref{AB_upp_ind_curv} leads us to the equality
\beq\label{trajectory_parity}
\alpha=0
\eeq
as it should be. Substituting $\alpha$ from \eqref{trajectory_parity} into \eqref{AB_low_ind_curv} and \eqref{AB_upp_ind_curv} we obtain  $A$'s and $B$'s in the LO
\begin{align}\label{AB_ind_curv}
& A_a=\(\e,\e,\frac{i\e}{2},\frac{i\e}{2}\)+\cO\(\e^3\), \quad A^a=\(\frac{i\e}{2},-\frac{i\e}{2},-\e,\e\)+\cO\(\e^3\), \\
& B_i=\(\frac{i\e}{2},\e,-\frac{i\e}{2},-\e\)+\cO\(\e^3\), \quad B^i=\(-\e,-\frac{i\e}{2},-\e,-\frac{i\e}{2}\)+\cO\(\e^3\). \notag
\end{align}

As we consider the case when the spin $S_1$ is not integer and the spin $S_2$ is integer, we can use the gluing conditions \eqref{gluing_matrix_integer_conformal_spin} from the Section \ref{num_calc_QSC}, which were also mentioned in the Section \ref{ext_QSC_nonint_sp}, thus
\begin{align}\label{gluing_conditions_curvature}
& \tilde{\bQ}^1=M_1^{11}\bar{\bQ}_1+M_1^{12}\bar{\bQ}_2+\(M_1^{13}+M_2^{13}e^{2\pi u}+M_3^{13}e^{-2\pi u}\)\bar{\bQ}_3\;, \\
& \tilde{\bQ}^2=\bar{M}_1^{12}\bar{\bQ}_1\;, \notag \\
& \tilde{\bQ}^3=\(\bar{M}_1^{13}+\bar{M}_2^{13}e^{2\pi u}+\bar{M}_3^{13}e^{-2\pi u}\)\bar{\bQ}_1+M_1^{33}\bar{\bQ}_3+M_1^{34}\bar{\bQ}_4\;, \notag \\
& \tilde{\bQ}^4=\bar{M}_1^{34}\bar{\bQ}_3\;, \notag
\end{align}
keeping in mind that $M_1^{11}$ and $M_1^{33}$ are real.

As now we have the matrix $\eta_c$ from \eqref{chi}, which relates the $\bQ$-functions with lower and upper indices with $\Delta$ replaced by $-\Delta$, it is possible to obtain an additional constraint on the gluing matrix. Plugging the relation \eqref{LR_sym_curv} into the first gluing condition \eqref{gluing_conditions} we derive
\beq
\tilde{\bQ}_i(-\Delta,u)=(\eta_c)_{ik}M^{kl}(\Delta,u)(\bar{\eta}_c)_{lj}\bar{\bQ}^j(-\Delta,u)\;,
\eeq
which after comparison with the second gluing condition from \eqref{gluing_conditions} leads us to $(M^{-t})_{ij}(-\Delta,u)=(\eta_c)_{ik}M^{kl}(\Delta,u)(\eta_c)_{lj}$ and after multiplication of the both sides by $(M^t)^{mj}(-\Delta,u)$ we obtain an additional constraint for the gluing matrix
\beq\label{M_constraint_curvature}
(\eta_c)_{ik}M^{kl}(\Delta,u)(\bar{\eta}_c)_{lm}(M^t)^{mj}(-\Delta,u)=\delta_i^j\;.
\eeq

Substitution of \eqref{gluing_matrix_ansatz_noninteger} into \eqref{M_constraint_curvature} leads us to the following equations
\begin{align}
& (\eta_c)_{ik}M_{2,3}^{kl}(\Delta)(\bar{\eta}_c)_{lm}(M^t_{2,3})^{mj}(-\Delta)=0\;, \label{M_constraint_curv1} \\
& (\eta_c)_{ik}M_{1}^{kl}(\Delta)(\bar{\eta}_c)_{lm}(M^t_{2,3})^{mj}(-\Delta)+(\eta_c)_{ik}M_{2,3}^{kl}(\Delta)(\bar{\eta}_c)_{lm}(M^t_1)^{mj}(-\Delta)=0\;, \label{M_constraint_curv2} \\
& (\eta_c)_{ik}M_1^{kl}(\Delta)(\bar{\eta}_c)_{lm}(M_1^t)^{mj}(-\Delta)+(\eta_c)_{ik}M_2^{kl}(\Delta)(\bar{\eta}_c)_{lm}(M_3^t)^{mj}(-\Delta)+ \label{M_constraint_curv3} \\
& +(\eta_c)_{ik}M_3^{kl}(\Delta)(\bar{\eta}_c)_{lm}(M_2^t)^{mj}(-\Delta)=\delta_i^j\;. \notag
\end{align}
It should be noted that the first equation \eqref{M_constraint_curv1} is satisfied for the gluing matrix \eqref{gluing_conditions_curvature} (the same as in \eqref{gluing_conditions_slope_to_intercept}).

To start solving the constraint \eqref{M_constraint_curvature} order by order in $\e$ analogously to the case of the slope-to-intercept function we use the following expansion of the gluing matrix, which is motivated by the scaling of the $\bQ$-functions \eqref{AB_low_ind_curv} and \eqref{AB_upp_ind_curv}
\beq\label{M_scaling_curv}
M^{ij}(u)=\sum\limits_{k=0}^{+\infty}M^{(k)ij}(u)\e^{2k}\;,
\eeq
where
\beq\label{Mk_exp_curv}
M^{(k)ij}(u)=M_1^{(k)ij}+M_2^{(k)ij}e^{2\pi u}+M_3^{(k)ij}e^{-2\pi u}\;.
\eeq

In the LO in $\e$ the constraints \eqref{omega_antisymmetry_noninteger_charges}, if we take into account \eqref{AB_low_ind_curv} and $S(\Delta,1)=-1+\gamma(g)\Delta^2+\cO(\Delta^4)$, where $\gamma(g)$ is the curvature function, lead us to
\begin{align}\label{omega_antisymmetry_noninteger_charges_LO_curv}
& M_2^{(0)31}=-M_2^{(0)13}\;, \\
& M_3^{(0)31}=-M_3^{(0)13}\;, \notag
\end{align}
which together with the hermiticity of the gluing matrix \eqref{glm_hermiticity} means that $M_{2,3}^{(0)13}$ are pure imaginary. In addition, from \eqref{commutativity_condition} in the LO in $\e$ we obtain
\beq\label{commutativity_condition_LO_curv}
M_3^{(0)13}=M_2^{(0)13}\;.
\eeq

Now we have to apply the constraint \eqref{M_constraint_curvature} in the LO in $\e$. Substitution of \eqref{omega_antisymmetry_noninteger_charges_LO_curv} and \eqref{commutativity_condition_LO_curv} into \eqref{M_constraint_curv2} allows us to fix
\beq\label{intermediate_fix_curv}
M_1^{(0)12}=-M_1^{(0)34}\;,
\eeq
given that $M_2^{(0)13}$ is non-zero. Combination of \eqref{M_constraint_sltint3}, \eqref{omega_antisymmetry_noninteger_charges_LO_curv}, \eqref{commutativity_condition_LO_curv} and \eqref{intermediate_fix_curv} allows us to fix the following elements of the gluing matrix on the LO
\beq\label{intermediate_fix_curv2}
\left|M_1^{(0)12}\right|=1\;, \quad M_1^{(0)11}=M_1^{(0)33}=0\;, \quad M_1^{(0)31}=-M_1^{(0)13}\;.
\eeq

In the LO after some calculations we obtain under the assumptions that $M_2^{(0)13}$ is not zero the solution which is
\beq\label{glm_curv_LO}
M^{(0)ij}=\begin{pmatrix}
0 & M_1^{(0)12} & M_1^{(0)13}+2M_2^{(0)13}\cosh(2\pi u) & 0 \\
\bar{M}_1^{(0)12} & 0 & 0 & 0 \\
-M_1^{(0)13}-2M_2^{(0)13}\cosh(2\pi u) & 0 & 0 & -M_1^{(0)12} \\
0 & 0 & -\bar{M}_1^{(0)12} & 0
\end{pmatrix}\;,
\eeq
where $M_{1,2}^{(0)13}$ are pure imaginary.

Let us start solving these equations in the LO. We have already found $\cQ_{a|i}$ and thus
\beq
\cQ^{a|i}=-\(\cQ_{a|i}\)^{-t}\;.
\eeq
As the scaling of the $\bP$- and $\bQ$-functions is determined by the scaling \eqref{AB_low_ind_curv} and \eqref{AB_upp_ind_curv} of the leading coefficient at large $u$, we are left with the following expansions of these quantities in the small $\e$ limit
\beq\label{PQ_scaling_curv}
    \bP_a=\e\sum\limits_{k=0}^{+\infty}\bP^{(k)}_a \e^{2k}\;, \quad \bP^a=\e\sum\limits_{k=0}^{+\infty}\bP^{(k)a} \e^{2k}\;, \quad \bQ_i=\e\sum\limits_{k=0}^{+\infty}\bQ^{(k)}_i \e^{2k}\;, \quad \bQ^i=\e\sum\limits_{k=0}^{+\infty}\bQ^{(k)i} \e^{2k}\;.
\eeq
Then, the correspondence between $\bQ$- and $\bP$-functions \eqref{uP} and \eqref{uQ} in the LO taking into account \eqref{Qai_LO_curv} looks as follows
\begin{align}\label{LO_QP_rel_curv}
& \bQ^{(0)}_1=-\bP^{(0)2}=\bP^{(0)}_3=-\bQ^{(0)2}\;, \\
& \bQ^{(0)}_2=-\bP^{(0)3}=\bP^{(0)}_2=-\bQ^{(0)1}\;, \notag \\
& \bQ^{(0)}_3=-\bP^{(0)1}=-\bP^{(0)}_4=\bQ^{(0)4}\;, \notag \\
& \bQ^{(0)}_4=-\bP^{(0)4}=-\bP^{(0)}_1=\bQ^{(0)3}\;. \notag
\end{align}
First of all we substitute \eqref{LO_QP_rel_curv} to the gluing conditions \eqref{gluing_conditions_curvature} written in the LO with the gluing matrix \eqref{glm_curv_LO} and take into account the conjugacy properties of the $\bP$-functions \eqref{conjugation_P_functions}
\begin{align}\label{gluing_conditions_curvature_LO}
& \tilde{\bP}^{(0)}_1=-M_1^{(0)12}\bP^{(0)}_1-\(M_1^{(0)13}+2M_2^{(0)13}\cosh(2\pi u)\)\bP^{(0)}_3\;, \\
& \tilde{\bP}_2^{(0)}=-M_1^{(0)12}\bP_2^{(0)}-\(M_1^{(0)13}+2M_2^{(0)13}\cosh(2\pi u)\)\bP_4^{(0)}\;, \notag \\
& \tilde{\bP}^{(0)}_3=\bar{M}_1^{(0)12}\bP^{(0)}_3\;, \notag \\
& \tilde{\bP}^{(0)}_4=\bar{M}_1^{(0)12}\bP^{(0)}_4\;. \notag
\end{align}

Let us consider the third equation from \eqref{gluing_conditions_curvature_LO}. According to \eqref{P_x_expansions_twist2} the LHS contains only the non-negative powers of Zhukovsky variable $x(u)$, whereas the RHS has only the non-positive powers of $x(u)$. Therefore $\bP^{(0)}_3$ is given only by the constant term from its expansion in the powers of $x(u)$ and taking into account the scaling of the $\bP$-functions \eqref{AB_ind_curv} we see that in the LO
\beq\label{M1_12_LO}
M_1^{(0)12}=1
\eeq
and
\beq\label{P3_LO_curv}
\bP^{(0)}_3=\frac{i}{2}\;.
\eeq

Substituting $\bP_3$ from \eqref{P3_LO_curv} and $M_1^{(0)12}$ from \eqref{M1_12_LO} into the first equation of \eqref{gluing_conditions_curvature_LO}, we have
\beq\label{eqP1_curv}
\tilde{\bP}^{(0)}_1+\bP^{(0)}_1=-\(M_1^{(0)13}+2M_2^{(0)13}I_0+2M_2^{(0)13}\(\cosh_++\cosh_-\)\)\frac{i}{2}\;.
\eeq
After inserting the $x$-expansion of $\bP^{(0)}_1$ from \eqref{P_x_expansions_twist2}, we see that the LHS of \eqref{eqP1_curv} contains all even powers of $x(u)$ except for $x^0$, thus the same term in the $x$-expansion of the RHS of \eqref{eqP1_curv} has to vanish
\beq\label{M1_13_M2_13_LO_curv}
M_1^{(0)13}=-2M_2^{(0)13} I_0
\eeq
and then we obtain
\beq\label{eqP1_curv3}
\tilde{\bP}^{(0)}_1+\bP^{(0)}_1=-M_2^{(0)13}\(\cosh_++\cosh_-\)i\;.
\eeq
Of course, the solution of \eqref{eqP1_curv} is defined modulo the solution of the homogeneous equation (\eqref{eqP1_curv} with zero RHS), but as the solution to this equation has to contain the positive powers of $x(u)$ it cannot contribute to $\bP^{(0)}_1$. Therefore recalling the expansion \eqref{P_x_expansions_twist2} it follows that the solution for $\bP^{(0)}_1$ is
\beq
\bP^{(0)}_1=-iM_2^{(0)13}\cosh_-\;.
\eeq
As we know the leading coefficient of $\bP^{(0)}_1$ then, remembering \eqref{AB_ind_curv}, we obtain
\beq\label{M2_13_LO_curv}
M_2^{(0)13}=\frac{i}{g^2 I_2}
\eeq
and write down the solution
\beq\label{P1_LO_curv}
\bP^{(0)}_1=\frac{\cosh_-}{g^2 I_2}\;.
\eeq

Considering the fourth equation from \eqref{gluing_conditions_curvature_LO} with the substituted \eqref{M1_12_LO}
\beq\label{eqP1_curv4}
\tilde{\bP}^{(0)}_4=\bP^{(0)}_4
\eeq
we see that in accordance with the expansion \eqref{P_x_expansions_twist2} the RHS has the odd powers of $x(u)$ less or equal to $1$ and the LHS has the odd powers of $x(u)$ greater or equal to $-1$. This means that $\bP^{(0)}_4$ contains only the terms with $x(u)$ and $1/x(u)$ and the only combination satisfying \eqref{eqP1_curv4} is
\beq\label{P4_LO_curv}
\bP^{(0)}_4=\frac{i}{2}g\(x+\frac{1}{x}\)\;.
\eeq

Now let us substitute \eqref{P3_LO_curv}, \eqref{P1_LO_curv} and \eqref{P4_LO_curv} into the second equation of \eqref{gluing_conditions_curvature_LO}. It gives the following expression
\beq\label{eqP2_curv}
\tilde{\bP}_2^{(0)}+\bP_2^{(0)}=-\(M_1^{(0)13}+2M_2^{(0)13}I_0+2M_2^{(0)13}\(\cosh_++\cosh_-\)\)\frac{i}{2}g\(x+\frac{1}{x}\)\;.
\eeq
Remembering \eqref{M1_13_M2_13_LO_curv} and \eqref{M2_13_LO_curv} we obtain
\beq\label{eqP2_curv2}
\tilde{\bP}_2^{(0)}+\bP_2^{(0)}=\frac{1}{gI_2}\(x+\frac{1}{x}\)\(\cosh_++\cosh_-\)\;.
\eeq
Analogously to the equation \eqref{eqP1_curv} for $\bP_1$, the solution to \eqref{eqP2_curv2} for $\bP^{(0)}_2$ may potentially include the solution of the homogeneous equation. However, this solution of \eqref{eqP2_curv2} with the zero RHS inevitably has positive powers of $x(u)$ in it, thus it cannot contribute to $\bP^{(0)}_2$. Then, from \eqref{eqP2_curv2} we can extract the LO solution for $\bP^{(0)}_2$
\beq\label{P2_LO_curv}
\bP^{(0)}_2=\frac{1}{gI_2}\(x+\frac{1}{x}\)\cosh_-\;.
\eeq

To summarize this part, we started from the gluing conditions for integer conformal spin $n$ \eqref{gluing_conditions_curvature} and we managed to solve the constraints on it in the LO getting \eqref{glm_curv_LO} and then, using the connection between the $\bQ$- and $\bP$-functions in the LO formulated the system of equations \eqref{gluing_conditions_curvature_LO} for the $\bP$-functions in the LO. Then from this system we found all the $\bP$-functions. In what follows using these functions we are going to show how to find the NLO solution.

\subsubsection{NLO solution}
\label{sec:NLOsolution}
Since the function $S(\Delta,n)$ is even, we have to consider the  next-to-leading order. Our strategy is to find the $\bP$-functions in the NLO order and extract the quantity of interest -- the curvature function -- from the leading coefficients of these functions. Indeed, from \eqref{Adu_symmetry_fixed_curv} and \eqref{AA_BB_general} we derive
\begin{align}\label{Adu_curv_NLO}
& A_a=\(\e,\e,\frac{i}{2}\e+i\(\frac{\gamma}{2}+\frac{3}{8}\)\e^3,\frac{i}{2}\e-i\(\frac{\gamma}{2}-\frac{7}{12}\)\e^3\)+\cO\(\e^5\)\;, \\
& A^a=\(\frac{i}{2}\e+i\(\frac{\gamma}{2}-\frac{7}{12}\)\e^3,-\frac{i}{2}\e+i\(\frac{\gamma}{2}+\frac{3}{8}\)\e^3,-\e,\e\)+\cO\(\e^3\)\;, \notag
\end{align}
where
\beq\label{curv_func_def}
\gamma=\left.\frac{\partial^2 S_1}{\partial \Delta^2}\right|_{\Delta=0 \atop S_2=1}
\eeq
is the curvature function.

We begin by finding the correction to $\cQ^{(0)}_{a|i}$. The equation \eqref{dif_eq3}, taking into account \eqref{uP} and the scaling \eqref{PQ_scaling_curv} and expanding it in the NLO, takes the form
\beq\label{Qai1_eq}
\cQ^{(1)+}_{a|i}-\cQ^{(1)-}_{a|i}=-i\bP^{(0)}_a \cQ^{(0)}_{b|i} \chi_c^{bd} \bP^{(0)}_d\;. 
\eeq
As the functions $\bP^{(0)}_a$ were completely fixed in the previous calculations we are able to determine all $\cQ^{(1)}_{a|i}$ up to a constant. In solving \eqref{Qai1_eq} we are going to act in a way similar to the one presented in \cite{Gromov:2014bva}. We have to find an UHPA solution to the equation of the form
\beq\label{fin_dif_eq}
f^{++}(u)-f(u)=h(u)\;,
\eeq
where the RHS has one cut on the real axis and it can be represented as a series in the Zhukovsky variable $x(u)$, whose powers are bounded from above. To build the solution of such an equation let us rewrite the RHS in the following form
\beq\label{RHS_rewriting}
h(u)=h_{\textrm{pol}}(u)+h_{-}(u)\;,
\eeq
where $h_{\textrm{pol}}(u)$ is a polynomial in $u$ and $h_{-}(u)$ is a series in the Zhukovsky variable $x(u)$ starting from the power not greater than $x^{-1}(u)$. Since $h(u)$ is a series in $x(u)$ bounded from above, we can always rewrite $x^a(u)$ with $a>0$ as
\beq\label{xpospow}
x^a=\(x^a+\frac{1}{x^a}\)-\frac{1}{x^a}\;,
\eeq
where $x^a+1/x^a$ is a polynomial in $u$. Thus, using \eqref{xpospow} we are able to replace any positive power of $x(u)$ in $h(u)$ with the difference of polynomial in $u$ and the negative power of $x(u)$ which justifies the form \eqref{RHS_rewriting}.

The UHPA solution of \eqref{fin_dif_eq} is given by
\beq
f(u)=\Sigma\(h_{\textrm{pol}}\)(u)+\(\Gamma_U \cdot h_-\)(u-i)+c\;,
\eeq
where $\Sigma\(h_{\textrm{pol}}\)(u)$ is the solution of
\beq
f_{\textrm{pol}}^{++}(u)-f_{\textrm{pol}}(u)=h_{\textrm{pol}}(u)
\eeq
subject to the condition
\beq
\Sigma\(h_{\textrm{pol}}\)\(\frac{i}{2}\)=0
\eeq
and $\(\Gamma_U \cdot h_-\)(u)$ is the solution of
\beq
f_U^{++}(u)-f_U(u)=h_-(u)
\eeq
and $c$ is a constant.
The operator $\Gamma_U$ is determined by
\beq\label{Gamma_U}
(\Gamma_U \cdot h)(u)=\oint\limits_{-2g}^{2g}\frac{dv}{2\pi i}i\psi^{(0)}(-i(u-v)+1)h(v)\;,
\eeq
where the integration contour around the cut goes clockwise.
Also we determine the operator $\Gamma_D$ which we will use to obtain the LHPA solution
\beq\label{Gamma_D}
(\Gamma_D \cdot h)(u)=\oint\limits_{-2g}^{2g}\frac{dv}{2\pi i}i\psi^{(0)}(i(u-v)+1)h(v)\;.
\eeq
It is not hard to check that
\begin{align}
& (\Gamma_U \cdot h_-)(u+i)-(\Gamma_U \cdot h_-)(u)=h_-(u+i)\;, \\
& (\Gamma_D \cdot h_-)(u+i)-(\Gamma_D \cdot h_-)(u)=h_-(u)\;. \notag
\end{align}

Rewriting the RHS of \eqref{Qai1_eq} in the form \eqref{RHS_rewriting} we can find the solution of it
\beq\label{Qai_NLO_curv}
\cQ^{(1)+}_{a|i}=\(\cQ^{(1)+}_{a|i}\)_{\textrm{pol}}+\(\cQ^{(1)+}_{a|i}\)_U+c_{a|i}\;,
\eeq
where $(\cQ^{(1)+}_{a|i})_{\textrm{pol}}$ and $(\cQ^{(1)+}_{a|i})_U$ are given in Appendix \ref{Qai1_solution} by the formulas \eqref{Qai1_pol} and \eqref{Qai1_U} respectively. Using the equation \eqref{Qai_coeff_P_coeff}, it is possible to fix the coefficients $c_{a|i}$ in terms of the curvature function $\gamma$ \eqref{curv_func_def} and the NLO coefficient $c^{(1)}_{4,1}$ of the $x$-series of the function $\bP_4$ (see the formula \eqref{cai_NLO} in the Appendix \ref{Qai1_solution}).

In what follows we will also need the complex conjugated function $\bar{\cQ}^{(1)}_{a|i}$, for which we have the equation
\beq\label{Qai1_eq_conj}
\bar{\cQ}^{(1)+}_{a|i}-\bar{\cQ}^{(1)-}_{a|i}=-i\bar{\bP}^{(0)}_a \cQ^{(0)}_{b|i} \bar{\chi}_c^{bd} \bar{\bP}^{(0)}_d\;. 
\eeq
Conjugating the solution \eqref{Qai_NLO_curv} and noting that this operation transforms the UHPA part of it with the kernel $\Gamma_U$ into the LHPA one with the kernel $\Gamma_D$ we get the solution of the previous equation \eqref{Qai1_eq_conj}
\beq\label{Qai1_conj_sol}
\bar{\cQ}^{(1)-}_{a|i}=\(\bar{\cQ}^{(1)-}_{a|i}\)_{\textrm{pol}}+\(\bar{\cQ}^{(1)-}_{a|i}\)_D+c^*_{a|i}\;,
\eeq
where the functions $(\bar{\cQ}^{(1)-}_{a|i})_{\textrm{pol}}$ and $(\bar{\cQ}^{(1)-}_{a|i})_D$ are given by the formulas \eqref{Qai_pol_conj} and \eqref{Qai_D_conj} respectively (see Appendix \ref{Qai1_solution}).

To proceed we are to solve the equations for the gluing matrix in the NLO in $\Delta$. The constraints \eqref{omega_antisymmetry_noninteger_charges} in the NLO in $\e$ lead us to the following equalities
\beq\label{omega_antisymmetry_noninteger_charges_NLO_curv}
M_2^{(1)31}=-M_2^{(1)13}\;, \quad M_3^{(1)31}=-M_3^{(1)13}-2i\pi M_3^{(0)13}\;,
\eeq
whereas the constraint \eqref{commutativity_condition} gives
\beq\label{commutativity_condition_NLO_curv}
M_3^{(1)13}=M_2^{(1)13}-i\pi M_2^{(0)13}\;,
\eeq
where the elements of the LO gluing matrix $M^{(0)ij}(u)$ are known to us. After this it remains to solve the constraints \eqref{M_constraint_curv2} and \eqref{M_constraint_curv3} in the NLO order. The solution can be compactly described by the formula
\beq
M_2^{(1)13}=\frac{i\(M_1^{(1)12}-M_1^{(1)34}\)}{2g^2 I_2}\;,
\eeq
while the other matrix elements of $M_{2,3}^{(0)ij}$ are determined by the relations \eqref{omega_antisymmetry_noninteger_charges_NLO_curv} and \eqref{commutativity_condition_NLO_curv} holds and $M_1^{(1)11}$, $M_1^{(1)12}$, $M_1^{(1)33}$ together with $M_1^{(1)34}$ are real.

To proceed let us slightly rewrite the gluing conditions expressing the $\bQ$-functions in terms of the $\bP$-functions as
\beq\label{P_gluing_condition}
\tilde{\bP}_a=\cQ_{a|i}^+ M^{ij}\bar{\cQ}_{b|j}^- \bar{\bP}^b\;.
\eeq
The same equation \eqref{P_gluing_condition} written in the NLO
\begin{multline}\label{P_gluing_condition_NLO}
\tilde{\bP}^{(1)}_a=-\cQ_{a|i}^{(0)+}M^{(0)ij}\bar{\cQ}_{b|j}^{(0)-}i\chi_c^{be}\bar{\bP}^{(1)}_e+ \\
+\(\cQ_{a|i}^{(1)+}M^{(0)ij}\bar{\cQ}_{b|j}^{(0)-}+\cQ_{a|i}^{(0)+}M^{(0)ij}\bar{\cQ}_{b|j}^{(1)-}\)i\chi_c^{be}\bar{\bP}^{(0)}_e+\cQ_{a|i}^{(0)+}M^{(1)ij}\bar{\cQ}_{b|j}^{(0)-}i\chi_c^{be}\bar{\bP}^{(0)}_e\;.
\end{multline}
Recalling the LO gluing matrix \eqref{glm_curv_LO} we are able to rewrite \eqref{P_gluing_condition_NLO} in the following way
\begin{align}\label{P_eq_curv_NLO}
& \tilde{\bP}^{(1)}_{1,2}-\bP^{(1)}_{1,2}=\frac{2i}{g^2 I_2}\(\cosh_++\cosh_-\)\bP^{(1)}_{3,4}+R_{1,2}\;, \\
& \tilde{\bP}^{(1)}_{3,4}+\bP^{(1)}_{3,4}=R_{3,4}\;, \notag
\end{align}
where
\beq\label{R_a_rewriting}
R_a=\(\cQ_{a|i}^{(1)+}M^{(0)ij}\bar{\cQ}_{b|j}^{(0)-}+\cQ_{a|i}^{(0)+}M^{(0)ij}\bar{\cQ}_{b|j}^{(1)-}\)i\chi_c^{be}\bar{\bP}^{(0)}_e+\cQ_{a|i}^{(0)+}M^{(1)ij}\bar{\cQ}_{b|j}^{(0)-}i\chi_c^{be}\bar{\bP}^{(0)}_e\;.
\eeq

In finding the solution of \eqref{P_eq_curv_NLO} we follow the same method as in \cite{Gromov:2014bva}. Let us briefly sketch its main points. One can notice that the equations for the $\bP$-functions in the NLO \eqref{P_eq_curv_NLO} have one of the two forms
\beq\label{FFG_eq}
\tilde{F}(u)+F(u)=G(u) \quad \textrm{and} \quad \tilde{F}(u)-F(u)=G(u)\;,
\eeq
where $F(u)$ is a power series in Zhukovsky variable $x(u)$ and the function $G(u)$ can be represented as Laurent series in Zhukovsky variable $x(u)$ in the vicinity of the point $x(u)=0$. As it can be seen, the equations \eqref{FFG_eq} are self-consistent only if, respectively, the conditions
\beq\label{consistency_conditions}
\tilde{G}(u)-G(u)=0 \quad \textrm{and} \quad \tilde{G}(u)+G(u)=0
\eeq
are satisfied. As it was explained in \cite{Gromov:2014caa}, the unique solutions to the equations \eqref{FFG_eq} on the classes of functions non-growing and decaying at infinity are respectively
\beq\label{int_ker_sol}
F(u)=\(H \cdot G\)(u)\equiv\oint\limits_{-2g}^{2g}dv H(u,v)G(v) \; \textrm{and} \; F(u)=\(K \cdot G\)(u)\equiv\oint\limits_{-2g}^{2g}dv K(u,v)G(v)\;,
\eeq
where the integral kernels are given by
\beq\label{HK_kernels}
H(u,v)=-\frac{1}{4\pi i}\frac{\sqrt{u^2-4g^2}}{\sqrt{v^2-4g^2}}\frac{1}{u-v}\;, \quad K(u,v)=\frac{1}{4\pi i}\frac{1}{u-v}\;.
\eeq
It should be noted that if the asymptotic of $F(u)$ is not specified to be non-growing of decaying respectively the solutions of \eqref{FFG_eq} may include the zero-modes, i.e. the solutions of the homogeneous equations \eqref{FFG_eq} with zero RHS.

With the usage of the integral kernels $H(u,v)$ defined in \eqref{HK_kernels} first the 3rd and 4th equations of \eqref{P_eq_curv_NLO} are solved, then the obtained $\bP^{(1)}_3$ and $\bP^{(1)}_4$ are substituted into the 1st and 2nd equations of \eqref{P_eq_curv_NLO}, which are solved by utilizing the integral kernel $K(u,v)$ and we derive the answer for $\bP^{(1)}_1$ and $\bP^{(1)}_2$. But before starting to solve \eqref{P_eq_curv_NLO} we are to check the validity of the conditions \eqref{consistency_conditions} in the case in question. First of all, from the form of the 3rd and 4th equations in \eqref{P_eq_curv_NLO} it follows that the equations
\beq
\tilde{R}_{3,4}-R_{3,4}=0
\eeq
are satisfied exactly and do not fix any constants in the RHS of \eqref{P_eq_curv_NLO}. But from the 1st and 2nd equations of \eqref{P_eq_curv_NLO} we obtain
\beq\label{P_eq_curv_NLO_tnott}
\tilde{R}_{1,2}+R_{1,2}+\frac{2i}{g^2 I_2}\(\cosh_++\cosh_-\)R_{3,4}=0\;,
\eeq
which is solved fixing $M_1^{(1)13}$ in terms of $M_1^{(1)12}$ and $M_1^{(1)34}$ together with $M_1^{(1)11}$ and $M_1^{(1)33}$. One can find the details of this calculation and the values of these constants in Appendix \ref{Fix_glm_NLO}. Therefore we checked that the RHS of \eqref{P_eq_curv_NLO} satisfy \eqref{consistency_conditions} and we can apply \eqref{int_ker_sol}.

As $\bP^{(1)}_3$ has the constant asymptotic at infinity, the unique solution for $\bP^{(1)}_3$ is given by
\beq\label{P3_NLO_curv}
\bP^{(1)}_3=H \cdot R_3
\eeq
and contains the constants $M_1^{(1)12}$, $c^{(1)}_{4,1}$ and the curvature function $\gamma(g)$ defined in \eqref{curv_func_def}, which we have to fix. With the usage of \eqref{Adu_symmetry_fixed_curv} and \eqref{Adu_curv_NLO} we can express $M_1^{(1)12}$ and $c^{(1)}_{4,1}$ in terms of $\gamma$
\begin{align}\label{b1c41_NLO_fixed}
& M_1^{(1)12}=2\gamma-\frac{1}{2}+\frac{iH_0\(u\(\Gamma \cdot \cosh_-^v\)(u)\)}{g^2 I_2}-\frac{iH_0\(\(\Gamma \cdot v\cosh_-^v\)(u)\)}{g^2 I_2}\;, \\
& c^{(1)}_{4,1}=\(\frac{I_4}{6I_2}+\frac{1}{3}\)ig+\frac{i}{24g}+\frac{iH_2\(u\(\Gamma \cdot \cosh_-^v\)(u)\)}{g^3 I_2}-\frac{iH_2\(\(\Gamma \cdot v\cosh_-^v\)(u)\)}{g^3 I_2}\;, \notag
\end{align}
where the integral kernel $\Gamma$ is defined as
\beq
\(\Gamma \cdot h(v)\)(u)=\(\Gamma_U \cdot h(v)\)(u)+\(\Gamma_D \cdot h(v)\)(u)
\eeq
and $H_k$ is the $k$-th coefficient in the large $u$ expansion of the convolution of the integral kernel $H(u,v)$ with the function on which it acts.

The asymptotic of $\bP^{(1)}_4(u)$ is growing as $u$, then the solution \eqref{int_ker_sol} with the kernel $H(u,v)$ is not unique and we have to add the solution of the homogeneous equation (zero-mode) to it. To find this zero-mode we first notice that according to the asymptotic of $\bP^{(1)}_4(u)$ it can include only the powers $x(u)$ and $1/x(u)$ and thus it is proportional to $x(u)-1/x(u)$. The coefficient in front of this zero-mode is determined by the coefficients \eqref{Adu_curv_NLO}, then $\bP^{(1)}_4$ is equal to
\beq\label{P4_NLO_curv}
\bP^{(1)}_4=\(\frac{7}{24}-\frac{\gamma}{2}\)ig\(x-\frac{1}{x}\)+H \cdot R_4
\eeq
and contains the constants $M_1^{(1)34}$ and $\gamma$, which are not fixed yet.

To proceed with the solution of the 3rd and 4th equations of \eqref{P_eq_curv_NLO} we take each equation and subtract the same equation conjugated
\begin{align}\label{P1P2_eq_NLO}
& \tilde{\bP}^{(1)}_1-\bP^{(1)}_1=-\frac{i}{g^2 I_2}\(\cosh_++\cosh_-\)\(R_3-2\bP^{(1)}_3\)+\frac{R_1-\tilde{R}_1}{2}\;, \\
& \tilde{\bP}^{(1)}_2-\bP^{(1)}_2=-\frac{i}{g^2 I_2}\(\cosh_++\cosh_-\)\(R_4-2\bP^{(1)}_4\)+\frac{R_2-\tilde{R}_2}{2}\;. \notag
\end{align}
As both the functions $\bP^{(1)}_1(u)$ and $\bP^{(1)}_2(u)$ have the decaying at infinity asymptotics, the solutions of \eqref{P1P2_eq_NLO} are uniquely determined by the formula \eqref{int_ker_sol} with the kernel $K(u,v)$. Calculating $R_3$ with the usage of \eqref{P3_NLO_curv} and \eqref{b1c41_NLO_fixed}, we find the solution for $\bP^{(1)}_1$
\beq\label{P1_NLO_curv}
\bP^{(1)}_1=-\frac{i}{g^2 I_2}K \cdot \(\(\cosh_++\cosh_-\)\(R_3-2\bP^{(1)}_3\)\)+K \cdot \frac{R_1-\tilde{R}_1}{2}\;.
\eeq
Doing the same for $R_4$ with the usage of \eqref{P4_NLO_curv}, we find the solution for $\bP^{(1)}_2$
\beq\label{P2_NLO_curv}
\bP^{(1)}_2=-\frac{i}{g^2 I_2}K \cdot \(\(\cosh_++\cosh_-\)\(R_4-2\bP^{(1)}_4\)\)+K \cdot \frac{R_2-\tilde{R}_2}{2}\;.
\eeq
From \eqref{Adu_curv_NLO} we conclude that
\beq
A^{(1)}_1=A^{(1)}_2=0\;,
\eeq
which leads to fixing the following constants
\begin{multline}\label{M134_NLO_fixed}
M_1^{(1)34}=\frac{7}{6}+\frac{iH_0\(\(\Gamma \cdot v\cosh_-^v\)(u)-u\(\Gamma \cdot \cosh_-^v\)(u)\)}{g^2 I_2}+ \\
+\frac{2iK_2\(\cosh_-^u \(\Gamma \cdot v\cosh_-^v\)(u)-u\cosh_-^u \(\Gamma \cdot \cosh_-^v\)(u)\)}{g^4 I_2}\;,
\end{multline}
where $K_k$ is the $k$-th coefficient in the large $u$ expansion of the convolution of the kernel $K(u,v)$ with the function on which it acts.

To sum up, we managed to solve the constraints for the gluing matrix in the NLO, which allowed us to write down the system of equations for the $\bP$-functions in the NLO \eqref{P_gluing_condition_NLO}. In the NLO some $Q$-functions contain infinite series of cuts, which led us to the usage of the integral kernels \eqref{Gamma_U}, \eqref{Gamma_D} and \eqref{HK_kernels}. After finding the $\bP$-functions in the NLO by solving the system for them we are ready to fix the curvature function by utilizing the values of the coefficients found in the previous calculations.

\subsubsection{Result for the curvature function}

To obtain the curvature function, we have to remember that $c^{(1)}_{4,1}$ from \eqref{b1c41_NLO_fixed} determines the coefficient in front of $1/x$ in $\bP^{(1)}_4$. Comparing this coefficient found from \eqref{P4_NLO_curv} with the usage of \eqref{M134_NLO_fixed} and the one from \eqref{b1c41_NLO_fixed} we find the answer
\begin{multline}
\gamma(g)=\frac{iH_0\(\(\Gamma \cdot v \cosh_-^v\)(u)-u\(\Gamma \cdot \cosh_-^v\)(u)\)}{2g^2 I_2}+\frac{iH_1\(u\(\Gamma \cdot v \cosh_-^v\)(u)-\(\Gamma \cdot v^2 \cosh_-^v\)(u)\)}{4g^2 I_2}+ \\
\\ +\frac{iH_2\(\(\Gamma \cdot v \cosh_-^v\)(u)-u\(\Gamma \cdot \cosh_-^v\)(u)\)}{4g^2 I_2}+\frac{iK_2\(\cosh_-^u\(\(\Gamma \cdot v\cosh_-^v\)(u)-u\(\Gamma \cdot \cosh_-^v\)(u)\)\)}{g^4 I_2}\;.
\end{multline}
By expanding the integral kernels \eqref{HK_kernels} at large $u$ we can rewrite the curvature function in a more concise form
\begin{empheq}[box=\fbox]{multline}\label{curvature_function}
\gamma(g)=\frac{1}{4\pi g^4 I_2^2}\oint\limits_{-2g}^{2g}dv(v\cosh^v_- \(\Gamma \cdot u\cosh^u_- \)(v)-v^2 \cosh^v_- \(\Gamma \cdot \cosh^u_-\)(v))+ \\
+\frac{1}{16\pi g^5 I_2}\oint\limits_{-2g}^{2g}dv\(\frac{v^3 \(\Gamma \cdot \cosh^u_-\)(v)-2v^2 \(\Gamma \cdot u\cosh^u_- \)(v)+v \(\Gamma \cdot u^2 \cosh^u_- \)}{x_v-\frac{1}{x_v}}\)\;,
\end{empheq}
where
\beq
\(\Gamma \cdot h(v)\)(u)=\oint\limits_{-2g}^{2g}\frac{dv}{2\pi i} \partial_u \log\frac{\Gamma\left[i(u-v)+1\right]}{\Gamma\left[-i(u-v)+1\right]}h(v)\;.
\eeq
After obtaining \eqref{curvature_function} we can compare it with the other known results. In the first two orders we know the BFKL Pomeron eigenvalues for arbitrary conformal spin including $n=1$, therefore we are able to calculate the curvature from these eigenvalues in these two first orders. Comparing with the weak coupling expansion of the formula \eqref{curvature_function} for the curvature function
\begin{multline}\label{curv_func_pert_exp}
\gamma(g)=2\zeta_3 g^2+\(-\frac{2\pi^2}{3}\zeta_3-35\zeta_5 \)g^4+\(\frac{16\pi^4}{45}\zeta_3+\frac{22\pi^2}{3}\zeta_5+504 \zeta_7 \)g^6+\(-\frac{28 \pi ^6}{135}\zeta_3-\frac{8\pi^4}{3}\zeta_5-\right. \\
\left.-56\pi^2 \zeta_7-6930\zeta_9 \)g^8+\(\frac{136\pi^8}{2835}\zeta_3+\frac{668\pi^6}{189}\zeta_5-\frac{112\pi^4}{3}\zeta_7+508\pi^2 \zeta_9+93720\zeta_{11}\) g^{10}+ \\
+\(\frac{754\pi^{10}}{42525}\zeta_3-\frac{1402\pi^8}{567}\zeta_5-\frac{73\pi^6}{45}\zeta_7+\frac{4618\pi^4}{3}\zeta_9-12969\pi^2 \zeta_{11}-1234233\zeta_{13}\) g^{12}+O\(g^{14}\)
\end{multline}
we find that the first two terms of the expansion \eqref{curv_func_pert_exp} coincide with the values obtained from the BFKL Pomeron eigenvalues, which represents a check of our result. Having computed the weak coupling expansion \eqref{curv_func_pert_exp}, in the next part we analyze the other interesting limit, i.e. the strong coupling expansion, which is a separate computational task in the case of the curvature function.

\subsubsection{Strong coupling expansion of the curvature function}

For the calculation of the strong coupling expansion of the curvature function \eqref{curvature_function} we utilize the same method as the one used in \cite{Gromov:2014bva}. Let us briefly sketch the scheme of this calculation. The curvature function can be represented in the following form
\beq\label{gen_int_form_curv_func}
\gamma(g)=\oint\limits_{-2g}^{2g}du\oint\limits_{-2g}^{2g}dv F\(x_u,x_v\) \partial_u \log\frac{\Gamma\(i(u-v)+1\)}{\Gamma\(-i(u-v)+1\)}\;.
\eeq
Integrating by parts \eqref{gen_int_form_curv_func} and changing the integration variables to $x_u$ and $x_v$ respectively, we obtain
\beq\label{curv_func_int_over_parts}
\gamma(g)=\oint dx_u \oint dx_v G(x_u,x_v) \log\frac{\Gamma\(i(u-v)+1\)}{\Gamma\(-i(u-v)+1\)}\;,
\eeq
where the integrals go over the unit circle and $G(x_u,x_v)$ is a polynomial in the variables $x_u$, $x_v$, $1/x_u$, $1/x_v$, $\cosh_-^u$, $\cosh_-^v$ and $\sinh_-^u$. Then, expanding $\cosh_-$ and $\sinh_-$ in a series in the Zhukovsky variable $x$, we can  represent \eqref{curv_func_int_over_parts} as an infinite series in the coefficients of the BES dressing phase \cite{Beisert:2006ez,Dorey:2007xn,Beisert:2006ib,Vieira:2010kb}, which admits the large $g$ expansion in the form of asymptotic series \cite{Beisert:2006ez}.

Using the computed series we calculated the numerical value of the curvature function for a number of points in the range $5 \leq g \leq 40$ and obtained the first coefficients of the strong coupling expansion of the curvature function with high precision by fitting the data with inverse powers of $g$. The first several coefficients appear to be simple rational numbers. In the 6th coefficient it was expected to have $\zeta$-function, which we managed to fit utilizing the \verb"EZFace" \cite{EZFace} webpage. As in the case of the slope-to-slope function \cite{Gromov:2014bva}, usage of the exact value fit in the order $1/g^k$ increases accuracy of the fit in the next order $1/g^{k+1}$, which represents a non-trivial check of the validity of the used method.

Application of the numerical method described above yields us the large $g$ expansion of the curvature function
\begin{multline}\label{cur_str_coupling}
\gamma=\frac{1}{2\lambda^{1/2}}-\frac{1}{4\lambda}-\frac{33}{16\lambda^{3/2}}-\frac{81}{16\lambda^2}- \\
-\frac{2265}{256\lambda^{5/2}}+\(\frac{1440\zeta_5}{64}-\frac{765}{64}\)\frac{1}{\lambda^3}+\(\frac{207360\zeta_5}{2048}-\frac{22545}{2048}\)\frac{1}{\lambda^{7/2}}+\mathcal{O}\(\frac{1}{\lambda^4}\),
\end{multline}
where $\lambda$ is given by \eqref{lambda_constant}. Together with the curvature function \eqref{curvature_function} and its weak coupling expansion \eqref{curv_func_pert_exp} the formula \eqref{cur_str_coupling}, containing the strong coupling expansion of this function, concludes the list of the results of the present Section.

\section{Intercept function at strong coupling}\label{intercept_strong_coupling}

The other interesting limit of the intercept function besides the small coupling limit considered in the Section \ref{weak_coup_exp} is the strong coupling one. For $n=0$ case the intercept function in the strong coupling limit was analyzed in \cite{Costa:2012cb,Kotikov:2013xu,Brower:2013jga,Janik:2013pxa} and then extended to the next orders by the QSC method in \cite{Gromov:2014bva}. As we have already the numerical data for the intercept for the different values of conformal spin $n$, then using the numerical algorithm described in the Section \ref{num_calc_QSC} and assuming that the coefficients are some simple rational numbers and extrapolating the high precision numerical data by the inverse powers of $\lambda^{1/2}$ we extract
\beqa\label{strong_coupling_fit}
S(0,3)&=&-3 +
\frac{10}{\lambda^{1/2}}
- \frac{25}{\lambda}
 + \frac{175}{4}\frac{1}{\lambda^{3/2}}+{\cal O}(\lambda^{-2})\;, \\
S(0,2)&=&-2 +
 \frac{4}{\lambda^{1/2}}
 - \frac{6}{\lambda}
 + \frac{9}{2}\frac{1}{\lambda^{3/2}}+{\cal O}(\lambda^{-2})\;, \notag \\
S(0,1)&=&-1\;, \notag \\
S(0,0)&=&0 -
\frac{2}{\lambda^{1/2}}
- \frac{1}{\lambda}
+ \frac{1}{4}\frac{1}{\lambda^{3/2}}+{\cal O}(\lambda^{-2})\;, \notag
\eeqa
where $\lambda$ is given by \eqref{lambda_constant} and the result for $n=0$ is taken from \cite{Gromov:2014bva}. We see that the leading term is linear in $n$, sub-leading is quadratic and so on. This pattern is quite typical at strong coupling (see for example \cite{Gromov:2017cja}). Assuming this polynomial pattern from the above data \eqref{strong_coupling_fit} we get
\begin{multline}\la{strongres}
S(0,n)=-n
+\frac{(n-1)(n+2)}{{\lambda}^{1/2}}- \\
-\frac{(n-1)(n+2)(2n-1)}{2{\lambda}}
+\frac{(n-1)(n+2)(7n^2-9n-1)}{8{\lambda}^{3/2}}+\mathcal{O}\(\frac{1}{\lambda^2}\),
\end{multline}
which we can cross-check with our slope-to-intercept function \eq{slope} by differentiating it at $n=1$! Comparison with \eqref{slope} shows us complete agreement. We also verified our result numerically by taking $n=1.5$ and fitting the data with inverse powers of $\lambda^{1/2}$ we reproduced precisely the coefficnets from \eq{strongres}.

\section{Conclusions and future directions}

BFKL regime is traditionally one of the ``hard" problems of high energy theoretical physics. QSC method allowed to make progress in this direction when the traditional perturbative methods become too complex to make one’s way through. In our paper we extended the area of applications of QSC even further, adding an additional parameter -- conformal spin $n$. Importantly, we can deal with the situation when conformal spin takes an arbitrary real value. QSC can now handle the situation when all three global charges corresponding to $AdS_5$ -- $S_1,S_2$ and $\Delta$ are non-integer. This required us to allow four components (or two independent if we take into account the hermiticity) of the gluing matrix to have exponential asymptotics. Gluing conditions are thus the main ingredient of the analytical continuation of QSC presented in this paper.

The knowledge of the gluing matrix allows us to implement an efficient numerical algorithm which solves the Pomeron eigenvalue problem numerically for any values of $\Delta,\;n$ and at any `t Hooft coupling $g$, limited only by the computing time\footnote{With reasonable starting points, our method converges in several iterations, each taking around $1$ minute on a laptop, producing precision of $20$ digits. It is also easy to get much higher precision by changing parameters in the beginning of the notebook \verb"code_for_arxiv.nb".  We also provide a few starting points with this submission, which can be used to generate new data. More data can be provided by request.}. Our \verb"Mathematica" implementation of this method is attached to this arXiv submission. 

As an illustration of our method we have computed two all-loop quantities -- slope of BFKL intercept at $\Delta=0$ with respect to $n$ around $n=1$ and curvature of the length-2 operator trajectory in the vicinity of the point $\Delta=0$ and $n=1$. We generated analytical perturbative and numerical data using the iterative procedures described in \cite{Gromov:2015vua}, which we had to modify to take into account exponential asymptotics for non-integer global charges. The iterative perturbative data allowed us to fix the NNLO intercept in terms of binomial harmonic sums and partially fix the NNNLO BFKL intercept in terms of binomial and ordinary nested harmonic sums.

Several further directions of work come to mind immediately. First, the basis of nested harmonic sums seems to describe well the perturbative expansion of BFKL eigenvalue and in particular its intercept. Since the iterative procedure can be used to arbitrary high order in $g$ and for arbitrary odd $n$, it is just a question of time and computational power to fix BFKL eigenvalues at higher orders. The first task would be to find a closed analytic expression for the NNNLO BFKL Pomeron eigenvalue for arbitrary $n$ and $\Delta$. Since as we saw the reciprocity in this order is broken, it is an interesting open problem to understand the nature of this.

The second direction one can pursue is exploring the neighbourhood of the BPS point $n=1$, $\Delta=0$. After computing the slope of intercept with respect to $n$, one can compute the next order $(n-1)^2$. The computation should be similar in spirit of the computation of $S^2$ correction to the twist operator anomalous dimension \cite{Gromov:2014bva}.

Furthermore, one should study in a similar way the Odderon spectrum, which is expected to correspond to the length parameter to be taken $L=3$. Most of the steps described in this paper should be applicable for arbitrary $L$ and it would be very interesting to reproduce previously known perturbative results 
and extend them to finite coupling.

Finally, there are indications \cite{Cavaglia:2018lxi} that the structure constants can be also governed by the QSC Q-functions, which were evaluated in this paper in various regimes. So it would be interesting to compare and extend to finite coupling the results on the triple Pomeron vertex \cite{Bartels:2002au}.

\section*{Acknowledgements}

We thank I.~Balitsky, B.~Basso, M.~Bershtein, S.~Caron-Huot, V.~Kazakov, G.~Korchemsky, A.~Leonidov, F.~Levkovich-Maslyuk and C.-I.~Tan for discussions and useful comments during our work. The research of M.A. has received funding from the People Programme (Marie Curie Actions) of the European Union’s Seventh Framework Programme FP7/2007- 2013/ under REA Grant Agreement No 317089 (GATIS) and from the European Research Council (Programme “Ideas” ERC-2012-AdG 320769 AdS-CFT-solvable). Also M.A. is grateful to Prof. Ilmar Gahramanov for his kind hospitality during the visit to the Mimar Sinan G\"uzel Sanatlar \"Universitesi in Istanbul, Turkey, where part of this work was done. N.G. wishes to thank STFC for support from Consolidated grant number ST/J002798/1.

\newpage

\appendix

\addtocontents{toc}{\protect\setcounter{tocdepth}{0}}
\section{Details of QSC construction}\label{QSC_construction_app}

To show that the matrix $\Omega_i^j=\bar{\cQ}_{a|i}^- C^a_b \cQ^{b|j-}$ is $i$-periodic let us apply the conjugated equation \eqref{Qai_eq_UHPA} and the same equation for $\cQ^{a|i}$ which is valid now in the whole complex plane as we know the functions $\bar{\bQ}_i$ and $\bQ^i$ on the sheet with the short cuts. We have
\begin{multline}
\Omega_i^{j++}-\Omega_i^j=-\bar{\bP}_a \bar{\bQ}_i C^a_b \cQ^{b|j-}+\bar{\cQ}_{a|i}^- C^a_b \bP^b \bQ^j= \\
=-\bar{\bQ}_i \bP_b \cQ^{b|j-}-\bar{\cQ}_{a|i}^- \bar{\bP}^a \bQ^j=-\bar{\bQ}_i \bQ^j+\bar{\bQ}_i \bQ^j=0\;.
\end{multline}
To show that the matrix $\Theta_i^j(u)=(-1)^{a+1}\cQ_{a|i}^-(-u) \cQ^{b|j-}(u)$ is $i$-periodic let us apply the equation \eqref{Qai_eq_UHPA} with $u$ replaced by $-u$ and the same equation for $\cQ^{a|i}$ which is valid now in the whole complex plane as we know the functions $\bQ_i(-u)$ and $\bQ^i(u)$ on the sheet with the short cuts. We have
\begin{multline}
\Theta_i^{j++}-\Theta_i^j=-(-1)^{a+1} \bP_a(-u) \bQ_i(-u) \cQ^{b|j-}(u)+(-1)^{a+1} \cQ_{a|i}^-(-u) \bP^b(u) \bQ^j(u)= \\
=-\bQ_i(-u) \bP_b(u) \cQ^{b|j-}(u)-\cQ_{a|i}^-(-u) \bP^a(-u) \bQ^j(u)= \\
=-\bQ_i(-u) \bQ^j(u)+\bQ_i(-u) \bQ^j(u)=0\;.
\end{multline}

\section{Weak coupling solution of the QSC for length-2 operators with nonzero \texorpdfstring{$S_2=n$}{Lg}}\label{weak_coupling_QSC}

In this Appendix we consider BFKL limit of the QSC with non-zero conformal spin. Let us first of all briefly remind what BFKL limit is. We are going to study the regime when at the same time the coupling constant $g \rightarrow 0$ and one of the spins $S_1=S \rightarrow -1$ while keeping the ratio $g^2/(S+1)$ finite. LO BFKL in this limit corresponds to resumming all the contributions of the form $(g^2/(S+1))^k$, NLO BFKL -- to the contributions of the form $(S+1)(g^2/(S+1))^k$ and so on. However, in the present Subsection we take the second spin (conformal spin) $S_2=n \neq 0$ and there appear some differences from the BFKL regime with zero conformal spin.

In order to find the BFKL kernel eigenvalue we are going to utilize the old fashioned method of $\bP\mu$-system. The $\bP\mu$-system consists of the functions $\bP_a(u)$, $\bP^a(u)$, which we introduced before and of an antisymmetric matrix $\mu_{ab}(u)$ (see \cite{Gromov:2013pga,Gromov:2014caa} for the detailed description). They satisfy the following equations
\begin{align}\label{Pmu_equations}
& \tilde{\mu}_{ab}-\mu_{ab}=\bP_a \tilde{\bP}_b-\bP_b \tilde{\bP}_a\;, \quad \tilde{\bP}_a=\mu_{ab}\bP^b\;, \\
& \tilde{\mu}^{ab}-\mu^{ab}=\bP^a \tilde{\bP}^b-\bP^b \tilde{\bP}^a\;, \quad \tilde{\bP}^a=\mu^{ab}\bP_b\;, \notag \\
& \bP_a \bP^a=0\;, \quad \mu_{ab}\mu^{bc}=\delta_a^c\;, \quad \tilde{\mu}_{ab}=\mu_{ab}^{++}\;, \quad \tilde{\mu}^{ab}=\mu^{ab++}\;. \notag
\end{align}
Before proceeding we are going to introduce a couple of new notations. Our notation for the BFKL scaling parameter is $w=S+1$. It is also convenient to introduce the notation $\Lambda=g^2/w$.

To start solving the $\bP\mu$-system in the BFKL regime we have to determine the scaling of the $\bP$- and $\mu$-functions in the limit $w \rightarrow 0$. In what follows we are going to use the arguments from \cite{Alfimov:2014bwa}, where the left-right symmetric case with zero conformal spin was considered.
First, we assume that the scaling of the $\bP$-functions coincides with the scaling of their leading coefficients in the large $u$ asymptotics. Thus as from \eqref{AA_BB_general} for the length-2 state in question in the BFKL limit $A_a A^a=\cO(w^0)$ for $a=1,\ldots,4$ these functions can be chosen to scale as $w^0$
\beq\label{P_BFKL_scaling}
\bP_a=\sum\limits_{k=0}^{+\infty}\bP^{(k)}_a w^k\;, \quad \bP^a=\sum\limits_{k=0}^{+\infty}\bP^{(k)a}w^k\;.
\eeq
Second, because the asymptotic of the function $\bP_1(u)$ is the same as in the left-right symmetric case, for $S_2=n \neq 0$ the argument from \cite{Alfimov:2014bwa} about the scaling of the $\mu$-functions is applicable and they scale as $w^{-2}$
\beq
\mu_{ab}=\sum\limits_{k=0}^{+\infty}\mu^{(k)}_{ab}w^{k-2}\;, \quad \mu^{ab}=\sum\limits_{k=0}^{+\infty}\mu^{(k)ab}w^{k-2}\;.
\eeq
Additionally, as all the $\bP$-functions for the length-2 states being considered possess the certain parity from the $\bP\mu$-system equations \eqref{Pmu_equations} we can conclude that the functions $\mu_{ab}^+(u)$ have the certain parity, which will be specified below.

Finally, we are to determine the asymptotics of the $\mu$-functions for the case of non-integer $S_1=S$ for non-zero $S_2=n$. Let us restrict ourselves from now on in this Section to the case of integer conformal spin $S_2=n$. The $\mu$-functions with the lower indices are given by the formula from \cite{Gromov:2014caa} combined with \eqref{omega_definition} from the Section \ref{ext_QSC_nonint_sp}
\beq\label{mu_Q_omega}
\mu_{ab}=\frac{1}{2}\cQ_{ab|ij}^- \omega^{ij}=\frac{1}{2}\cQ_{ab|ij}^- M^{ik}\Omega_k^j\;.
\eeq
Substituting the matrices \eqref{gluing_matrix_integer_conformal_spin} and \eqref{Omega_matrix_asymptotic}, which are applicable to the case of integer conformal spin, into \eqref{mu_Q_omega}, we are able to determine the asymptotics of the $\omega$-functions. Taking into account also the antisymmetry of $\omega^{ij}=M^{ik}\Omega_j^k$ from \eqref{omega_antisymmetry} we find that they are
\beq\label{uomega_asymptotics_integer_conformal_spin}
\omega^{ij}\sim\(\textrm{const},e^{2\pi |u|},e^{-2\pi |u|},e^{-2\pi |u|},e^{-2\pi|u|},\textrm{const}\)\;.
\eeq
Using \eqref{uomega_asymptotics_integer_conformal_spin} we see that the leading asymptotics of the $\mu$-functions with the lower indices at $u \rightarrow \pm\infty$ are
\beq
\mu_{ab} \sim (u^{-S-1},u^{-S},u^{-S+1},u^{-S+1},u^{-S+2},u^{-S+3})e^{2\pi |u|}\;,
\eeq
while the $\mu$-functions with the upper indices have the same asymptotics but in the reverse order.

In addition, we know that the $\bP$-functions have only one cut on one of the sheets, therefore they can be written as a series in the Zhukovsky variable. The parametrization of the $\bP$-functions for the case $J_1=2$ and $J_2=J_3=0$ is already known to us and can be taken from the formula \eqref{P_x_expansions_twist2}. Utilizing the $\bP$-functions rescaling \eqref{Q_rescaling} we can set the coefficients $A_1=1$, $A_2=1$, $A^3=-1$ and $A^4=1$. Then, we are also allowed to apply the certain $H$-transformation of the $\bP$-functions, which do not alter their asymptotics and parity
\beq
\bP_a \rightarrow \(h_B\)_a^b \bP_b\;, \quad \bP^a \rightarrow \(h_B^{-t}\)_b^a \bP^b\;,
\eeq
where
\beq\label{P_rescaling_parity_asymptotic}
\(h_B\)_a^b=
\begin{pmatrix}
1 & 0 & 0 & 0 \\
0 & 1 & 0 & 0 \\
\alpha_1 & 0 & 1 & 0 \\
0 & \alpha_2 & 0 & 1
\end{pmatrix}\;.
\eeq
As \eqref{P_rescaling_parity_asymptotic} has two parameters then, applying this transformation, we can set the coefficients $c_{3,1}$ and $c^{2,1}$ to zero. Thus, we arrive to the formulas
\begin{align}\label{P_pert_ansatz}
& \bP_1=\frac{1}{\Lambda w x^2}+\sum_{k=1}^{+\infty}\frac{c_{1,k}}{x^{2k+2}}\;, \quad \bP^1=A^1 \sqrt{\Lambda w}\(x+\frac{1}{x}\)+\sum_{k=1}^{+\infty}\frac{c^{1,k}}{x^{2k-1}}\;, \\
& \bP_2=\frac{1}{\sqrt{\Lambda w}x}+\sum_{k=2}^{+\infty}\frac{c_{2,k}}{x^{2k+1}}\;, \quad \bP^2=A^2+\sum_{k=2}^{+\infty}\frac{c^{2,k}}{x^{2k}}\;, \notag \\
& \bP_3=A_3+\sum_{k=2}^{+\infty}\frac{c_{3,k}}{x^{2k}}\;, \quad \bP^3=-\frac{1}{\sqrt{\Lambda w} x}+\sum_{k=1}^{+\infty}\frac{c^{3,k}}{x^{2k+1}}\;, \notag \\
& \bP_4=A_4 \sqrt{\Lambda w}\(x+\frac{1}{x}\)+\sum_{k=1}^{+\infty}\frac{c_{4,k}}{x^{2k-1}}\;, \quad \bP^4=\frac{1}{\Lambda w x^2}+\sum_{k=1}^{+\infty}\frac{c^{4,k}}{x^{2k+2}}\;. \notag
\end{align}
The scaling of $\bP$-functions \eqref{P_BFKL_scaling} suggests that the coefficients should have the expansions
\beq\label{coef_exp}
c_{m,n}=\(\sqrt{\Lambda w}\)^{2n-m-1}\sum_{k=0}^{+\infty}c_{n,m}^{(k)}w^k\;, \quad c^{m,n}=\(\sqrt{\Lambda w}\)^{2n+m-6}\sum_{k=0}^{+\infty}c^{n,m(k)}w^k\;.
\eeq

In what follows for the sake of convenience we change the numeration of the $\mu$-functions to $\mu_{12}=\mu_1$, $\mu_{13}=\mu_2$, $\mu_{14}=\mu_3$, $\mu_{23}=\mu_4$, $\mu_{24}=\mu_5$ and $\mu_{34}=\mu_6$ and the same for $\mu$ with the upper indices. Having set the scaling and the expansions of $\bP$- and $\mu$-functions in the scaling parameter together with their asymptotics we are able to proceed with the solution of the $\bP\mu$-system order by order in the scaling parameter.

\subsection{LO solution}

In the present part we will find the LO solution of the $\bP\mu$-system in the BFKL regime. Taking into account that $\tilde{\bP}$ scale as $1/w^2$, in the LO in $w$ we obtain for the $\bP$-functions from \eqref{P_pert_ansatz} and \eqref{coef_exp}
\begin{align}\label{P_BFKL_LO}
& \bP^{(0)}_1=\frac{1}{u^2}\;, \quad \bP^{(0)}_2=\frac{1}{u}\;, \quad \bP^{(0)}_3=A_3^{(0)}\;, \quad \bP^{(0)}_4=A_4^{(0)}u+\frac{c_{4,1}^{(1)}}{\Lambda u}\;, \\
& \bP^{(0)1}=A^{1(0)}u+\frac{c^{1,1(1)}}{\Lambda u}\;, \quad \bP^{(0)2}=A^{2(0)}\;, \quad \bP^{(0)3}=-\frac{1}{u}\;, \quad \bP^{(0)4}=\frac{1}{u^2}\;, \notag
\end{align}
where we implied $c_{4,1}=\frac{1}{\(\Lambda w\)^{3/2}}\(c_{4,1}^{(1)}w+\mathcal{O}(w^2)\)$, $c^{1,1}=\frac{1}{\(\Lambda w\)^{3/2}}\(c^{1,1(1)}w+\mathcal{O}(w^2)\)$ and
\beq\label{A3A4_BFKL_LO}
A_3^{(0)}=-\frac{i\(\(\Delta-n\)^2-1\)\(\(\Delta+n\)^2-9\)}{32}\;, \; A_4^{(0)}=-\frac{i\(\(\Delta+n\)^2-1\)\(\(\Delta-n\)^2-25\)}{96}\;.
\eeq

Let us now examine the $\mu$-functions in the vicinity of the point $S=-1$. In the LO, when $S=-1$, $\mu$-functions with the lower indices at $u \rightarrow \infty$ have the asymptotics
\beq\label{mu_asymptotics_LO}
(\mu_1,\mu_2,\mu_3,\mu_4,\mu_5,\mu_6)\sim(u^0,u^1,u^2,u^2,u^3,u^4)e^{2\pi |u|}\;.
\eeq
To find the $\mu$-functions with the lower indices in the LO analogously to the left-right symmetric case we notice that $\bP^{(0)}_{1,2}$ and $\bP^{(0)3,4}$ are singular, while $\tilde{\bP}_{1,2}$ and $\tilde{\bP}^{3,4}$ are regular at $u=0$. This can be guaranteed if $\mu^{(0)}_{ab}$ are regular at $u=0$ and have a zero of sufficient order at this point. Therefore, as in the left-right symmetric case we can assume that in the LO the only solution which contributes in $\mu^{(0)}_{ab}$ is the one with the asymptotics \eqref{mu_asymptotics_LO}. Consequently, we conjecture $\mu^{(0)}_{ab}$ to possess the form of a polynomial times some $i$-periodic factor. As for the length-2 operators the functions $\mu_{ab}^+$ have the certain parity, we are led to the ansatz
\begin{align}\label{ansatz_mu_LO}
& \mu_1^{(0)+}=\mathcal{P}(u)b_{1,1}\;, \quad \mu_2^{(0)+}=\mathcal{P}(u)b_{2,1}u\;, \\
& \mu_3^{(0)+}=\mathcal{P}(u)(b_{3,1}u^2+b_{3,2})\;, \quad \mu_4^{(0)+}=\mathcal{P}(u)(b_{4,1}u^2+b_{4,2})\;, \notag \\
& \mu_5^{(0)+}=\mathcal{P}(u)(b_{5,1}u^3+b_{5,2}u)\;, \quad 
\mu_6^{(0)+}=\mathcal{P}(u)(b_{6,1}u^4+b_{6,2}u^2+b_{6,3})\;. \notag
\end{align}
where $\mathcal{P}(u)$ is an $i$-periodic function. Let us first determine only the polynomial part of the solution. Plugging the ansatz \eqref{ansatz_mu_LO} into the equations $\mu_{ab}^{++}=\mu_{ab}+\bP_a \mu_{bc} \tilde{\bP}^c-\bP_b \mu_{ac} \tilde{\bP}^c$ we fix all the coefficients besides $b_{1,1}$ and $c^{1,1(1)}=-c_{4,1}^{(1)}$. The latter equality automatically leads to the validity of the requirement
\beq
\bP^{(0)}_a \bP^{(0)a}=0
\eeq
in the LO.

Now let us return to the general solution. Because in the leading order the equations for $\mu$'s are homogeneous, the solution is determined up to some $i$-periodic function. This periodic function grows not faster than $e^{2\pi |u|}$, then the most general ansatz for it is
\beq
B_{1,1}e^{2\pi u}+B_{1,2}e^{-2\pi u}+B_{1,3}\;.
\eeq

Also we have to remember about the requirements of analyticity, i.e. that the following expressions have no cuts
\beq\label{analyticity_constraints}
\bP+\tilde{\bP}\;, \quad \frac{\bP-\tilde{\bP}}{\sqrt{u^2-4\Lambda w}}\;, \quad \mu+\tilde{\mu}\;, \quad \frac{\mu-\tilde{\mu}}{\sqrt{u^2-4\Lambda w}}\;.
\eeq
The constraints \eqref{analyticity_constraints} lead us to the results
\beq\label{coefficients_analyticity_constraints}
B_{1,2}=B_{1,1}\;, \quad B_{1,3}=-2B_{1,1}\;, \quad c_{4,1}^{(1)}=-\frac{i\Lambda}{96}\(\(\Delta^2-1\)^2-2\(\Delta^2+1\)n^2+n^4\)\;.
\eeq
With these conditions \eqref{coefficients_analyticity_constraints} the parity requirement for $\mu^+$ is automatically satisfied.

Summarizing the obtained results and setting $B_{1,1}=-B_1/4$, all the requirements give us the unique solution (up to a multiplicative constant) given by
\begin{align}\label{mu_BFKL_LO}
& \mu_1^{(0)+}=B_1 \cosh^2(\pi u)\;, \\
& \mu_2^{(0)+}=-B_1 \frac{i}{16}\(\(\Delta^2-1\)^2-2\(\Delta^2+1\)n^2+n^4\)u\cosh^2(\pi u)\;, \notag \\
& \mu_3^{(0)+}=\mu_4^{(0)+}=-B_1 \frac{i}{128}\(\(\Delta^2-1\)^2-2\(\Delta^2+1\)n^2+n^4\)\(4u^2+1\)\cosh^2(\pi u)\;, \notag \\
& \mu_5^{(0)+}=-B_1 \frac{i}{192}\(\(\Delta^2-1\)^2-2\(\Delta^2+1\)n^2+n^4\)u\(4u^2+1\)\cosh^2(\pi u)\;, \notag \\
& \mu_6^{(0)+}=-B_1 \frac{1}{49152}\(\(\Delta^2-1\)^2-2\(\Delta^2+1\)n^2+n^4\)^2\(4u^2-3\)\(4u^2+1\)\cosh^2(\pi u)\;. \notag
\end{align}

Having obtained the LO solution given by \eqref{P_BFKL_LO}, \eqref{mu_BFKL_LO} and \eqref{A3A4_BFKL_LO} we are able to proceed with finding the NLO solution of the $\bP\mu$-system together with the coefficient $B_1$, which will be done in the next Subsection.

\subsection{NLO solution}

Now it remains to fix the coefficient $B_1$ and find the $\bP$-functions in the next order in $w$. To do this we need to calculate $\tilde{\bP}$. First, we have to understand, which coefficients it is necessary to find in order to calculate $\bP$ in the NLO order $w^1$ using the scaling or the $\bP$- \eqref{P_BFKL_scaling} and $\tilde{\bP}$-functions ($w^{-2}$)
\begin{align}\label{P_BFKL_NLO}
& \bP^{(1)}_1=\frac{2\Lambda}{u^4}\;, \quad \bP^{(1)1}=A^{1(1)}u+\frac{c^{1,1(2)}}{\Lambda u}+\frac{c^{1,1(1)}+\Lambda c^{1,2(0)}}{u^3}\;, \\
& \bP^{(1)}_2=\frac{1}{u}+\frac{\(1+c_{2,2}^{(0)}\)\Lambda w}{u^3}+\mathcal{O}(w^2)\;, \quad \bP^{(1)2}=A^{2(0)}+A^{2(1)}w+\mathcal{O}(w^2)\;, \notag \\
& \bP^{(1)}_3=A_3^{(1)}\;, \quad \bP^{(1)3}=\frac{\(-1+c^{3,2(0)}\)\Lambda}{u^3}\;, \notag \\
& \bP^{(1)}_4=A_4^{(1)}u+\frac{c_{4,1}^{(2)}}{\Lambda u}+\frac{c_{4,1}^{(1)}+\Lambda c_{4,2}^{(0)}}{u^3}\;, \quad \bP^{(1)4}=\frac{2\Lambda}{u^4}\;. \notag
\end{align}
Notice, that we already know $c_{4,1}^{(1)}=-c^{1,1(1)}$ given by \eqref{coefficients_analyticity_constraints} and $A_3^{(1)}$, $A_4^{(1)}$, $A^{1(1)}$, $A^{2(1)}$
\begin{align}
& A_3^{(1)}=-\frac{i}{4}\(\Delta^2+\Delta n+n^2-3\), \quad A_4^{(1)}=-\frac{i}{12}\(\Delta^2+3\Delta n+n^2+5\), \\
& A^{1(1)}=\frac{i}{12}\(\Delta^2-3\Delta n+n^2+5\),\quad A^{2(1)}=-\frac{i}{4}\(\Delta^2-\Delta n+n^2-3\). \notag
\end{align}
Thus, we have to find $c_{2,2}^{(0)}$, $c_{4,1}^{(2)}$, $c_{4,2}^{(0)}$, $c^{1,1(2)}$, $c^{1,2(0)}$ and $c^{3,2(0)}$. Comparing the $\tilde{\bP}$-functions calculated from the LO result and the ones found with the equation $\tilde{\bP}_a=\mu_{ab}\bP^b$ we can fix some unknown coefficients. On the one hand we get
\begin{align}
& \tilde{\bP}_1=\frac{1}{\Lambda^2 w^2}\(u^2+\mathcal{O}\(u^4\)\)+\mathcal{O}\(w^{-1}\)\;, \\
& \tilde{\bP}_2=\frac{1}{\Lambda^2 w^2}\(c_{2,3}^{(0)}u^3+\mathcal{O}\(u^5\)\)+\mathcal{O}\(w^{-1}\)\;, \notag \\
& \tilde{\bP}_3=\frac{1}{\Lambda^2 w^2}\(c_{3,4}^{(0)}u^4+\mathcal{O}\(u^6\)\)+\mathcal{O}\(w^{-1}\)\;, \notag \\
& \tilde{\bP}_4=\frac{1}{\Lambda^2 w^2}\(c_{4,3}^{(0)}u^5+\mathcal{O}\(u^7\)\)+\mathcal{O}\(w^{-1}\)\;. \notag
\end{align}
On the other hand
\begin{align}
& \tilde{\bP}_1=-\frac{B_1}{w^2}\frac{i}{4}\(\(\Delta-n\)^2-1\)\sinh^2(\pi u)+\mathcal{O}\(w^{-1}\)\;, \\
& \tilde{\bP}_2=-\frac{B_1}{w^2}\frac{i}{4}\(\(\Delta-n\)^2-1\)u\sinh^2(\pi u)+\mathcal{O}\(w^{-1}\)\;, \notag \\
& \tilde{\bP}_3=-\frac{B_1}{w^2}\frac{1}{128}\(\(\Delta-n\)^2-1\)^2\(\(\Delta+n\)^2-1\)u^2\sinh^2(\pi u)+\mathcal{O}\(w^{-1}\)\;, \notag \\
& \tilde{\bP}_4=-\frac{B_1}{w^2}\frac{1}{384}\(\(\Delta-n\)^2-1\)^2\(\(\Delta+n\)^2-1\)u(u^2+1)\sinh^2(\pi u)+\mathcal{O}\(w^{-1}\)\;. \notag
\end{align}
The result is the following
\begin{align}
& B_1=\frac{4i}{\pi^2\(\(\Delta-n\)^2-1\)\Lambda^2}\;, \quad c_{2,2}^{(0)}=1\;, \\
& c_{4,1}^{(0)}=0\;, \quad c_{4,2}^{(0)}=-\frac{i}{96}\(\(\Delta^2-1\)^2-2\(\Delta^2+1\)n^2+n^4\)\;. \notag
\end{align}
Thus, the final answer for the LO $\mu$-functions is
\begin{align}\label{mu_BFKL_conformal_spin_LO}
& \mu_1^{(0)+}=\frac{4i}{\pi^2\Lambda^2}\frac{\cosh^2(\pi u)}{\(\Delta-n\)^2-1}\;, \\
& \mu_2^{(0)+}=\frac{1}{4\pi^2\Lambda^2}\(\(\Delta+n\)^2-1\)u\cosh^2(\pi u)\;, \notag \\
& \mu_3^{(0)+}=\mu_4^{(0)+}=\frac{1}{32\pi^2\Lambda^2}\(\(\Delta+n\)^2-1\)\(4u^2+1\)\cosh^2(\pi u)\;, \notag \\
& \mu_5^{(0)+}=\frac{1}{48\pi^2\Lambda^2}\(\(\Delta+n\)^2-1\)u\(4u^2+1\)\cosh^2(\pi u)\;, \notag \\
& \mu_6^{(0)+}=-\frac{i}{12288\pi^2\Lambda^2}\(\(\Delta-n\)^2-1\)\(\(\Delta+n\)^2-1\)^2\(4u^2-3\)\(4u^2+1\)\cosh^2(\pi u)\;. \notag
\end{align}

Using the equation
\beq
\mu_{ab}\mu^{bc}=\delta_a^c\;,
\eeq
we are able to determine the $\mu$-functions with upper indices in the LO up to a multiplication constant. Then, inserting them into $\bP\mu$-system equations, we fix this constant obtaining
\begin{align}
& \mu^{(0)1+}=\frac{i}{12288\pi^2\Lambda^2}\(\(\Delta+n\)^2-1\)\(\(\Delta-n\)^2-1\)^2\(4u^2-3\)\(4u^2+1\)\cosh^2(\pi u)\;, \\
& \mu^{(0)2+}=\frac{1}{48\pi^2\Lambda^2}\(\(\Delta-n\)^2-1\)u\(4u^2+1\)\cosh^2(\pi u)\;, \notag \\
& \mu^{(0)3+}=\mu^{(0)4+}=-\frac{1}{32\pi^2\Lambda^2}\(\(\Delta-n\)^2-1\)\(4u^2+1\)\cosh^2(\pi u)\;, \notag \\
& \mu^{(0)5+}=\frac{1}{4\pi^2\Lambda^2}\(\(\Delta-n\)^2-1\)u\cosh^2(\pi u)\;, \notag \\
& \mu^{(0)6+}=-\frac{4i}{\pi^2\Lambda^2}\frac{\cosh^2(\pi u)}{\(\Delta+n\)^2-1} \notag
\end{align}
and some other coefficients in the $\bP$-functions with upper indices
\beq
c^{1,2(0)}=\frac{i}{96}\(\(\Delta^2-n^2\)^2-2\(\Delta^2+n^2\)+1\)\;, \quad c^{3,2(0)}=-1\;.
\eeq

We are able even to go further and calculate completely the $\mu$-functions in the NLO and partly fix the expansion coefficients in the $\bP$-functions in the NNLO order. Let us now fix the other coefficients in the NLO order in $w$. To find the unknown coefficient $c_{4,1}^{(2)}$, we have to build an ansatz for $\mu$ in the NLO. The asymptotics of the NLO $\mu$ are $\{ u^0,u^3,u^2,u^1,u^5 \}\log u$. First we introduce a new function
\beq
\Psi_0(u)=\psi^{(0)}\(\frac{1}{2}+iu\)+\psi^{(0)}\(\frac{1}{2}-iu\)+2\gamma\;.
\eeq
This function has the infinite series of poles in the points $\frac{i}{2}+i\mathbb{Z}$. As the case of general integer conformal spin $n$ has to be consistent with the known left-right symmetric case $n=0$, it is natural to assume that the ansatz for the NLO $\mu$-function is essentially the same as for $n=0$, but relaxing the requirement that $\mu_3$ is equal to $\mu_4$. So, we use the following ansatz
\begin{align}\label{ansatz_mu_NLO}
& \mu_1^{(1)+}=(B_1 \Psi_0(u)+b_1)\cosh^2(\pi u)+K_1\;, \\
& \mu_4^{(1)+}=(B_2 u \Psi_0(u)+b_{21}u)\cosh^2(\pi u)+K_2 u\;, \notag\\
& \mu_3^{(1)+}=(B_3 (4u^2+1) \Psi_0(u)+b_{31}u^2+b_{32})\cosh^2(\pi u)+K_3 (4u^2+1)\;, \notag \\
& \mu_4^{(1)+}=(B_4 (4u^2+1) \Psi_0(u)+b_{41}u^2+b_{42})\cosh^2(\pi u)+K_4 (4u^2+1)\;, \notag \\
& \mu_5^{(1)+}=(B_5 u(4u^2+1) \Psi_0(u)+b_{51}u^3+b_{52}u)\cosh^2(\pi u)+K_5 u(4u^2+1)\;, \notag \\
& \mu_6^{(1)+}=(B_6 (4u^2-3)(4u^2+1) \Psi_0(u)+b_{61}u^4+b_{62}u^2+b_{63})\cosh^2(\pi u)+ \notag \\
& +K_5(4u^2-3)(4u^2+1)\;. \notag
\end{align}
Substituting this ansatz \eqref{ansatz_mu_LO} into the $\bP\mu$-system equations, first, we get the unknown coefficient
\beq
c_{4,1}^{(2)}=-\frac{i\Lambda}{24}(\Delta^2-1)(2(\Delta^2-1)\Lambda-1)
\eeq
and obtain $\mu$ with the lower indices in the NLO order
\begin{align}
& \mu_1^{(1)+}=\frac{1}{(\Delta-n)^2-1}\(\(-\frac{2i\Psi(u)}{\pi^2 \Lambda^2}-\frac{16i\(\pi^2 \Lambda\((\Delta-n)^2-1\)+\frac{3}{2}\)}{3 \pi^2 \Lambda^2}\)\cosh^2 (\pi u)+\frac{8i}{\Lambda}\)\;, \\
& \mu_2^{(1)+}=((\Delta+n)^2-1)\(\(-\frac{u\Psi(u)}{8\pi^2 \Lambda^2}-\frac{u}{3\Lambda}+\frac{u}{2\pi^2 \Lambda^2 ((\Delta+n)^2-1)}\)\cosh^2(\pi u)+\frac{u}{2\Lambda}\)\;, \notag \\
& \mu_3^{(1)+}=((\Delta+n)^2-1)\(\(-\frac{\(4u^2+1\)\Psi(u)}{64\pi^2 \Lambda^2}-\frac{\pi^2 \left(4u^2+1\)-6}{24\pi^2 \Lambda}-\right.\right. \notag \\
& \left.\left.-\frac{4u^2+1}{16\pi^2 \Lambda^2 ((\Delta+n)^2-1)}\)\cosh^2 (\pi u)+\frac{4u^2+1}{16\Lambda}\)\;, \notag \\
& \mu_4^{(1)+}=((\Delta+n)^2-1)\(\(-\frac{\(4u^2+1\)\Psi(u)}{64\pi^2 \Lambda^2}-\frac{\pi^2 \left(4u^2+1\)-6}{24\pi^2 \Lambda}-\frac{4u^2+1}{16\pi^2 \Lambda^2 ((\Delta+n)^2-1)}\right.\right. \notag \\
& \left.\left.-\frac{\(4u^2+1\)\Delta n}{2\pi^2 \Lambda^2 \((\Delta-n)^2-1\)((\Delta+n)^2-1)}\)\cosh^2 (\pi u)+\frac{4u^2+1}{16\Lambda}\)\;, \notag \\
& \mu_5^{(1)+}=((\Delta+n)^2-1)\(\(-\frac{u\(4u^2+1\)\Psi(u)}{96\pi^2 \Lambda^2}-\frac{u\(4u^2+1\)}{36\Lambda}-\frac{u\(4u^2+1\)}{8\pi^2 \Lambda^2 \((\Delta-n)^2-1\)}\right.\right. \notag \\
& \left.\left.+\frac{u\(4u^2+1\)\Delta n}{6\pi^2 \Lambda^2 ((\Delta+n)^2-1)\((\Delta-n)^2-1\)}\)\cosh^2 (\pi u)+\frac{u\(4u^2+1\)}{24\Lambda}\)\;, \notag \\
& \mu_6^{(1)+}=((\Delta+n)^2-1)^2 \((\Delta-n)^2-1\)\(\(\frac{i\(4u^2-3\)\(4u^2+1\)\Psi(u)}{24576\pi^2 \Lambda^2}+\right.\right. \notag \\
& \left.\left.+\frac{i\(\pi^2 \(4 u^2-3\)+36\)\(4u^2+1\)}{9216\pi^2 \Lambda}+\frac{i\(4u^2-3\)\(4u^2+1\)}{6144\pi^2 \Lambda^2 \((\Delta-n)^2-1\)}\)\cosh^2 (\pi u)-\right. \notag \\
& \left.-\frac{i\(4u^2-3\) \(4u^2+1\)}{6144\Lambda}\)\;. \notag
\end{align}
The $\mu$-functions with the upper indices in the NLO are given by the formula
\beq
\mu^{(1)ab}(n,u)=\chi^{ac}\mu^{(1)}_{cd}(-n,u)\chi^{db}\;.
\eeq
where the matrix $\chi$ is given by \eqref{chi}.
Usage of the equation $\tilde{\bP}_a=\mu_{ab}\bP^b$ allows to partially fix the $\bP$-functions in the NNLO order. They are written in the formulas below
\begin{align}\label{P_BFKL_NNLO}
& \bP^{(2)}_1=\frac{\pi^2 \Lambda^2}{3u^4}+\frac{5\Lambda^2}{u^6}\;, \quad \bP^{(2)1}=-\frac{i\(\Delta ^2+n^2-3\)}{48}u+\frac{c^{1,1(3)}}{\Lambda  u}- \\
& -i\Lambda\(\(\Delta+n\)^2-1\)\(\frac{24-\(51+\pi^2\)\(\(\Delta-n\)^2-1\)}{288}-\frac{\Delta n}{6\(\(\Delta+n\)^2-1\)}\)\frac{1}{u^3}+ \notag \\
& +\frac{5i\Lambda^2 \(\(\Delta+n\)^2-1\)\(\(\Delta-n\)^2-1\)}{96u^5}\;, \notag \\
& \bP^{(2)}_2=\(\frac{\pi^2 \Lambda^2}{3}-\frac{4\Lambda}{(\Delta-n)^2-1}\)\frac{1}{u^3}+\frac{5\Lambda^2}{u^5}\;, \notag \\
& \bP^{(2)2}=\frac{i\(\Delta^2+n^2-11\)}{16}-\frac{i\Lambda^2 \(\(\Delta+n\)^2-1\)\(\(\Delta-n\)^2+1\)}{32u^4}\;, \notag \\
& \bP^{(2)}_3=\frac{i\(\Delta^2+n^2-11\)}{16}-\frac{i\Lambda^2 \(\(\Delta+n\)^2-1\)\(\(\Delta-n\)^2-1\)}{32u^4}\;, \notag \\
& \bP^{(2)3}=\(-\frac{\pi^2 \Lambda^2}{3}+\frac{4\Lambda}{((\Delta+n)^2-1)}\)\frac{1}{u^3}-\frac{5\Lambda^2}{u^5}\;, \notag \\
& \bP^{(2)}_4=\frac{i\(\Delta^2+n^2-3\)}{48}u+\frac{c_{4,1}^{(3)}}{\Lambda  u}+ \notag \\
& +i\Lambda\(\(\Delta-n\)^2-1\)\(\frac{24-\(51+\pi^2\)\(\(\Delta+n\)^2-1\)}{288}+\frac{\Delta n}{6\(\(\Delta-n\)^2-1\)}\)\frac{1}{u^3}- \notag \\
& -\frac{5i\Lambda^2 \(\(\Delta+n\)^2-1\)\(\(\Delta-n\)^2-1\)}{96u^5}\;, \notag \\
& \bP^{(2)4}=\frac{\pi^2 \Lambda^2}{3u^4}+\frac{5\Lambda^2}{u^6}\;. \notag
\end{align}

Also, from the obtained results one can notice that the $\bP$- and $\mu$-functions possess the following symmetry
\beq\label{gen_LR_sym}
\bP^a(n,u)=\chi^{ab}\bP_b(-n,u)\;, \quad \mu^{ab}(n,u)=\chi^{ac}\mu_{cd}(-n,u)\chi^{db}\;.
\eeq
where the matrix $\chi$ is given by \eqref{chi}. In what follows we will assume that this symmetry is present in all orders of the perturbative expansion.

\subsection{Passing to $\bQ\omega$-system}\label{Qomega_system}

To proceed we need to use the system which is dual to $\bP\mu$-system -- $\bQ\omega$-system. The equations of this system look as follows \cite{Gromov:2013pga,Gromov:2014caa}
\begin{align}
& \tilde{\omega}_{ij}-\omega_{ij}=\bQ_i \tilde{\bQ}_j-\bQ_j \tilde{\bQ}_i\;, \quad \tilde{\bQ}_i=\omega_{ij}\bQ^j\;, \\
& \tilde{\omega}^{ij}-\omega^{ij}=\bQ^i \tilde{\bQ}^j-\bQ^j \tilde{\bQ}^i\;, \quad \tilde{\bQ}^i=\omega^{ij}\bQ_j\;, \notag \\
& \bQ_i \bQ^i=0\;, \quad \omega_{ij}\omega^{jk}=\delta_i^k\;, \quad \omega_{ij}^{++}=\omega_{ij}\;. \notag
\end{align}
We assume the scaling of the $\omega$-functions is the same as in the case $n=0$ considered in \cite{Alfimov:2014bwa}, which is motivated by the fact that the case of general $n$ has to be consistent with the known case of $n=0$. Thus, the function $\omega^{13}$ scales as $w^{-2}$, $\omega^{12}$, $\omega^{14}$, $\omega^{23}$ and $\omega^{34}$ scale as $w^0$ and $\omega^{24}$ scales as $w^2$. For the lower indices we have $\omega_{24}$ as $w^{-2}$, $\omega_{12}$, $\omega_{14}$, $\omega_{23}$ and $\omega_{34}$ as $w^0$ and $\omega_{13}$ as $w^2$. Let us remind the connection between $\mu$- and $\omega$-functions
\beq
\mu_{ab}^+=\frac{1}{2}\cQ_{ab|ij}\omega^{ij+}\;, \quad \mu^{ab+}=\frac{1}{2}\cQ^{ab|ij}\omega_{ij}^+\;.
\eeq
With the obtained $\bP$- and $\mu$-functions in the LO and NLO we can extract the functions $\cQ_{ab|13}$ and $\cQ^{ab|24}$ in the LO and NLO as well. Let us proceed with further calculations taking $\cQ_{ab|13}$ only as the actions with $\cQ^{ab|24}$ are completely the same but with the exchange $n \rightarrow -n$.

One of the QQ-relations says
\beq
\cQ_{a|1}\cQ_{b|3}-\cQ_{a|3}\cQ_{b|1}=\cQ_{ab|13}
\eeq
and it allows to express $\cQ_{3|1}$ and $\cQ_{4|1}$ in terms of $\cQ_{1|1}$, $\cQ_{2|1}$ and $\cQ_{ab|13}$. Using this fact and the equation
\beq
\cQ_{a|i}^+-\cQ_{a|i}^-=-\bP_a \bP^b \cQ_{a|i}^+
\eeq
we are able to derive the following second order Baxter equation for $\cQ_{1|1}$ in the LO
\beq
\(u^2+\frac{1}{4}\) \cQ_{1|1}^{(0)++}+\(u^2+\frac{1}{4}\) \cQ_{1|1}^{(0)--}+\(-2u^2+\frac{(\Delta-n)^2-3}{4}\)\cQ^{(0)}_{1|1}=0\;.
\eeq
Then, utilizing
\beq
\cQ_{1|1}^+-\cQ_{1|1}^-=\bP_1 \bQ_1
\eeq
we obtain the second order Baxter equation
\beq\label{lQ13_Baxter_LO}
\bQ_{1,3}^{(0)++}+\bQ_{1,3}^{(0)--}+\(-2+\frac{(\Delta-n)^2-1}{4u^2}\)\bQ^{(0)}_{1,3}=0\;,
\eeq
where an additional index 3 appears because the Baxter equation for $\bQ_3$ is the same.

Repeating analogous calculations with $\cQ^{ab|24}$ we get the similar equation for $\bQ^2$ and $\bQ^4$ in the LO
\beq\label{uQ24_Baxter_LO}
\bQ^{(0)2,4++}+\bQ^{(0)2,4--}+\(-2+\frac{(\Delta+n)^2-1}{4u^2} \)\bQ^{(0)2,4}=0\;.
\eeq
As we know the $\bP$-functions completely in the LO and NLO orders the equations \eqref{lQ13_Baxter_LO} and \eqref{uQ24_Baxter_LO} are known to us up to NLO order and will be used below.

It was said in the Section \ref{ext_QSC_nonint_sp} that we should find the solutions to the Baxter equation with pure asymptotics. Let us consider $n=0$ in \eqref{lQ13_Baxter_LO} and \eqref{uQ24_Baxter_LO} for simplicity (we always can restore them by shifting $\Delta \rightarrow \Delta-n$ for the lower indices or $\Delta \rightarrow \Delta+n$ for the upper indices). The Baxter equation
\beq\label{Baxter_equation}
\bq^{++}+\bq^{--}+\(-2+\frac{\Delta^2-1}{4u^2}\)\bq=0
\eeq
is known to have two independent solutions, which are UHPA
\begin{align}
& \bq_{I}(\Delta,u)=2iu \, _3F_2\( \frac{1-\Delta}{2},\frac{1+\Delta}{2},1+iu;1,2;1\)\;, \\
& \bq_{II}(\Delta,u)=-i\coth(\pi u)\bq_{I}(\Delta,u)+\frac{\bq_{I}(\Delta,-u)}{\cos \frac{\pi \Delta}{2}}\;. \notag
\end{align}
We found two solutions $\bq_1$ and $\bq_2$ with the pure asymptotics $u^{\frac{\Delta+1}{2}}$ and $u^{-\frac{\Delta-1}{2}}$ respectively
\begin{align}
& \bq_1(\Delta,u)=-\tan \frac{\pi \Delta}{2}\bq_{I}(\Delta,u)+\bq_{II}(\Delta,u)=\left(-i\coth(\pi u)-\tan\frac{\pi \Delta}{2}\right)\bq_{I}(\Delta,u)+\frac{\bq_{I}(\Delta,-u)}{\cos \frac{\pi \Delta}{2}}\;, \notag \\
& \bq_2(\Delta,u)=\tan \frac{\pi \Delta}{2}\bq_{I}(\Delta,u)+\bq_{II}(\Delta,u)=\left(-i\coth(\pi u)+\tan\frac{\pi \Delta}{2}\right)\bq_{I}(\Delta,u)+\frac{\bq_{I}(\Delta,-u)}{\cos \frac{\pi \Delta}{2}}\;.
\end{align}
Then, the solutions of \eqref{lQ13_Baxter_LO} and \eqref{uQ24_Baxter_LO} with the pure asumptotics can be expressed as follows
\begin{align}\label{Baxter_equation_solutions}
& \bQ^{(0)}_{1,3}(u)=\bq_{1,2}(\Delta-n,u)\;, \\
& \bQ^{(0)2,4}(u)=\bq_{2,1}(\Delta+n,u)\;. \notag
\end{align}

Let us now turn back to the gluing conditions for the integer conformal spin \eqref{gluing_matrix_integer_conformal_spin} and write down two of them containing only the $\bQ$-functions from \eqref{Baxter_equation_solutions} in the LO in the scaling parameter $w$
\begin{align}
& \tilde{\bQ}^2=\bar{M}_1^{(0)12}\bar{\bQ}_1\;, \\
& \tilde{\bQ}^4=\bar{M}_1^{(0)34}\bar{\bQ}_3\;. \notag
\end{align}
To find $M_1^{(0)12}$ and $M_1^{(0)34}$ we can use the continuity of the functions $\bQ^2$ and $\bQ^4$ on the cut $\tilde{\bQ}^2(0)=\bQ^2(0)$ and $\tilde{\bQ}^4(0)=\bQ^4(0)$. The result is
\beq
M_1^{(0)12}=M_1^{(0)34}=\frac{\cos\frac{\pi(\Delta+n)}{2}}{\cos\frac{\pi(\Delta-n)}{2}}\frac{(\Delta-n)^2-1}{(\Delta+n)^2-1}\;.
\eeq
In terms of the $\bQ$-functions with the pure asymptotics, we derive the following gluing conditions in the LO
\beq\label{gluing_conditions_perturbative_regime_LO}
\tilde{\bQ}^{(0)2,4}(u)=\frac{\cos\frac{\pi(\Delta+n)}{2}}{\cos\frac{\pi(\Delta-n)}{2}}\frac{(\Delta-n)^2-1}{(\Delta+n)^2-1}\bar{\bQ}^{(0)}_{1,3}(u)\;.
\eeq

\subsection{LO BFKL eigenvalue}

Here we will obtain the LO BFKL eigenvalue in a way similar to \cite{Alfimov:2014bwa}. To begin with, let us write down the Baxter equation for $\bQ^{2,4}$ in the NLO
\begin{multline}\label{uQ24_Baxter_NLO}
\bQ^{(1)2,4++}+\bQ^{(1)2,4--}+\(-2+\frac{(\Delta+n)^2-1}{4u^2}\)\bQ^{(1)2,4}= \\
=-\frac{i}{2(u+i)}\bQ^{(0)2,4++}+\frac{i}{2(u-i)}\bQ^{(0)2,4--}+\frac{u^2-\Lambda(\Delta+n)^2-1)}{2u^4}\bQ^{(0)2,4}\;.
\end{multline}
From one side from the Baxter equation \eqref{uQ24_Baxter_NLO} it follows that
\beq\label{pole_Q_Baxter}
\frac{\bQ^{(1)j}(u)}{\bQ^{(0)j}(u)}=\frac{i}{2u}+\mathcal{O}(u^0)\;, \quad j=2,4\;.
\eeq
On the other side, we can apply the following trick to find the singular part of $\bQ^j$ in the NLO
\beq
\bQ^j=\frac{\bQ^j+\tilde{\bQ}^j}{2}+\frac{\bQ^j-\tilde{\bQ}^j}{2\sqrt{u^2-4\Lambda w}}\sqrt{u^2-4\Lambda w}\;.
\eeq
For $\bQ^2$ we obtain
\beq\label{Qdif_sqrt}
\frac{\bQ^2-\tilde{\bQ}^2}{2\sqrt{u^2-4\Lambda w}}=\frac{1}{2u}\(\bQ^{(0)2}(u)-\frac{\cos\frac{\pi(\Delta+n)}{2}}{\cos\frac{\pi(\Delta-n)}{2}}\frac{(\Delta-n)^2-1}{(\Delta+n)^2-1}\bar{\bQ}^{(0)}_1(u)\)+\mathcal{O}(w)\;.
\eeq
Combining \eqref{Qdif_sqrt} and the previously obtained results \eqref{Baxter_equation_solutions}, we get
\beq\label{pole_Q_analytic_structure}
\bQ^{(1)2}(u)=\(-\frac{i\bQ^{(0)2}(0)(\Psi(\Delta+n)+\Psi(\Delta-n))\Lambda}{u}+\mathcal{O}(u^0)\)w+{\cal O}(w^2)\;,
\eeq
where
\beq
\Psi(\Delta)\equiv \psi\left(\frac{1}{2}-\frac{\Delta}{2}\right)+\psi\left(\frac{1}{2}+\frac{\Delta}{2}\right)-2\psi(1)\;.
\eeq
Thus, comparing two independent results \eqref{pole_Q_Baxter} and \eqref{pole_Q_analytic_structure}, we have the relation
\beq
-2(\Psi(\Delta+n)+\Psi(\Delta-n))\Lambda=1\;.
\eeq
After some calculations, we obtain for the integer $n$
\begin{multline}
\frac{1}{4\Lambda}=\frac{1}{2}\left(\Psi(\Delta+n)+\Psi(\Delta-n)\right)+\mathcal{O}(w)= \\
=-\psi\left(\frac{1+n-\Delta}{2}\right)-\psi\left(\frac{1+n+\Delta}{2}\right)+2\psi(1)+\mathcal{O}(w)\;.
\end{multline}
Rewriting the result in terms of the usual expansion parameters, we obtain the well-known LO BFKL kernel eigenvalue for nonzero integer conformal spin $n$ \cite{Kotikov:2000pm}
\beq
S=-1-4g^2 \(\psi\(\frac{1+n-\Delta}{2}\)+\psi\(\frac{1+n+\Delta}{2}\)-2\psi(1)\)+\mathcal{O}(g^4)\;.
\eeq

\section{NLO BFKL eigenvalue in terms of nested harmonic sums}\label{app:NLOBFKL}

In this Appendix we show how to rewrite the NNLO BFKL Pomeron eigenvalue given in \cite{Kotikov:2001sc,Kotikov:2002ab} in terms of nested harmonic sums. In the notations of the present work we have
\beq
S(\Delta,n)=-1+g^2 \chi_{LO}(\Delta,n)+g^4 \chi_{NLO}(\Delta,n)+\cO(g^6)\;,
\eeq
where the LO BFKL Pomeron eigenvalue is known in terms of the harmonic sums
\beq
\chi_{LO}(\Delta,n)=-4\(S_1\(\frac{1-\Delta+n}{2}-1\)+S_1\(\frac{1-\Delta+n}{2}-1\)\)\;.
\eeq
Upon identification with the notations of \cite{Kotikov:2001sc,Kotikov:2002ab} of the BFKL Pomeron eigenvalues $\chi_{LO,NLO}(\Delta,n)=4\chi_{1,2}(n,(1+\Delta)/2)$ we write down the original expression, which reads as
\begin{multline}\label{NLOoriginal}
\chi_2(n,\gamma)=-\frac{1}{4}\[2\Phi(n,\gamma)+2\Phi(n,1-\gamma)+2\zeta_2 \chi_1(n,\gamma)-\right. \\
\\ \left.-6\zeta_3-\psi''\(\gamma+\frac{n}{2}\)-\psi''\(1-\gamma+\frac{n}{2}\)\]\;,
\end{multline}
where $\gamma=(1+\Delta)/2$ and
\begin{align}
& \Phi(n,\gamma)=\sum\limits_{k=0}^\infty\frac{(-1)^{k+1}}{k+\gamma+n/2}\[\psi'(k+n+1)-\psi'(k+1)+\right. \\
& \left. +(-1)^{k+1}(\beta'(k+n+1)+\beta'(k+1))+\frac{1}{k+\gamma+n/2}\(\psi(k+n+1)-\psi(k+1)\)\] \notag
\end{align}
and 
\beq
\beta'(z)=\frac{1}{4}\[\psi'\(\frac{z+1}{2}\)-\psi'\(\frac{z}{2}\)\]=\sum_{k=0}^{+\infty}\frac{(-1)^{k+1}}{(z+k)^2}\;.
\eeq

Let us consider the following function
\beq
\tilde{\Phi}(x)=\sum_{k=0}^{+\infty}\frac{(-1)^k}{(k+x)^2}(\psi(k+x+1)-\psi(1))\;.
\eeq
We can express it in terms of harmonic sums
\beq
\tilde{\Phi}(x)=S_{-2,1}(x-1)+\frac{5}{8}\zeta(3)\;.
\eeq
Then we use the following identity for $\Phi(0,x)$\footnote{In our notations $\Phi(0,x)$ coicides with $\Phi(x)$ from \cite{Costa:2012cb}.} which can be found in \cite{Costa:2012cb}
\beq\label{Phi_Phitilde}
\Phi(0,-x)+\Phi(0,1+x)=2\tilde{\Phi}(-x)+2\tilde{\Phi}(1+x)+\frac{\pi^3}{2\sin(\pi x)}\;.
\eeq
Taking into account the relation
\beq
\frac{\pi}{\sin(\pi x)}=-S_{-1}(x)-S_{-1}(-x-1)-2\log 2\;,
\eeq
we obtain the following equation
\begin{multline}
\Phi(0,-x)+\Phi(0,1+x)=2S_{-2,1}(x)+2S_{-2,1}(-x-1)+ \\
+\frac{5}{2}\zeta(3)-\frac{\pi^2}{2}(S_{-1}(x)+S_{-1}(-x-1)+2\log 2)\;.
\end{multline}
Thus, the NLO BFKL Pomeron eigenvalue
\begin{multline}
\chi_{NLO}(\Delta,n)=-2\Phi\(n,\frac{1-\Delta}{2}\)-2\Phi\(n,\frac{1+\Delta}{2}\)-2\zeta(2)\chi_{LO}(\Delta,n)+ \\
+6\zeta(3)+\psi''\(\frac{1+n-\Delta}{2}\)+\psi''\(\frac{1+n+\Delta}{2}\)
\end{multline}
for $n=0$
\begin{multline}\label{NLO_BFKL_LR_sym_ev}
\chi_{NLO}(\Delta,0)=-2\Phi\(0,\frac{1-\Delta}{2}\)-2\Phi\(0,\frac{1+\Delta}{2}\)-2\zeta(2)\chi_{LO}(\Delta,0)+ \\
+6\zeta(3)+\psi''\(\frac{1-\Delta}{2}\)+\psi''\(\frac{1+\Delta}{2}\)
\end{multline}
can be rewritten as follows
\beq
\chi_{NLO}(\Delta,0)=F_2\(\frac{1-\Delta}{2}\)+F_2\(\frac{1+\Delta}{2}\)\;,
\eeq
where
\beq\label{F2x}
F_2(x)=-\frac{3}{2}\zeta(3)+\pi^2 \log 2+\frac{\pi^2}{3}S_1(x-1)+\pi^2 S_{-1}(x-1)+2S_3(x-1)-4S_{-2,1}(x-1)\;.
\eeq
Using these results, we are able to rewrite the NLO BFKL Pomeron eigenvalue for non-zero conformal spin in the following way
\beq\label{NLO_BFKL_ev}
\chi_{NLO}(n,\Delta)=\frac{1}{2}(\chi_{NLO}(0,\Delta+n)+\chi_{NLO}(0,\Delta-n))+R_n\(\frac{1-\Delta}{2}\)+R_n\(\frac{1+\Delta}{2}\)\;.
\eeq
After some calculations, we derive
\begin{multline}
R_n(\gamma)+R_n(1-\gamma)=2\Phi(n,\gamma)+2\Phi(n,1-\gamma)-\Phi\(0,\gamma+\frac{n}{2}\)- \\
-\Phi\(0,\gamma-\frac{n}{2}\)-\Phi\(0,1-\gamma+\frac{n}{2}\)-\Phi\(0,1-\gamma-\frac{n}{2}\)\;.
\end{multline}
To proceed we rewrite $\Phi(n,\gamma)$ in the following way
\beq
\Phi(n,\gamma)=-\beta'\left(\gamma+\frac{n}{2}\right)\sum_{p=1}^{n}\frac{1}{p-\gamma-\frac{n}{2}}+\sum_{p=1}^{n}\frac{\beta'(p)}{p-\gamma-\frac{n}{2}}+\sum_{k=n}^{+\infty}\frac{\beta'(k+1)}{k+\gamma-\frac{n}{2}}+\sum_{k=0}^{+\infty}\frac{\beta'(k+1)}{k+\gamma+\frac{n}{2}}\;.
\eeq
Then, after simple calculations, we obtain the result
\begin{multline}\label{Rng}
R_n(\gamma)=-2\beta'\left(\gamma+\frac{n}{2}\right)\sum_{p=1}^{n}\frac{1}{p-\gamma-\frac{n}{2}}= \\
=-2\(S_{-2}\(\gamma+\frac{n}{2}-1\)+\frac{\pi^2}{12}\)\(S_1\(\gamma+\frac{n}{2}-1\)-S_1\(\gamma-\frac{n}{2}-1\)\)\;.
\end{multline}

To sum up, substituting into \eqref{NLO_BFKL_ev} the expressions from \eqref{NLO_BFKL_LR_sym_ev} with \eqref{F2x} and \eqref{Rng}, we are able to write the NLO BFKL Pomeron eigenvalue with non-zero conformal spin in terms of nested harmonic sums.

\section{Values of the NNNLO intercept}\label{NNNLO_intercept_app}

Using the iterative procedure we managed to calculate the values of NNNLO intercept for a set of values of the conformal spin $n$. They are written below
\begin{align}
& j_{N^3 LO}(1)=0\;, \\
& j_{N^3 LO}(3)=-128\zeta_3^2+\(\frac{32\pi^4}{3}-64\pi^2 \) \zeta_3+\(\frac{320\pi^2}{3}-704\)\zeta_5+2208\zeta_7+ \frac{544\pi^6}{189}-\frac{32\pi^4}{3}-\frac{128\pi^2}{3}\;, \notag \\
& j_{N^3 LO}(5)=\(48 \pi ^2-864\right)\zeta_3+1008\zeta_5+\frac{56\pi^6}{45}+\frac{4\pi^4}{5}-12\pi^2+2832\;, \notag \\
& j_{N^3 LO}(7)=-\frac{704\zeta_3^2}{3}+\(\frac{16976}{9}-\frac{976\pi^2}{9}+\frac{32\pi^4}{3}\)\zeta_3+\(\frac{320\pi^2}{3}-\frac{15536}{9}\)\zeta_5+ \notag \\
& +2208\zeta_7+\frac{2992\pi^6}{567}+\frac{8476\pi^2}{729}-\frac{644\pi^4}{81}-\frac{5440444}{2187}\;, \notag \\
& j_{N^3 LO}(9)=\(\frac{2500\pi^2}{27}-\frac{196750}{81}\)\zeta_3+\frac{17500\zeta_5}{9}+\frac{140\pi^6}{81}+\frac{95\pi^4}{54}-\frac{291275 \pi^2}{5832}+ \frac{76123175}{17496}\;, \notag \\
& j_{N^3 LO}(11)=-\frac{4384\zeta_3^2}{15}+\(\frac{157729}{45}-\frac{33076\pi^2}{225}+\frac{32\pi^4}{3}\)\zeta_3+\(\frac{320\pi^2}{3}-\frac{573836}{225}\)\zeta_5+ \notag \\
& +2208\zeta_7+\frac{18632\pi^6}{2835}+\frac{959892043\pi^2}{18225000}-\frac{145853\pi^4}{20250}-\frac{20803468134991}{5467500000}\;. \notag
\end{align}

\section{Solution of the NLO finite difference equation for $\cQ_{a|i}$}\label{Qai1_solution}

Rewriting the equation \eqref{Qai1_eq} in the form \eqref{RHS_rewriting} we have
\beq
\cQ^{(1)+}_{a|i}-\cQ^{(1)-}_{a|i}=
\left(
\begin{array}{cccc}
\frac{i\cosh_-}{2 g^2 I_2} & \frac{u\cosh_-^2}{g^4 I_2^2} & -\frac{iu\cosh_-}{2g^2 I_2} & -\frac{\cosh_-^2}{g^4 I_2^2} \\
\frac{iu\cosh_-}{2g^2 I_2} & \frac{u^2 \cosh_-^2}{g^4 I_2^2} & -\frac{i}{2}-\frac{i\(u^2 \cosh_- -g^2 I_2\)}{2g^2 I_2} & -\frac{u\cosh_-^2}{g^4 I_2^2} \\
-\frac{1}{4} & \frac{iu\cosh_-}{2g^2 I_2} & \frac{u}{4} & -\frac{i\cosh_-}{2 g^2 I_2} \\
-\frac{u}{4} & \frac{i}{2}+\frac{i\(u^2 \cosh_- -g^2 I_2\)}{2g^2 I_2} & \frac{u^2}{4} & -\frac{iu\cosh_-}{2g^2 I_2} \\
\end{array}
\right).
\eeq
The parts of the solution are given by
\beq\label{Qai1_pol}
\(\cQ^{(1)+}_{a|i}\)_{\textrm{pol}}=\(\begin{array}{cccc}
0 & 0 & 0 & 0 \\
0 & 0 & -\frac{u}{2} & 0 \\
\frac{iu}{4} & 0 & -\frac{iu^2}{8}+\frac{u}{8} & 0 \\
\frac{iu^2}{8}-\frac{u}{8} & \frac{u}{2} & -\frac{iu^3}{12}+\frac{u^2}{8}+\frac{iu}{24} & 0
\end{array}\)
\eeq
and
\beq\label{Qai1_U}
\(\cQ^{(1)+}_{a|i}\)_U=\frac{i}{2 g^2 I_2}\Gamma_U \cdot \left(
\begin{array}{cccc}
\cosh_-^v & -\frac{2i}{g^2 I_2}v\(\cosh_-^v\)^2 & -v\cosh_-^v & \frac{2i}{g^2 I_2}\(\cosh_-^v\)^2 \\
v\cosh_-^v & -\frac{2i}{g^2 I_2}v^2 \(\cosh_-^v\)^2 & -v^2 \cosh_-^v 
& \frac{2i}{g^2 I_2}v\(\cosh_-^v\)^2 \\
0 & v\cosh_-^v & 0 & -\cosh_-^v \\
0 & v^2 \cosh_-^v 
& 0 & -v\cosh_-^v \\
\end{array}
\right)\;,
\eeq
where we used the fact that $\(\Gamma_U \cdot \textrm{const}\)=0$.

Using the large $u$ expansion of the kernel $\Gamma_U$, it is possible to fix the coefficients $c_{a|i}$
\beq\label{cai_NLO}
c_{a|i}=\(\begin{array}{cccc}
0 & 0 & \frac{7}{12}-\frac{i\pi}{4} & 0 \\
\frac{3}{4}+\frac{i\pi}{4} & 0 & -\frac{i}{4} & 0 \\
-\frac{1}{8} & \frac{3}{4}+\frac{i\pi}{4} & -\frac{ig^2 I_4}{24I_2}+\frac{ig^2}{12}-\frac{g}{2}c^{(1)}_{4,1}+\frac{i}{48} & 0 \\
\frac{ig^2 I_4}{24I_2}-\frac{ig^2}{12}+\frac{g}{2}c^{(1)}_{4,1}-\frac{i}{48} & \frac{i}{4} & 0 & \frac{7}{12}-\frac{i\pi}{4}
\end{array}\).
\eeq

Rewriting equation \eqref{Qai1_eq_conj} in the form \eqref{RHS_rewriting} we have
\beq
\bar{\cQ}^{(1)+}_{a|i}-\bar{\cQ}^{(1)-}_{a|i}=
\left(
\begin{array}{cccc}
\frac{i\cosh_-}{2 g^2 I_2} & -\frac{u\cosh_-^2}{g^4 I_2^2} & -\frac{iu\cosh_-}{2g^2 I_2} & \frac{\cosh_-^2}{g^4 I_2^2} \\
\frac{iu\cosh_-}{2g^2 I_2} & -\frac{u^2 \cosh_-^2}{g^4 I_2^2} & -\frac{i}{2}-\frac{i\(u^2 \cosh_- -g^2 I_2\)}{2g^2 I_2} & \frac{u\cosh_-^2}{g^4 I_2^2} \\
\frac{1}{4} & \frac{iu\cosh_-}{2g^2 I_2} & -\frac{u}{4} & -\frac{i\cosh_-}{2 g^2 I_2} \\
\frac{u}{4} & \frac{i}{2}+\frac{i\(u^2 \cosh_- -g^2 I_2\)}{2g^2 I_2} & -\frac{u^2}{4} & -\frac{iu\cosh_-}{2g^2 I_2} \\
\end{array}
\right).
\eeq
Parts of the solution are as follows
\beq\label{Qai_pol_conj}
\(\bar{\cQ}^{(1)-}_{a|i}\)_{\textrm{pol}}=\(\begin{array}{cccc}
0 & 0 & 0 & 0 \\
0 & 0 & -\frac{u}{2} & 0 \\
-\frac{iu}{4} & 0 & \frac{iu^2}{8}+\frac{u}{8} & 0 \\
-\frac{iu^2}{8}-\frac{u}{8} & \frac{u}{2} & \frac{iu^3}{12}+\frac{u^2}{8}-\frac{iu}{24} & 0
\end{array}\)
\eeq
and
\beq\label{Qai_D_conj}
\(\bar{\cQ}^{(1)-}_{a|i}\)_D=\frac{i}{2 g^2 I_2}\Gamma_D \cdot \left(
\begin{array}{cccc}
\cosh_-^v & \frac{2i}{g^2 I_2}v\(\cosh_-^v\)^2 & -v\cosh_-^v & -\frac{2i}{g^2 I_2}\(\cosh_-^v\)^2 \\
v\cosh_-^v & \frac{2i}{g^2 I_2}v^2 \(\cosh_-^v\)^2 & -v^2 \cosh_-^v 
& -\frac{2i}{g^2 I_2}v\(\cosh_-^v\)^2 \\
0 & v\cosh_-^v & 0 & -\cosh_-^v \\
0 & v^2 \cosh_-^v 
& 0 & -v\cosh_-^v \\
\end{array}
\right),
\eeq
where we used the fact that $\Gamma_D\(\textrm{const}\)=0$.

\section{Fixing the constants $M_1^{(1)13}$, $M_1^{(1)11}$ and $M_1^{(1)33}$.}\label{Fix_glm_NLO}

Let us rewrite the equation \eqref{P_eq_curv_NLO_tnott} once more introducing a new designation for the LHS of \eqref{R_a_rewriting}
\beq\label{RHS_app}
Z_{1,2}\equiv\tilde{R}_{1,2}+R_{1,2}+\frac{2i}{g^2 I_2}\(\cosh_++\cosh_-\)R_{3,4}=0\;,
\eeq
where $R_a$, $a=1,\ldots,4$ are given by \eqref{R_a_rewriting}. After the substitution of the $\bP$-functions, Q-functions with two indices and the gluing matrix we see that the integral kernels $\Gamma_U$ and $\Gamma_D$ in \eqref{RHS_app} combine into the difference
\beq\label{GammaUD_dif}
(\Gamma_U \cdot h)(u)-(\Gamma_D \cdot h)(u)=\oint\limits_{-2g}^{2g}\frac{dv}{2\pi i}\(\frac{1}{v-u}+\pi\coth(\pi(u-v))\)h(v)\;.
\eeq
As we are dealing in \eqref{RHS_app} with the functions $h(v)$ being the series in $x(v)$ with negative powers, the first integral in \eqref{GammaUD_dif} with the kernel $1/(u-v)$ can be taken by using the residue theorem and we obtain
\beq
(\Gamma_U \cdot h)(u)-(\Gamma_D \cdot h)(u)=h(u)+\oint\limits_{-2g}^{2g}\frac{dv}{2\pi i}\pi\coth(\pi(u-v))h(v)\;.
\eeq
Rewriting $\coth(2\pi(u-v))=(e^{2\pi u}+e^{2\pi v})/(e^{2\pi u}-e^{2\pi v})$ and changing the integration variable from $v$ to Zhukovsky variable $x(v)$, we are able to derive relatively compact expressions for $Z_1$ and $Z_2$
\begin{align}\label{Z12}
& Z_1=-\frac{i\pi}{g^2 I_2}\sinh(2\pi u)-iM_1^{(1)33}u+iM_1^{(1)13}+\frac{iI_0}{g^2 I_2}\(i\(M_1^{(1)12}-M_1^{(1)34}\)+\pi\)+T_1\;, \\
& Z_2=\frac{i\pi}{g^2 I_2}u\sinh(2\pi u)-\(iM_1^{(1)13}+\frac{iI_0}{g^2 I_2}\(i\(M_1^{(1)12}-M_1^{(1)34}\)+\pi\)\)u+iM_1^{(1)11}+T_2\;, \notag
\end{align}
where
\begin{align}\label{T12_integrals}
& T_1=-\frac{i\pi}{g^2 I_2^2}\oint\limits_{|x_v|=1}\frac{dx_v}{2\pi i}\frac{e^{2\pi u}+e^{2\pi v}}{e^{2\pi u}-e^{2\pi v}}\(x_v-\frac{1}{x_v}\)\(x_u-\frac{1}{x_v}\)\(\frac{1}{x_v}-\frac{1}{x_u}\)\times \\
& \times\(I_0+\cosh_-^v-\sinh(2\pi u)\)\cosh_-^v\;, \notag \\
& T_2=\frac{i\pi}{g I_2^2}\oint\limits_{|x_v|=1}\frac{dx_v}{2\pi i}\frac{e^{2\pi u}+e^{2\pi v}}{e^{2\pi u}-e^{2\pi v}}\(x_v^2-\frac{1}{x_v^2}\)\(x_u-\frac{1}{x_v}\)\(\frac{1}{x_v}-\frac{1}{x_u}\)\times \notag \\
& \times \(I_0(1-I_2 x_u x_v^2)+\cosh_-^v-(1-I_2 x_u x_v^2)\sinh(2\pi u)\)\cosh_-^v\;, \notag
\end{align}
where the integration contour goes clockwise. To calculate the integrals \eqref{T12_integrals} we utilize the following trick. If in both integrals we make the inversion of the integration variable $x_v$ this does not change their value. Thus, taking the half-sum of $T_{1,2}$ and $T_{1,2}$ with $x_v$ replaced by $1/x_v$ leaves the integrals intact, but gives us the integrands which are much simpler to work with
\begin{align}\label{T12_rewritten}
& T_1=-\frac{i\pi}{2g^2 I_2^2}\oint\limits_{|x_v|=1}\frac{dx_v}{2\pi i}\(x_v-\frac{1}{x_v}\)\(x_u-\frac{1}{x_v}\)\(\frac{1}{x_v}-\frac{1}{x_u}\)\times \\
& \times\(\cosh_+^v-\cosh_-^v\)\(\sinh(2\pi u)+\sinh(2\pi v)\)\;, \notag \\
& T_2=-\frac{i\pi}{2g I_2^2}\oint\limits_{|x_v|=1}\frac{dx_v}{2\pi i}\(x_v^2-\frac{1}{x_v^2}\)\(x_u-\frac{1}{x_v}\)\(\frac{1}{x_v}-\frac{1}{x_u}\)\times \notag \\
& \times\(\cosh_+^v-\cosh_-^v\)\(\sinh(2\pi u)+\sinh(2\pi v)\)\;. \notag
\end{align}
Calculation of the integrals \eqref{T12_rewritten} leads us to the result
\begin{align}\label{T12_calculated}
& T_1=\frac{i\pi}{g^2 I_2}\sinh(2\pi u)+\frac{i\pi}{gI_2^2}u\sum\limits_{k=1}^{+\infty}I_{2k}(I_{2k+1}-I_{2k-1})\;, \\
& T_2=-\frac{i\pi}{g^2 I_2}u\sinh(2\pi u)-\frac{i\pi}{gI_2^2}\sum\limits_{k=1}^{+\infty}I_{2k}(I_{2k+3}+I_{2k+1}-I_{2k-1}-I_{2k-3})\;. \notag
\end{align}
After the substitution of \eqref{T12_calculated} into \eqref{Z12}, we manage to fix the following coefficients of the gluing matrix in the NLO
\begin{align}
& M_1^{(1)13}=-\frac{I_0}{g^2 I_2}\(i\(M_1^{(1)12}-M_1^{(1)34}\)+\pi\)\;, \\
& M_1^{(1)11}=\frac{\pi}{g^2 I_2}\sum\limits_{k=1}^{+\infty}I_{2k}(I_{2k+3}+I_{2k+1}-I_{2k-1}-I_{2k-3})\;, \notag \\
& M_1^{(1)33}=\frac{\pi}{g^2 I_2}\sum\limits_{k=1}^{+\infty}I_{2k}(I_{2k+1}-I_{2k-1})\;. \notag
\end{align}

\newpage


\begin{thebibliography}{10}





\bibitem{Gromov:2015wca}
  N.~Gromov, F.~Levkovich-Maslyuk and G.~Sizov,
  ``Quantum Spectral Curve and the Numerical Solution of the Spectral Problem in AdS5/CFT4,''
  JHEP {\bf 1606} (2016) 036
  doi:10.1007/JHEP06(2016)036
  [arXiv:1504.06640 [hep-th]].





\bibitem{Kotikov:2001sc}
  A.~V.~Kotikov and L.~N.~Lipatov,
  ``DGLAP and BFKL evolution equations in the N=4 supersymmetric gauge theory,''
  hep-ph/0112346.





\bibitem{Kotikov:2002ab}
  A.~V.~Kotikov and L.~N.~Lipatov,
  ``DGLAP and BFKL equations in the $N=4$ supersymmetric gauge theory,''
  Nucl.\ Phys.\ B {\bf 661} (2003) 19
   Erratum: [Nucl.\ Phys.\ B {\bf 685} (2004) 405]
  doi:10.1016/S0550-3213(03)00264-5, 10.1016/j.nuclphysb.2004.02.032
  [hep-ph/0208220].





\bibitem{Lipatov:1993yb}
  L.~N.~Lipatov,
  ``Asymptotic behavior of multicolor QCD at high energies in connection with exactly solvable spin models,''
  JETP Lett.\  {\bf 59} (1994) 596
   [Pisma Zh.\ Eksp.\ Teor.\ Fiz.\  {\bf 59} (1994) 571]
  [hep-th/9311037].





\bibitem{Kotikov:2000pm}
  A.~V.~Kotikov and L.~N.~Lipatov,
  ``NLO corrections to the BFKL equation in QCD and in supersymmetric gauge theories,''
  Nucl.\ Phys.\ B {\bf 582} (2000) 19
  doi:10.1016/S0550-3213(00)00329-1
  [hep-ph/0004008].




\bibitem{Kepka:2010hu}
  O.~Kepka, C.~Marquet and C.~Royon,
  ``Gaps between jets in hadronic collisions,''
  Phys.\ Rev.\ D {\bf 83} (2011) 034036
  doi:10.1103/PhysRevD.83.034036
  [arXiv:1012.3849 [hep-ph]].





\bibitem{Bartels:1980pe}
  J.~Bartels,
  ``High-Energy Behavior in a Nonabelian Gauge Theory (II) : First Corrections to $T_{n\to m}$ Beyond the Leading $\ln s$ Approximation,''
  Nucl.\ Phys.\ B {\bf 175} (1980) 365.
  doi:10.1016/0550-3213(80)90019-X





\bibitem{Kwiecinski:1980wb}
  J.~Kwiecinski and M.~Praszalowicz,
  ``Three Gluon Integral Equation and Odd c Singlet Regge Singularities in QCD,''
  Phys.\ Lett.\  {\bf 94B} (1980) 413.
  doi:10.1016/0370-2693(80)90909-0





\bibitem{Faddeev:1994zg}
  L.~D.~Faddeev and G.~P.~Korchemsky,
  ``High-energy QCD as a completely integrable model,''
  Phys.\ Lett.\ B {\bf 342} (1995) 311
  doi:10.1016/0370-2693(94)01363-H
  [hep-th/9404173].





\bibitem{Braun:1998id}
  V.~M.~Braun, S.~E.~Derkachov and A.~N.~Manashov,
  Phys.\ Rev.\ Lett.\  {\bf 81} (1998) 2020
  doi:10.1103/PhysRevLett.81.2020
  [hep-ph/9805225].



\bibitem{Belitsky:2004sf}
  A.~V.~Belitsky, G.~P.~Korchemsky and D.~Mueller,
  ``Integrability in Yang-Mills theory on the light cone beyond leading order,''
  Phys.\ Rev.\ Lett.\  {\bf 94} (2005) 151603
  doi:10.1103/PhysRevLett.94.151603
  [hep-th/0412054].





\bibitem{Belitsky:2005bu}
  A.~V.~Belitsky, G.~P.~Korchemsky and D.~Mueller,
  ``Integrability of two-loop dilatation operator in gauge theories,''
  Nucl.\ Phys.\ B {\bf 735} (2006) 17
  doi:10.1016/j.nuclphysb.2005.11.015
  [hep-th/0509121].





\bibitem{Belitsky:2006av}
  A.~V.~Belitsky, G.~P.~Korchemsky and D.~Mueller,
  ``Towards Baxter equation in supersymmetric Yang-Mills theories,''
  Nucl.\ Phys.\ B {\bf 768} (2007) 116
  doi:10.1016/j.nuclphysb.2007.01.024
  [hep-th/0605291].





\bibitem{Minahan:2002ve}
  J.~A.~Minahan and K.~Zarembo,
  ``The Bethe ansatz for N=4 superYang-Mills,''
  JHEP {\bf 0303} (2003) 013
  doi:10.1088/1126-6708/2003/03/013
  [hep-th/0212208].





\bibitem{Beisert:2010jr}
  N.~Beisert {\it et al.},
  ``Review of AdS/CFT Integrability: An Overview,''
  Lett.\ Math.\ Phys.\  {\bf 99} (2012) 3
  doi:10.1007/s11005-011-0529-2
  [arXiv:1012.3982 [hep-th]].





\bibitem{Gromov:2013pga}
  N.~Gromov, V.~Kazakov, S.~Leurent and D.~Volin,
  ``Quantum Spectral Curve for Planar $\mathcal{N} = 4$ Super-Yang-Mills Theory,''
  Phys.\ Rev.\ Lett.\  {\bf 112} (2014) no.1,  011602
  doi:10.1103/PhysRevLett.112.011602
  [arXiv:1305.1939 [hep-th]].





\bibitem{Gromov:2014caa}
  N.~Gromov, V.~Kazakov, S.~Leurent and D.~Volin,
  ``Quantum spectral curve for arbitrary state/operator in AdS$_{5}$/CFT$_{4}$,''
  JHEP {\bf 1509} (2015) 187
  doi:10.1007/JHEP09(2015)187
  [arXiv:1405.4857 [hep-th]].





\bibitem{Gromov:2017blm}
  N.~Gromov,
  ``Introduction to the Spectrum of $N=4$ SYM and the Quantum Spectral Curve,''
  arXiv:1708.03648 [hep-th].





\bibitem{Kazakov:2018ugh}
  V.~Kazakov,
  ``Quantum Spectral Curve of $\gamma$-twisted ${\cal N}=4$ SYM theory and fishnet CFT,''
  arXiv:1802.02160 [hep-th].





\bibitem{Kotikov:2007cy}
  A.~V.~Kotikov, L.~N.~Lipatov, A.~Rej, M.~Staudacher and V.~N.~Velizhanin,
  ``Dressing and wrapping,''
  J.\ Stat.\ Mech.\  {\bf 0710} (2007) P10003
  doi:10.1088/1742-5468/2007/10/P10003
  [arXiv:0704.3586 [hep-th]].





\bibitem{Gromov:2014bva}
  N.~Gromov, F.~Levkovich-Maslyuk, G.~Sizov and S.~Valatka,
  ``Quantum spectral curve at work: from small spin to strong coupling in $ \mathcal{N} $ = 4 SYM,''
  JHEP {\bf 1407} (2014) 156
  doi:10.1007/JHEP07(2014)156
  [arXiv:1402.0871 [hep-th]].





\bibitem{Alfimov:2014bwa}
  M.~Alfimov, N.~Gromov and V.~Kazakov,
  ``QCD Pomeron from AdS/CFT Quantum Spectral Curve,''
  JHEP {\bf 1507} (2015) 164
  doi:10.1007/JHEP07(2015)164
  [arXiv:1408.2530 [hep-th]].





\bibitem{Gromov:2015vua}
  N.~Gromov, F.~Levkovich-Maslyuk and G.~Sizov,
  ``Pomeron Eigenvalue at Three Loops in $\mathcal N=$ 4 Supersymmetric Yang-Mills Theory,''
  Phys.\ Rev.\ Lett.\  {\bf 115} (2015) no.25,  251601
  doi:10.1103/PhysRevLett.115.251601
  [arXiv:1507.04010 [hep-th]].





\bibitem{GromovKazakovunpublished}
  N.~Gromov and V.~Kazakov, unpublished.
  
  
  

\bibitem{Janik:2013nqa}
  R.~A.~Janik,
  ``Twist-two operators and the BFKL regime - nonstandard solutions of the Baxter equation,''
  JHEP {\bf 1311} (2013) 153
  doi:10.1007/JHEP11(2013)153
  [arXiv:1309.2844 [hep-th]].





\bibitem{Alfimov:presentation}
M.~Alfimov, {\it {QCD Pomeron with conformal spin from AdS/CFT Quantum Spectral
  Curve}},  29.02.2016.
\newblock GATIS Training Event, DESY, Hamburg,
  \url{https://indico.desy.de/getFile.py/access?contribId=3&resId=0&materialId=slides&confId=8491}.





\bibitem{Caron-Huot:2016tzz}
  S.~Caron-Huot and M.~Herranen,
  ``High-energy evolution to three loops,''
  JHEP {\bf 1802} (2018) 058
  doi:10.1007/JHEP02(2018)058
  [arXiv:1604.07417 [hep-ph]].





\bibitem{Marboe:2014gma}
  C.~Marboe and D.~Volin,
  ``Quantum spectral curve as a tool for a perturbative quantum field theory,''
  Nucl.\ Phys.\ B {\bf 899} (2015) 810
  doi:10.1016/j.nuclphysb.2015.08.021
  [arXiv:1411.4758 [hep-th]].





\bibitem{Marboe:2014sya}
  C.~Marboe, V.~Velizhanin and D.~Volin,
  ``Six-loop anomalous dimension of twist-two operators in planar $ \mathcal{N}=4 $ SYM theory,''
  JHEP {\bf 1507} (2015) 084
  doi:10.1007/JHEP07(2015)084
  [arXiv:1412.4762 [hep-th]].





\bibitem{Marboe:2017dmb}
  C.~Marboe and D.~Volin,
  ``The full spectrum of AdS5/CFT4 I: Representation theory and one-loop Q-system,''
  arXiv:1701.03704 [hep-th].





\bibitem{Gromov:2017cja}
  N.~Gromov, V.~Kazakov, G.~Korchemsky, S.~Negro and G.~Sizov,
  ``Integrability of Conformal Fishnet Theory,''
  JHEP {\bf 1801} (2018) 095
  doi:10.1007/JHEP01(2018)095
  [arXiv:1706.04167 [hep-th]].





\bibitem{Gromov:2016rrp}
  N.~Gromov and F.~Levkovich-Maslyuk,
  ``Quark-anti-quark potential in $ \mathcal{N} =$ 4 SYM,''
  JHEP {\bf 1612} (2016) 122
  doi:10.1007/JHEP12(2016)122
  [arXiv:1601.05679 [hep-th]].





\bibitem{Beccaria:2011uz}
  M.~Beccaria and G.~Macorini,
  ``Quantum folded string in $S^5$ and the Konishi multiplet at strong coupling,''
  JHEP {\bf 1110} (2011) 040
  doi:10.1007/JHEP10(2011)040
  [arXiv:1108.3480 [hep-th]].





\bibitem{Brower:2014wha} 
  R.~C.~Brower, M.~S.~Costa, M.~Djurić, T.~Raben and C.~I.~Tan,
  ``Strong Coupling Expansion for the Conformal Pomeron/Odderon Trajectories,''
  JHEP {\bf 1502}, 104 (2015)
  doi:10.1007/JHEP02(2015)104
  [arXiv:1409.2730 [hep-th]].





\bibitem{Gromov:2015dfa}
  N.~Gromov and F.~Levkovich-Maslyuk,
  ``Quantum Spectral Curve for a cusped Wilson line in $ \mathcal{N}=4 $ SYM,''
  JHEP {\bf 1604} (2016) 134
  doi:10.1007/JHEP04(2016)134
  [arXiv:1510.02098 [hep-th]].





\bibitem{Hegedus:2016eop}
  Á.~Hegedűs and J.~Konczer,
  JHEP {\bf 1608} (2016) 061
  doi:10.1007/JHEP08(2016)061
  [arXiv:1604.02346 [hep-th]].



\bibitem{Leurent:2013mr}
  S.~Leurent and D.~Volin,
  ``Multiple zeta functions and double wrapping in planar $N=4$ SYM,''
  Nucl.\ Phys.\ B {\bf 875} (2013) 757
  doi:10.1016/j.nuclphysb.2013.07.020
  [arXiv:1302.1135 [hep-th]].





\bibitem{Anselmetti:2015mda}
  L.~Anselmetti, D.~Bombardelli, A.~Cavaglià and R.~Tateo,
  ``12 loops and triple wrapping in ABJM theory from integrability,''
  JHEP {\bf 1510} (2015) 117
  doi:10.1007/JHEP10(2015)117
  [arXiv:1506.09089 [hep-th]].




\bibitem{Lee:2017mhh}
  R.~N.~Lee and A.~I.~Onishchenko,
  ``ABJM quantum spectral curve and Mellin transform,''
  arXiv:1712.00412 [hep-th].





\bibitem{Duhr:2014woa}
  C.~Duhr,
  ``Mathematical aspects of scattering amplitudes,''
  doi:10.1142/9789814678766\_0010
  arXiv:1411.7538 [hep-ph].





\bibitem{Costa:2012cb}
  M.~S.~Costa, V.~Goncalves and J.~Penedones,
  ``Conformal Regge theory,''
  JHEP {\bf 1212} (2012) 091
  doi:10.1007/JHEP12(2012)091
  [arXiv:1209.4355 [hep-th]].





\bibitem{Kotikov:2005gr}
  A.~V.~Kotikov and V.~N.~Velizhanin,
  ``Analytic continuation of the Mellin moments of deep inelastic structure functions,''
  hep-ph/0501274.





\bibitem{Kazakov:1987jk}
  D.~I.~Kazakov and A.~V.~Kotikov,
  ``Total $\alpha^- s$ Correction to Deep Inelastic Scattering Cross-section Ratio, R = $\sigma^-$l / $\sigma^-$t in {QCD}. Calculation of Longitudinal Structure Function,''
  Nucl.\ Phys.\ B {\bf 307} (1988) 721
   Erratum: [Nucl.\ Phys.\ B {\bf 345} (1990) 299].
  doi:10.1016/0550-3213(88)90106-X





\bibitem{Lopez:1980dj}
  C.~Lopez and F.~J.~Yndurain,
  ``Behavior at $x = 0, 1$, Sum Rules and Parametrizations for Structure Functions Beyond the Leading Order,''
  Nucl.\ Phys.\ B {\bf 183} (1981) 157.
  doi:10.1016/0550-3213(81)90551-4





\bibitem{Kotikov:2004er}
  A.~V.~Kotikov, L.~N.~Lipatov, A.~I.~Onishchenko and V.~N.~Velizhanin,
  ``Three loop universal anomalous dimension of the Wilson operators in $N=4$ SUSY Yang-Mills model,''
  Phys.\ Lett.\ B {\bf 595} (2004) 521
   Erratum: [Phys.\ Lett.\ B {\bf 632} (2006) 754]
  doi:10.1016/j.physletb.2004.05.078, 10.1016/j.physletb.2005.11.002
  [hep-th/0404092].





\bibitem{Blumlein:2009ta}
  J.~Blumlein,
  ``Structural Relations of Harmonic Sums and Mellin Transforms up to Weight w = 5,''
  Comput.\ Phys.\ Commun.\  {\bf 180} (2009) 2218
  doi:10.1016/j.cpc.2009.07.004
  [arXiv:0901.3106 [hep-ph]].





\bibitem{Ablinger:2010kw}
  J.~Ablinger,
  ``A Computer Algebra Toolbox for Harmonic Sums Related to Particle Physics,''
  arXiv:1011.1176 [math-ph].





\bibitem{Ablinger:2013hcp}
  J.~Ablinger,
  ``Computer Algebra Algorithms for Special Functions in Particle Physics,''
  arXiv:1305.0687 [math-ph].





\bibitem{Ablinger:2013cf}
  J.~Ablinger, J.~Blümlein and C.~Schneider,
  ``Analytic and Algorithmic Aspects of Generalized Harmonic Sums and Polylogarithms,''
  J.\ Math.\ Phys.\  {\bf 54} (2013) 082301
  doi:10.1063/1.4811117
  [arXiv:1302.0378 [math-ph]].





\bibitem{Ablinger:2011te}
  J.~Ablinger, J.~Blumlein and C.~Schneider,
  ``Harmonic Sums and Polylogarithms Generated by Cyclotomic Polynomials,''
  J.\ Math.\ Phys.\  {\bf 52} (2011) 102301
  doi:10.1063/1.3629472
  [arXiv:1105.6063 [math-ph]].





\bibitem{Remiddi:1999ew}
  E.~Remiddi and J.~A.~M.~Vermaseren,
  ``Harmonic polylogarithms,''
  Int.\ J.\ Mod.\ Phys.\ A {\bf 15} (2000) 725
  doi:10.1142/S0217751X00000367
  [hep-ph/9905237].





\bibitem{Vermaseren:1998uu}
  J.~A.~M.~Vermaseren,
  ``Harmonic sums, Mellin transforms and integrals,''
  Int.\ J.\ Mod.\ Phys.\ A {\bf 14} (1999) 2037
  doi:10.1142/S0217751X99001032
  [hep-ph/9806280].





\bibitem{Dokshitzer:2005bf}
  Y.~L.~Dokshitzer, G.~Marchesini and G.~P.~Salam,
  ``Revisiting parton evolution and the large-x limit,''
  Phys.\ Lett.\ B {\bf 634} (2006) 504
  doi:10.1016/j.physletb.2006.02.023
  [hep-ph/0511302].





\bibitem{Dokshitzer:2006nm}
  Y.~L.~Dokshitzer and G.~Marchesini,
  ``N=4 SUSY Yang-Mills: three loops made simple(r),''
  Phys.\ Lett.\ B {\bf 646} (2007) 189
  doi:10.1016/j.physletb.2007.01.016
  [hep-th/0612248].





\bibitem{Basso:2006nk}
  B.~Basso and G.~P.~Korchemsky,
  ``Anomalous dimensions of high-spin operators beyond the leading order,''
  Nucl.\ Phys.\ B {\bf 775} (2007) 1
  doi:10.1016/j.nuclphysb.2007.03.044
  [hep-th/0612247].





\bibitem{Alday:2015eya}
  L.~F.~Alday, A.~Bissi and T.~Lukowski,
  ``Large spin systematics in CFT,''
  JHEP {\bf 1511} (2015) 101
  doi:10.1007/JHEP11(2015)101
  [arXiv:1502.07707 [hep-th]].





\bibitem{Gribov:1972ri}
  V.~N.~Gribov and L.~N.~Lipatov,
  ``Deep inelastic e p scattering in perturbation theory,''
  Sov.\ J.\ Nucl.\ Phys.\  {\bf 15} (1972) 438
   [Yad.\ Fiz.\  {\bf 15} (1972) 781].





\bibitem{Gribov:1972rt}
  V.~N.~Gribov and L.~N.~Lipatov,
  ``e+ e- pair annihilation and deep inelastic e p scattering in perturbation theory,''
  Sov.\ J.\ Nucl.\ Phys.\  {\bf 15} (1972) 675
   [Yad.\ Fiz.\  {\bf 15} (1972) 1218].





\bibitem{Beccaria:2007bb}
  M.~Beccaria, Y.~L.~Dokshitzer and G.~Marchesini,
  ``Twist 3 of the sl(2) sector of N=4 SYM and reciprocity respecting evolution,''
  Phys.\ Lett.\ B {\bf 652} (2007) 194
  doi:10.1016/j.physletb.2007.07.016
  [arXiv:0705.2639 [hep-th]].





\bibitem{Beccaria:2009eq}
  M.~Beccaria, V.~Forini, T.~Lukowski and S.~Zieme,
  ``Twist-three at five loops, Bethe Ansatz and wrapping,''
  JHEP {\bf 0903} (2009) 129
  doi:10.1088/1126-6708/2009/03/129
  [arXiv:0901.4864 [hep-th]].





\bibitem{Beccaria:2009vt}
  M.~Beccaria and V.~Forini,
  ``Four loop reciprocity of twist two operators in N=4 SYM,''
  JHEP {\bf 0903} (2009) 111
  doi:10.1088/1126-6708/2009/03/111
  [arXiv:0901.1256 [hep-th]].
  




\bibitem{Lukowski:2009ce}
  T.~Lukowski, A.~Rej and V.~N.~Velizhanin,
  ``Five-Loop Anomalous Dimension of Twist-Two Operators,''
  Nucl.\ Phys.\ B {\bf 831} (2010) 105
  doi:10.1016/j.nuclphysb.2010.01.008
  [arXiv:0912.1624 [hep-th]].





\bibitem{Velizhanin:2010cm}
  V.~N.~Velizhanin,
  ``Six-Loop Anomalous Dimension of Twist-Three Operators in N=4 SYM,''
  JHEP {\bf 1011} (2010) 129
  doi:10.1007/JHEP11(2010)129
  [arXiv:1003.4717 [hep-th]].





\bibitem{Velizhanin:2010vw}
  V.~N.~Velizhanin,
  ``Vanishing of the four-loop charge renormalization function in N=4 SYM theory,''
  Phys.\ Lett.\ B {\bf 696} (2011) 560
  doi:10.1016/j.physletb.2011.01.019
  [arXiv:1008.2198 [hep-th]].





\bibitem{Velizhanin:2011pb}
  V.~N.~Velizhanin,
  ``Double-logs, Gribov-Lipatov reciprocity and wrapping,''
  JHEP {\bf 1108} (2011) 092
  doi:10.1007/JHEP08(2011)092
  [arXiv:1104.4100 [hep-th]].





\bibitem{Velizhanin:2013vla}
  V.~N.~Velizhanin,
  ``Twist-2 at five loops: Wrapping corrections without wrapping computations,''
  JHEP {\bf 1406} (2014) 108
  doi:10.1007/JHEP06(2014)108
  [arXiv:1311.6953 [hep-th]].





\bibitem{Marboe:2016igj}
  C.~Marboe and V.~Velizhanin,
  ``Twist-2 at seven loops in planar $ \mathcal{N} $ = 4 SYM theory: full result and analytic properties,''
  JHEP {\bf 1611} (2016) 013
  doi:10.1007/JHEP11(2016)013
  [arXiv:1607.06047 [hep-th]].





\bibitem{Basso:2011rs} 
  B.~Basso,
  ``An exact slope for AdS/CFT,''
  arXiv:1109.3154 [hep-th].





\bibitem{Gromov:2012eg} 
  N.~Gromov,
  ``On the Derivation of the Exact Slope Function,''
  JHEP {\bf 1302}, 055 (2013)
  doi:10.1007/JHEP02(2013)055
  [arXiv:1205.0018 [hep-th]].





\bibitem{Gromov:2012eu}
  N.~Gromov and A.~Sever,
  ``Analytic Solution of Bremsstrahlung TBA,''
  JHEP {\bf 1211} (2012) 075
  doi:10.1007/JHEP11(2012)075
  [arXiv:1207.5489 [hep-th]].





\bibitem{Beisert:2006ez}
  N.~Beisert, B.~Eden and M.~Staudacher,
  ``Transcendentality and Crossing,''
  J.\ Stat.\ Mech.\  {\bf 0701} (2007) P01021
  doi:10.1088/1742-5468/2007/01/P01021
  [hep-th/0610251].





\bibitem{Dorey:2007xn}
  N.~Dorey, D.~M.~Hofman and J.~M.~Maldacena,
  ``On the Singularities of the Magnon S-matrix,''
  Phys.\ Rev.\ D {\bf 76} (2007) 025011
  doi:10.1103/PhysRevD.76.025011
  [hep-th/0703104 [HEP-TH]].





\bibitem{Beisert:2006ib}
  N.~Beisert, R.~Hernandez and E.~Lopez,
  ``A Crossing-symmetric phase for AdS(5) x S**5 strings,''
  JHEP {\bf 0611} (2006) 070
  doi:10.1088/1126-6708/2006/11/070
  [hep-th/0609044].





\bibitem{Vieira:2010kb}
  P.~Vieira and D.~Volin,
  ``Review of AdS/CFT Integrability, Chapter III.3: The Dressing factor,''
  Lett.\ Math.\ Phys.\  {\bf 99} (2012) 231
  doi:10.1007/s11005-011-0482-0
  [arXiv:1012.3992 [hep-th]].





\bibitem{EZFace}
{\it \url{http://wayback.cecm.sfu.ca/projects/EZFace/}}.





\bibitem{Kotikov:2013xu}
  A.~V.~Kotikov and L.~N.~Lipatov,
  ``Pomeron in the N=4 supersymmetric gauge model at strong couplings,''
  Nucl.\ Phys.\ B {\bf 874} (2013) 889
  doi:10.1016/j.nuclphysb.2013.06.018
  [arXiv:1301.0882 [hep-th]].





\bibitem{Brower:2013jga}
  R.~C.~Brower, M.~Costa, M.~Djuric, T.~Raben and C.~I.~Tan,
  ``Conformal Pomeron and Odderon in Strong Coupling,''
  arXiv:1312.1419 [hep-ph].





\bibitem{Janik:2013pxa}
  R.~A.~Janik and P.~Laskoś-Grabowski,
  ``Approaching the BFKL Pomeron via integrable classical solutions,''
  JHEP {\bf 1401} (2014) 074
  doi:10.1007/JHEP01(2014)074
  [arXiv:1311.2302 [hep-th]].





\bibitem{Cavaglia:2018lxi}
  A.~Cavaglià, N.~Gromov and F.~Levkovich-Maslyuk,
  ``Quantum Spectral Curve and Structure Constants in N=4 SYM: Cusps in the Ladder Limit,''
  arXiv:1802.04237 [hep-th].





\bibitem{Bartels:2002au}
  J.~Bartels, M.~G.~Ryskin and G.~P.~Vacca,
  ``On the triple Pomeron vertex in perturbative QCD,''
  Eur.\ Phys.\ J.\ C {\bf 27} (2003) 101
  doi:10.1140/epjc/s2002-01089-x
  [hep-ph/0207173].

\end{thebibliography}
\bibliographystyle{JHEP}

\providecommand{\href}[2]{#2}\begingroup\raggedright\endgroup

\end{document}